\documentclass[pre,aps,floats,superscriptaddress,floatfix,twocolumn]{revtex4-1}
\usepackage{amssymb,amsmath}
\usepackage{amsmath,amssymb}
\usepackage{graphicx}
\usepackage{psfrag}
\usepackage{color}
\usepackage{soul}
\usepackage{dcolumn}
\usepackage[utf8]{inputenc}
\usepackage{hyperref}

\usepackage{bm}
\usepackage[normalem]{ulem}
 \hypersetup{
    colorlinks=true,
    citecolor=blue,
    }
 \usepackage{amsmath}

\begin{document}

 \title{Distributed fixed resources exchanging particles: Phases of an asymmetric exclusion process connected to two reservoirs}
 \author{Sourav Pal}\email{isourav81@gmail.com}
\affiliation{Theory Division, Saha Institute of
Nuclear Physics, A CI of Homi Bhabha National Institute, 1/AF Bidhannagar, Calcutta 700064, West Bengal, India}
\author{Parna Roy}\email{parna.roy14@gmail.com}
\affiliation{Shahid Matangini Hazra Government General Degree College for Women, Purba Medinipore 721649, West Bengal, India}
\author{Abhik Basu}\email{abhik.123@gmail.com, abhik.basu@saha.ac.in}
\affiliation{Theory Division, Saha Institute of
Nuclear Physics, a CI of Homi Bhabha National Institute, 1/AF Bidhannagar, Calcutta 700064, West Bengal, India}

\begin{abstract}
 We propose and study a conceptual one-dimensional model to explore how the combined interplay between fixed resources  with unlimited carrying capacity and particle exchanges between different parts of an extended system can affect the stationary densities in a current carrying channel connecting different parts of the system.  Our model is composed of a totally asymmetric simple exclusion process (TASEP) connecting two particle reservoirs without any internal dynamics, but which can directly exchange particles between each other, ensuring nonvanishing currents in the steady-states. The total particle number in the system that defines the ``resources'' available, although held conserved by the model dynamics, can take any value giving unrestricted capacity.  We show how the resulting phase diagrams of the model are controlled by relevant the parameters,  together with the total available resources. These control parameters can be tuned to make the density on the TASEP lane globally uniform or piecewise continuous with localized domain walls, and can also control populations of the two reservoirs. In general, the phase diagrams are quite different from a TASEP with open boundaries. However, in the limit of large amount of resources, the phase diagrams in the plane of the control parameters become topologically identical to that for an open TASEP along with delocalization of domain walls.

\end{abstract}

\maketitle

 \section{INTRODUCTION}
 \label{introduction}

 %Far-from-equilibrium phenomena are ubiquitous in nature, which, in contrast to their equilibrium counterparts, are yet to be fully explored. Studies on driven diffusive systems are fundamentally important to understand the rich behavior of low dimensional collective particle transport in nonequilibrium systems.
 %A particularly important model is the
 Totally Asymmetric Simple Exclusion Process (TASEP), originally proposed to study protein synthesis in biological cells~\cite{macdonald}, has subsequently been reinvented as a paradigmatic one-dimensional (1D) open model that shows boundary-induced phase transitions~\cite{krug,krug1,krug2}.
 It consists of a 1D lattice with $L$ sites having open boundaries. The dynamics is stochastic in nature, and involves unidirectional particle hopping, subject to exclusion at all the sites, i.e., any site can contain at most one particle at a time. In a TASEP, a particle enters at  the left boundary ($i=1$) of the 1D lattice at a given rate $\alpha_T$,  hops unidirectionally to the following sites at rate unity subject to exclusion until it reaches the last site at the other end ($i=L$), from which it leaves the system at another specified rate $\beta_T$. Controlled by $\alpha_T$ and $\beta_T$, the TASEP exhibits three distinct phases: the steady-state densities in bulk can be either less than half, which is the low-density (LD) phase, or more than half, forming the high-density (HD) phase, or just half, which is the maximal current (MC) phase.  Furthermore, for parameters on the coexistence line in the $\alpha_T-\beta_T$ plane,  separating the LD and HD phases,  nonuniform densities in the form of a domain wall (DW) that is delocalized, can be found~\cite{tasep-rev}. %This DW is {\em delocalized}, a property that is attributed to the particle nonconservering dynamics governing the open TASEP~\cite{tasep-rev}.

 { Inspired by hosts of phenomena across biological and socially relevant systems, many generalizations of TASEP have been proposed and studied, leading to the discovery of novel phases and collective phenomena.
%~\cite{erwin-lk-prl,erwin-lk-pre,erwin-bottleneck} and close geometries~\cite{tirtha-lk1,tirtha-lk2}.
For instance, it has been discovered that  exchange of particles between the TASEP lane and the embedding environment gives rise to traffic jams~\cite{erwin-lk-prl,erwin-lk-pre}. It was further shown how molecular motors from the kinesin-8 family can regulate microtubule depolymerization dynamics and ultimately lead to length control of microtubules~\cite{melbing1,melbing2}.
Furthermore, unexpected and novel behaviors like domain wall delocalization and traffic jams are discovered in systems with two TASEP lanes, which are coupled through rare particle-switching events~\cite{tobias-ef1,tobias-ef2,tobias-ef3}.
Motivated by the dynamics of molecular motors along complex networks of microtubules in biological cells~\cite{howard}, TASEP has been explored on simple networks, studying the conditions for the existence of the various phases and domain walls~\cite{andrea1,rakesh1,rakesh2} on various segments of the networks.} 
 A  TASEP  on a ring%on a uniform ring necessarily has a constant mean density, owing to the translational invariance of the system. Macroscopically nonuniform densities can be expected only when the translational invariance is broken. This can be achieved by, 
 can display macroscopically nonuniform stationary densities in the presence of inhomogeneities, point~\cite{lebo,niladri-tasep} or extended~\cite{mustansir,hinsch,tirtha-niladri,parna-anjan,tirtha-qkpz}, or a combination of multiple defects of various kinds~\cite{sm-atri3}. How rare particle nonconserving events can couple with an inhomogeneous TASEP on a ring to produce space-dependent stationary densities and phase-coexistence has been studied in Refs.~\cite{tirtha-lk1,tirtha-lk2}.

  A particularly interesting issue is the consequence of coupling TASEP with diffusion, an equilibrium process. For instance, the consequences of the effects of diffusion on the steady-states of the filament (TASEP) has been explored in Refs.~\cite{klump-PRL,klump-JSTAT,ciandrini-2019,dauloudet-2019}, by generally modeling the system as a filament (microtubule) executing TASEP confined in a three-dimensional (3D) particle reservoir representing the cell cytoplasm, where the motors diffuse around. In an approach that is complementary to these 3D model studies, a two-lane system has been considered with one lane being a TASEP and the other executing simple exclusion process (SEP), a 1D representation of diffusion. This has been considered in closed~\cite{hinsch,parna-anjan}, half-open~\cite{frey-graf-PRL,bojer-graf-frey-PRE} and open geometries~\cite{sm}, each revealing the nontrivial steady-states borne by the interplay between TASEP and SEP. Further, Ref.~\cite{klump-PRE} studied a 1D open model with diffusive and driven segments connected serially, and explored the resulting modifications of the pure open TASEP phase diagram.

  In this work, we study a TASEP model   connected to a particle reservoir at each end, exchanging particles directly between them which ensures a finite stationary current in the system. The total number of particles in the combined system is conserved. The finite availability of particles effectively limits the particle supply to the TASEP lane, significantly affecting its phase behavior, making it different from a conventional open or a periodic TASEP.
See, e.g., Refs.~\cite{reser1,reser2,reser3} for similar earlier works.
%More than one
%TASEP connected to a reservoir has also been considered; see
%Ref.~\cite{reser3}.
These studies were motivated by the biological processes of protein synthesis in cells~\cite{reser1} and
also in the context of traffic~\cite{traffic1}; see also Ref.~\cite{limited}
for a similar study. Detailed studies, both numerical Monte Carlo simulation (MCS) and
analytical mean-field theory (MFT) reveal rich nonuniform steady-state density profiles including domain walls in these
models~\cite{reser1,reser2,reser3}.  %Notable experimental studies relevant to these model
%systems include studies on spindles in eukaryotic cells~\cite{lim-exp1,lim-exp2}.

%Although significant progress has been made in the research in TASEPs with finite resources, there are still unresolved questions. 
{ A key unresolved issue in this field is how the reservoir-TASEP couplings and the associated constraints on the model affect the ensuing steady states and phase diagrams of the TASEP.}    In a recent related study~\cite{sourav-1}, not only the system had finite resources, the reservoirs have finite carrying capacity as well, i.e., each reservoir can accommodate a fixed maximum number of particles, controlled by  the specific reservoir-TASEP couplings.  A consequence of this is that the model in Ref.~\cite{sourav-1} {\em never} reduces to an ordinary TASEP for any ``resources'' or total particle number.  Formally, the model in Ref.~\cite{sourav-1} is effectively a {\em two-constraint} model, with the finite resources and finite capacities being the two constraints. A pertinent question is the generality and degree of robustness of these results, and the role of constraints on the steady-states of the model. For instance, what happens when one of the constraints is lifted? Broadly speaking, these issues are the analogues of symmetry and universality in models for continuous transitions at critical points.  To that end, we construct and study a conceptual model whose dynamics necessarily {\em conserves the total particle number}, but has {\em no upper limit} on the admissible total particle number for any value of the dynamical model parameters, i.e., has {\em no bound} on the carrying capacity.

 {   Our model (i) strictly conserves the total particle number (covering the TASEP lane and the two reservoirs), but (ii) { the reservoir-TASEP couplings are such that it} can accommodate any number of particles (including a diverging number of particles approaching infinity) for {\em any values} of the model parameters which define the dynamical update rules.  Thus it has {\em only} one constraint, {\em viz.}, the global particle number conservation, but has no restrictions on the maximum number of particles that can be accommodated by the model, or has an {\em unrestricted capacity}, in contrast to previous studies~\cite{sourav-1}. This has a strong consequence: in the present study, the phase diagram of our model smoothly reduces to one that is topologically identical to that for an open TASEP in the limit of large (formally diverging) number of total particles for {\em any} finite choices of the model parameters that define the particle exchanges between the reservoirs, a feature absent in   the model in Ref.~\cite{sourav-1}. Lastly, with increasing total available particle number or ``resources'', a localized DW in this model gradually delocalizes, again in contrast to the results in Ref.~\cite{sourav-1}.} We study our model in two different limits: (i) when the particle exchanges between the reservoirs compete with the hopping along the TASEP, called weak coupling limit, and (ii) when the former overwhelms the latter, designated as the strong coupling limit of the model.

The remaining parts of this article are organized as follows. In Sec.~\ref{Model}, we define our model and give the dynamical update rules. It is then followed by setting up a mean-field analysis of the steady-state densities in the weak coupling limit given in Sec.~\ref{steady-state density model N0}. Next, in Sec.~\ref{mean-field phase diagrams} phase diagrams are obtained using analytical mean-field theory corroborated by Monte Carlo simulation. The steady-state densities and phase boundaries are derived in Sec.~\ref{ssd and pb}; wherein in Sec.~\ref{dw and pt} the nature of the phase transitions of the model is discussed. We summarize the findings in Sec.~\ref{conclusions}. Our results on the strong coupling limit are given in Appendix~\ref{mft strong coupling}.

 \section{THE MODEL}
 \label{Model}

 Our model consists of a 1D lattice $T$ having $L$ sites, connected to two particle reservoirs $R_{1}$ and $R_{2}$ at both ends. The reservoirs are considered to be point reservoirs without any internal structure or dynamics. The hard-core exclusion principle enforces that each site of $T$ can hold a maximum of one particle at any given moment. The lattice $T$ executes TASEP dynamics~\cite{sourav-1}. Provided the first site of $T$ is empty, a particle from $R_{1}$ enters $T$ with a rate $\alpha_\text{eff}$ that depends on the instantaneous population of $R_{1}$. The particle then hops clockwise to the next available site. The hopping rate in the bulk of $T$ is taken to be unity to set the time scale. Eventually, after reaching the last site of $T$, the particle jumps into $R_{2}$ with a rate $\beta_\text{eff}$ that depends on the instantaneous population of $R_{2}$. The closed geometry of the model ensures the overall particle number (including both the TASEP lane and reservoirs), $N_{0}$, to be conserved. This can be expressed as
 \begin{equation}
  N_{0}=N_{1}+N_{2}+\sum_{i=1}^{L}n_{i},
 \end{equation}
 where $N_{1},N_{2}$ are the number of particles in $R_{1},R_{2}$ respectively, and $n_{i}$'s are binary variables denoting the occupation number of sites $i=1,2,...,L$ in the TASEP lane. We choose  $n_{i}=0$ for vacant sites and $n_{i}=1$ for occupied ones. The dynamical dependence of effective rates $\alpha_\text{eff}$ and $\beta_\text{eff}$ on the instantaneous populations $N_{1}$ and $N_{2}$ of the reservoirs $R_{1}$ and $R_{2}$ are expressed as following:
 \begin{equation}
 \label{effective entry and exit rates}
 \alpha_\text{eff}=\alpha f(N_{1}); \hspace{5mm} \beta_\text{eff}=\beta g(N_{2}),
 \end{equation}
  where $f$ and $g$ are functions of $N_{1}$ and $N_{2}$ respectively. Inflow of particles from $R_1$ to $T$ increases with the rising population of $R_{1}$, whereas the outflow of particles from $T$ to $R_2$ should be hindered by the growing population in $R_{2}$. This suggests that $f$ and $g$ should be monotonically increasing and decreasing functions of $N_{1}$ and $N_{2}$ respectively. We consider the rate functions $f$ and $g$ to have the following limiting values: (i) $f(0)=0$, (ii) $f(\infty)=1$, (iii) $g(0)=1$, and (iv) $g(\infty)=0$. In general, the rate functions $f$ and $g$ have values between 0 and 1. Furthermore, the functions $f$ and $g$ should not impose any constraint on the upper limit on $N_0$. To proceed further for the subsequent explicit calculations,
  we choose the following simple functions which display the above properties:
 \begin{equation}
 \label{Model 1}
 f(N_{1})=\frac{N_{1}}{N_{0}}, \hspace{5mm} g(N_{2})=1-\frac{N_{2}}{N_{0}}.
 \end{equation}
 { Similar functions have been used elsewhere before; see, e.g., Refs.~\cite{astik-parna,arvind1,arvind2,arvind3}. Notice that the choices in (\ref{Model 1}) are {\em different} from those used in Refs.~\cite{reser1,reser2,reser3,sourav-1,erwin}. The rate functions in the latter studies have the TASEP size $L$ appearing as a normalization constant, whereas in the present study as   well as those in Refs.~\cite{astik-parna,arvind1,arvind2,arvind3}, $N_0$ appears as the normalization constant.  While this may appear as a minor technical matter, this is actually significant: Choice (\ref{Model 1}) {\em does not} impose any constraints on the carrying capacity, unlike in Ref.~\cite{sourav-1}.} %Indeed, as
 %our subsequent analyses show that depending upon these choices, the basic properties of the model and the results that follow can change significantly.}
 \begin{figure}[!h]
 \centerline{
 \includegraphics[width=\linewidth]{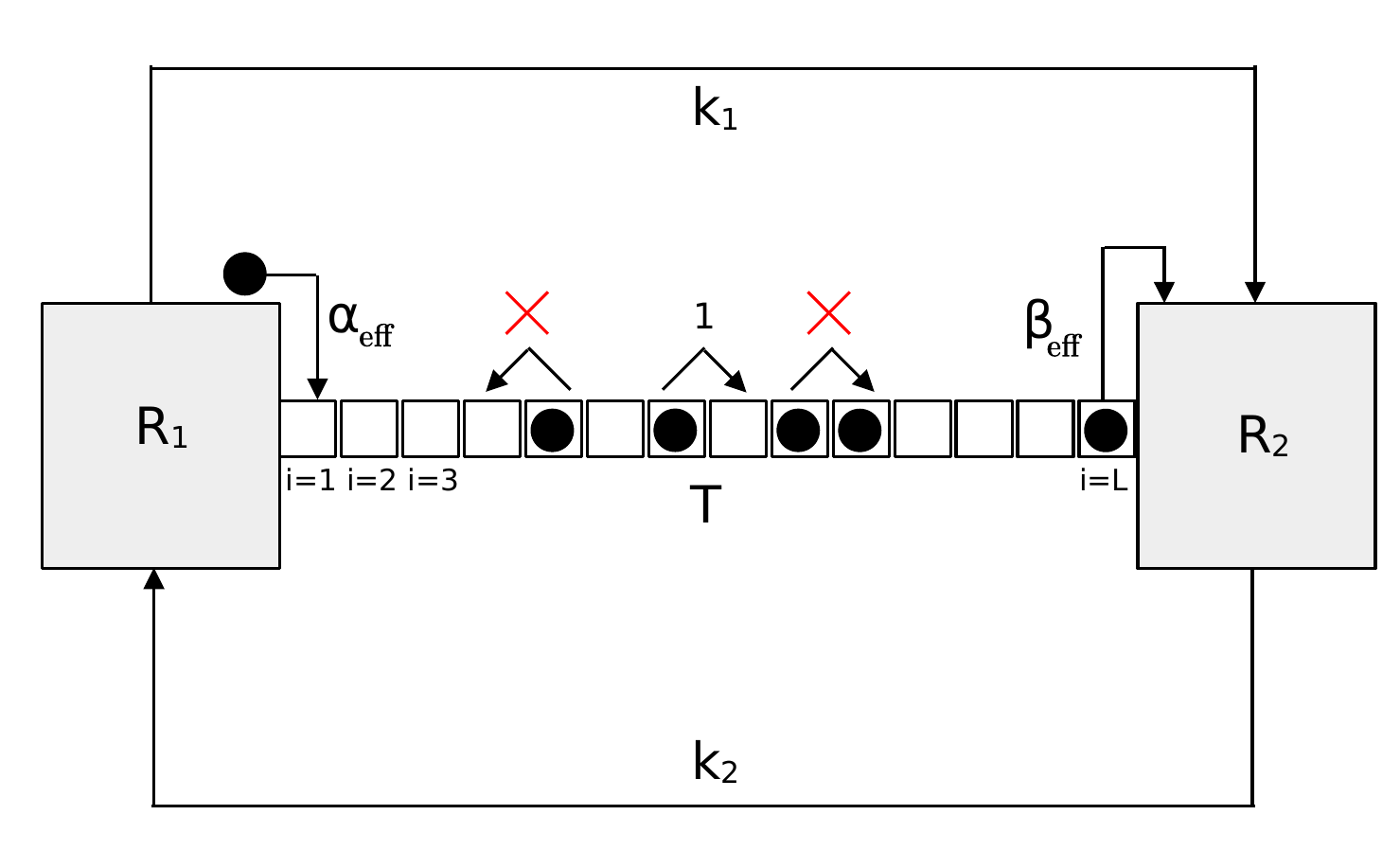}}
 \caption{Schematic diagram of the model: Here, $R_{1}$ and $R_{2}$ are two particle reservoirs, both connected to a TASEP lane $T$ with $L$ sites at its both ends. Particles, represented by black solid circles, enter the first site ($i=1$) of $T$ from $R_{1}$ with a rate $\alpha_\text{eff}$, followed by hopping to the sites on the right side with a rate of unity, subject to exclusion. Eventually, particles exit from the last site ($i=L$) of $T$ to go to $R_{2}$ with a  rate of $\beta_\text{eff}$. Furthermore, particle exchanges occur directly between $R_{1}$ and $R_{2}$ with rates $k_{1}$ and $k_{2}$, modeling diffusion in the system (see text).}
 \label{Schematic diagram of the model.}
 \end{figure}

 In our model, the upper limit on each of $N_1,\,N_2$ can be $N_0$, the total particle number, or the ``total available resources'' in the system. Further, $N_0$ itself is {\em unrestricted}, i.e., $N_0$ can take {\em any} value including approaching infinity. Thus $N_{1}$ and $N_{2}$ can also grow indefinitely, independent of the other model parameters. This is in stark contrast to the model studied in Ref.~\cite{sourav-1}. In addition to particle hopping motions in $T$, the reservoirs can directly exchange particles between them with certain rates to maintain a nonzero current in $T$: $R_{1}(R_{2})$ releases particles at rate $k_{1}(k_{2})$ which $R_{2}(R_{1})$ receives instantaneously. We define a parameter $\mu=N_{0}/L$ to denote the availability of particles in the system. Since the total particle number, $N_{0}$, has no upper limit, $\mu$ can be any positive real number: $0 \le \mu < \infty$. Taken together, our model has five control parameters: $\alpha$, $\beta$, $\mu$, $k_{1}$, and $k_{2}$.%, same as the model studied in Ref.~\cite{sourav-1}.

 \section{MEAN-FIELD THEORY}
 \label{steady-state density model N0}

 We now set up the mean-field theory (MFT) for our model by following the same line of reasoning as in previous studies~\cite{blythe}. MFT involves neglecting the correlation effects and replacing averages of products by the products of averages. The  ensuing MFT equations take the same form as those in Ref.~\cite{sourav-1}, which we revisit here for the convenience of the readers. Considering the density $\rho_{i}$  as the time-average of occupation number $n_{i}$ at site $i$, $\rho_{i} \equiv \langle n_{i} \rangle$, we have the following equations of motion for the bulk sites $1<i<L$ in MFT:
 \begin{eqnarray}
  \label{bulk}
  &&\frac{d\rho_i}{dt}=\rho_{i-1}(1-\rho_{i})-\rho_{i}(1-\rho_{i+1}).
  \end{eqnarray}
 %For the boundary sites ($i=1$ and $i=L$), we have
 %\begin{eqnarray}
 % &&\frac{d\rho_1}{dt}=\alpha_\text{eff}(1-\rho_{1})-\rho_{1}(1-\rho_{2}), \label{first site}\\
 % &&\frac{d\rho_L}{dt}=\rho_{L-1}(1-\rho_{L})-\beta_\text{eff}\rho_{L}. \label{last site}
 %\end{eqnarray}
 Similarly, the time-evolution of reservoir populations $N_{1}$ and $N_{2}$ are represented by the following equations:
 \begin{eqnarray}
  &&\frac{dN_1}{dt}=k_{2}N_{2}-k_{1}N_{1}-J_{T-in}, \label{mft4}\\
  &&\frac{dN_2}{dt}=k_{1}N_{1}-k_{2}N_{2}+J_{T-out}, \label{mft5}
   \end{eqnarray}
where $J_{T-in}=\alpha_\text{eff}(1-\rho_1)$ and $J_{T-out}=\beta_\text{eff}\rho_L$ are, respectively, the incoming and outgoing currents in the TASEP.

 %\begin{eqnarray}
 % &&n_{i} \leftrightarrow 1-n_{L-(i-1)}, \label{rho tr}\\
 % &&\alpha_\text{eff} \leftrightarrow \beta_\text{eff} \label{a b tr}
 %\end{eqnarray}

 %\noindent

 %To proceed further,  we now assume unit geometric length of the lattice without any loss of generality and
 Now, introduce a quasi-continuous variable $x=i/L \in [0,1]$ where $\epsilon\equiv 1/L$ is the lattice constant~\cite{erwin-lk-prl,sourav-1}. The density at a position $x$ at time $t$ is denoted by $\rho(x,t)\equiv \rho_i(t)$. %As shown in Ref.~\cite{sourav-1}, 
 The discrete equation of motion [Eq.~(\ref{bulk})] takes a conservation law form,
%\begin{eqnarray}
%  &&\frac{\partial \rho(x,t)}{\partial t}=-\epsilon\partial_x(\rho(x)[1-\rho(x)])+ {\cal O}(\epsilon^2), \label{mft1-x}
 % \end{eqnarray}
 % supplemented by the boundary conditions $\rho(0)= \alpha_\text{eff},\,\rho(1)=1-\beta_\text{eff}$. Equation~(\ref{mft1-x}) %has a conservation law form, reflecting the conserving nature of the TASEP dynamics in the bulk, and thus
  %allows us to 
  giving  a mean-field particle current $J_T$ in the bulk~\cite{niladri-tasep} %that in the MF has the form
  \begin{equation}
   J_T=\rho(1-\rho),\label{tasep-curr}
  \end{equation}
which becomes a constant in the steady-state, along with $J_{T-in}=J_{T-out}=J_T$ in the steady-states, supplemented by the boundary conditions $\rho(0)= \alpha_\text{eff},\,\rho(1)=1-\beta_\text{eff}$.

 %\noindent
 In the nonequilibrium steady-states, the reservoir populations are constants over time in MFT. %Hence,
 %\begin{equation}
 % \label{n1-n2-const}
 % \frac{dN_{1}}{dt}=\frac{dN_{2}}{dt}=0.
 %\end{equation}
 %Substituing Eq.~(\ref{n1-n2-const}) in
 Then using Eqs.~(\ref{mft4}) and (\ref{mft5}), we get
 \begin{equation}
 \label{Flux balance equation}
 k_{2}N_{2}=k_{1}N_{1}+J_{T}.%;
 \end{equation}
 %see also Ref.~\cite{sourav-1}, where a similar equation is obtained.
 %Eq.~(\ref{Flux balance equation}) is a flux balance equation relating the particle current $J_{T}$ with the reservoir populations $N_{1}$ and $N_{2}$.

 %In the steady-state, $\partial\rho/\partial t=0$, yielding the mean-field particle current in the bulk
 %\begin{equation}
 % \label{current-density}
 % J_{T}=\rho(1-\rho).
 %\end{equation}
 %to be constant, see Eq. (\ref{mft1-x}). Therefore, for a given current $J_{T}$,
 %As discussed in Ref.~\cite{sourav-1},
 The solution of Eq.~(\ref{tasep-curr}) gives the steady-state densities
 \begin{equation}
  \rho=\frac{1}{2}\left(1\pm \sqrt{1-4J_{T}}\right):=\rho_\pm.
 \end{equation}
 This gives a low-density (LD) phase with $\rho=\rho_{-}\le 1/2$ and a high-density (HD) phase with $\rho=\rho_{+}>1/2$. Additionally, a maximum current (MC) phase can arise with $J_{T}=1/4$ when $\rho=1/2$. Further,  macroscopic nonuniformity in density in the form of a domain wall connecting the LD and HD domains can be formed, giving the domain wall (DW) phase or a shock phase.

 When solved for $N_{1}$ and $N_{2}$, Eq.~(\ref{Flux balance equation}) along with particle number conservation, $N_{0}=N_{1}+N_{2}+N_{T}$, where $N_{T}\equiv L\int_0^1 \rho(x)\,dx$ is the number of particles in the TASEP lane, gives~\cite{sourav-1}
 \begin{subequations}
 \begin{equation}
 \label{N1}
 N_{1}=\frac{k_{2}(N_{0}-N_{T})-J_{T}}{k_{1}+k_{2}},
 \end{equation}
 \begin{equation}
 \label{N2}
 N_{2}=\frac{k_{1}(N_{0}-N_{T})+J_{T}}{k_{1}+k_{2}}.
 \end{equation}
 \end{subequations}
 Thus, $N_1,\,N_2$ depend upon $J_T$ explicitly, and also implicitly through the dependence of $N_0$ on $J_T$.

 We consider two distinct cases. We have $J_{T}<1$ ($J_\text{max}=1/4$), whereas $N_{1}$ and $N_{2}$ rise with $L$, since $N_0,\,N_T$ scale with $L$ for a given $\mu$. Thus, if $k_{1,2} \sim {\cal O}(1)$, we can neglect $J_{T}$ safely in Eq.~(\ref{Flux balance equation}) in the thermodynamic limit (TL). Similarly, we note that in Eqs.~(\ref{N1}) and (\ref{N2}), where $J_{T}<1$, but $N_{0}$ and $N_{T}$ should rise with $L$ indefinitely. Thus, if $k_{1,2} \sim {\cal O}(1)$, we can neglect $J_{T}$ in TL, giving $k_{1}N_{1}=k_{2}N_{2}$ asymptotically exact, independent of $J_{T}$. Therefore, in TL, the relative population of the two reservoirs is controlled by just the ratio $k_{2}/k_{1}$, and independent of $J_{T}$, $\alpha$, and $\beta$. This effectively eliminates one of the reservoirs, enabling us to describe the steady-state in terms of an effective single reservoir.  This is the diffusion dominated regime, named \textit{strong coupling limit} below. The more interesting case, named \textit{weak coupling limit}, is where diffusion competes with hopping. This is achieved by introducing a scaling $k_{1(2)}=k_{10(20)}/L$ where $k_{10,20} \sim \mathcal{O}(1)$.
 In this case, the general solutions for the reservoir populations $N_1$ and $N_2$ in the MFT are given by
\begin{subequations}
 \begin{equation}
 \label{N1w}
 N_{1}=\frac{k_{20}(N_{0}-N_{T})-LJ_{T}}{k_{10}+k_{20}},
 \end{equation}
 \begin{equation}
 \label{N2w}
 N_{2}=\frac{k_{10}(N_{0}-N_{T})+LJ_{T}}{k_{10}+k_{20}},
 \end{equation}
 \end{subequations}
 revealing their explicit $J_T$-dependence, which do not vanish in the limit $L\rightarrow \infty$, since both $N_0,\,N_T$ also scale linearly with $L$ for a fixed $\mu$. %{\bf why so? we need to say something here}
 This ensures competition between $J_T$ and the other contributions in (\ref{N1w}) and (\ref{N2w}). In the main text of the paper, phase diagrams and density profiles in weak coupling limit are discussed elaborately. We discuss the strong coupling limit case in Appendix~\ref{mft strong coupling}. Although these classifications of weak and strong coupling cases of our model are same as those in Ref.~\cite{sourav-1}, nonetheless, we show below that the results from the present study are significantly different from those obtained in Ref.~\cite{sourav-1} due to { the unrestricted carrying capacities of the reservoirs in the present problem}.

 %\vspace{2mm}

\section{PHASE DIAGRAMS IN THE WEAK COUPLING CASE}
\label{mean-field phase diagrams}

\begin{figure*}[htb]
 \includegraphics[width=\linewidth]{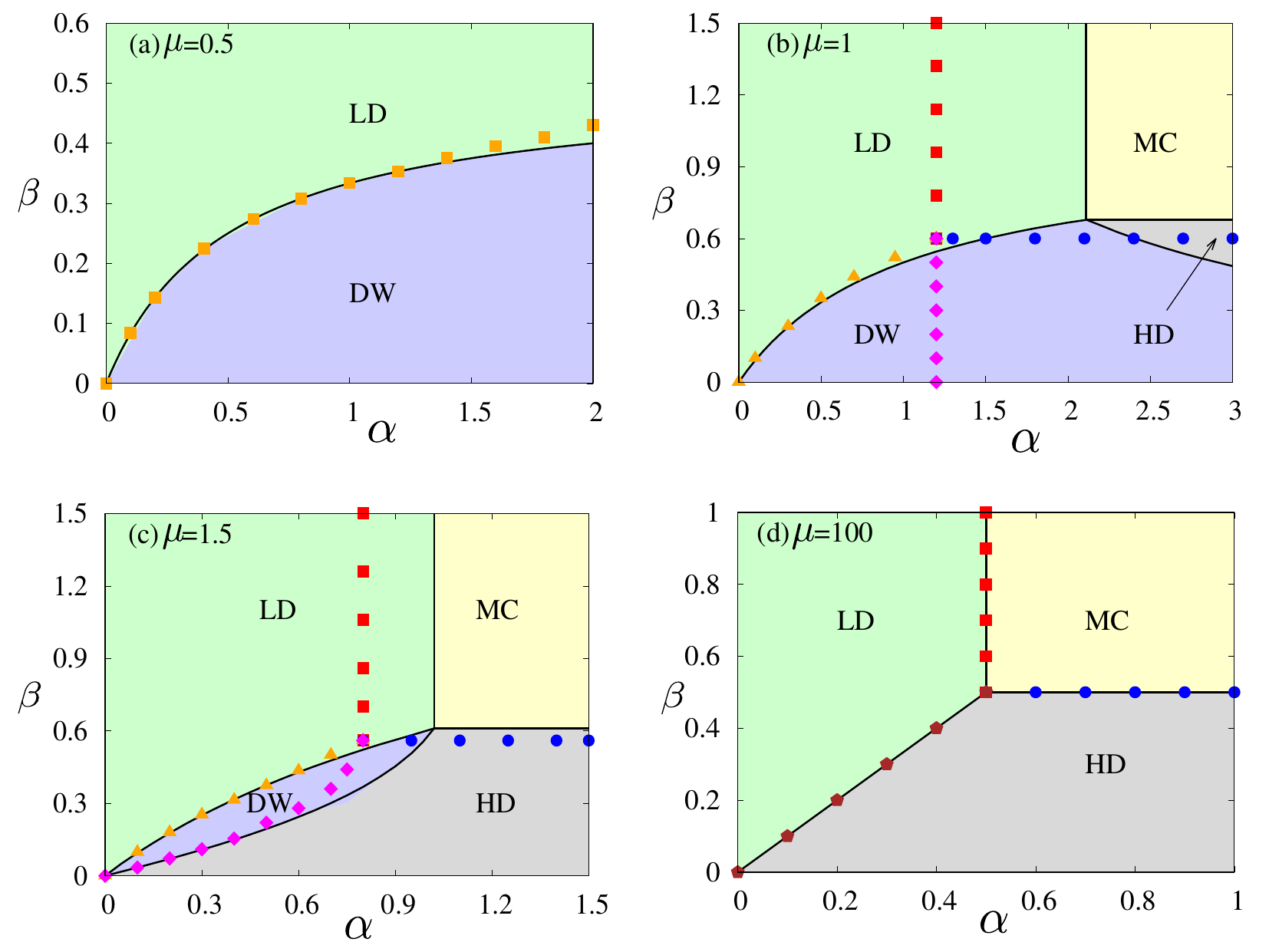}
 \\
\caption{Phase diagrams in the $\alpha-\beta$ plane with different values of the filling factor $\mu$, (a) $\mu=0.5$, (b) $\mu=1$, (c) $\mu=1.5$, and (d) $\mu=100$, in the weak coupling limit of the model and with fixed particle exchange rates  $k_{10}=0$ and $k_{20}=0.95$. The LD, HD, MC, and DW phases are denoted respectively by green, gray, yellow, and blue regions with solid black lines as the phase boundaries according to MFT; see Eqs.~(\ref{ld-mc boundary model 2 weak}), (\ref{hd-mc boundary model 2 weak}), (\ref{ld-sp boundary model 2 weak}), and (\ref{hd-sp boundary model 2 weak}). The phase boundaries obtained from MCS are represented by discrete colored points: red (LD-MC), blue (HD-MC), orange (LD-DW), and magenta (HD-DW). Noticeable deviations between the theoretical and simulated phase diagrams are observed for intermediate values of $\mu$. Either two or four phases can be simultaneously observed for small or intermediate values of $\mu$. Interestingly, for large values of $\mu$, e.g., $\mu=100$ in Fig.~\ref{phase-diagram-weak-coupling}(d), the phase diagram of our model exhibits a structure similar to an open TASEP without any noticeable discrepancies between MFT and MCS results, with LD, HD, and MC phase regions and the DW phase region contracts to an inclined line.}
\label{phase-diagram-weak-coupling}
\end{figure*}
 %\vspace{2mm}

 Before presenting the phase diagrams of the present model, we briefly revisit the usual open TASEP model~\cite{blythe}. %For a given combination of entry ($\alpha_T$) and exit ($\beta_T$) rates, the mean-field behavior of our model is characterized by mapping it onto a corresponding open model with appropriately adjusted effective rates.
 The phase diagram of an open boundary TASEP, with $\alpha_T,\,\beta_T<1$ as the entry and exit rates which are the control parameters, exhibits three distinct phases: an entry-limited low-density (LD) phase with bulk density $\rho_\text{LD} < 1/2$, an exit-limited high-density (HD) phase with bulk density $\rho_\text{HD}>1/2$, and a bulk-limited maximal current (MC) phase with bulk density $\rho_\text{MC}=1/2$.   Conditions for the occurrence of these phases are as follows: for the LD phase, we have $\rho_\text{LD}=\alpha_\text{T}<\beta_\text{T}$ and $\alpha_\text{T}<1/2$; for the HD phase, we have $1-\rho_\text{HD}=\beta_\text{T}<\alpha_\text{T}$ and $\beta_\text{T}<1/2$, and for the MC phase, we have $\alpha_\text{T},\beta_\text{T} \ge 1/2=\rho_\text{MC}$. On the coexistence line between (0,0) and (1/2,1/2), which corresponds to points with $\alpha_\text{T}=\beta_\text{T}<1/2$, the bulk density in $T$ exhibits a piecewise discontinuity, forming a domain wall (DW) that connects the LD and HD domains together. This domain wall spans  the entire lattice instead of being confined to a specific region and hence is delocalized. This is attributed to the particle number nonconservation in the model due to the stochastic entry-exit events. The conditions for transitions between different phases in an open TASEP can be obtained by equating the currents in the respective phases. Specifically, the transition from LD to HD phases occurs when $\alpha_\text{T}=\beta_\text{T}<1/2$. Likewise, the transitions from LD to MC and from HD to MC occur when $\alpha_\text{T}=1/2<\beta_\text{T}$ and $\beta_\text{T}=1/2<\alpha_\text{T}$, respectively. In the $\alpha_T-\beta_T$ plane, the three phases intersect at point $(1/2,1/2)$. The LD-HD transition manifests as a sudden change in the bulk density, indicating a first-order phase transition. In contrast, the transitions from either LD or HD to the MC phase involve a continuous variation in the bulk density, indicating a second-order phase transition.

   Figure~\ref{phase-diagram-weak-coupling} gives the phase diagrams of our model in the $\alpha-\beta$ plane in the weak coupling limit for various $\mu$ together with exchange rates $k_{10}=0$ and $k_{20}=0.95$,  where the different phases are delineated by comparing them with those found in an open TASEP.  The structure of these phase diagrams  in Fig.~\ref{phase-diagram-weak-coupling} is quite sensitive to the values of $\mu$. The steady-state bulk density in the LD phase is $\rho_\text{LD}=\alpha_\text{eff}$, while in the HD phase, $\rho_\text{HD}=1-\beta_\text{eff}$. In an open TASEP, phase transitions between different phases can be determined by equating the currents corresponding to each phase. Extending that principle to the present case, the transition from the LD to the HD phase occurs when $\alpha_\text{eff}=\beta_\text{eff} < 1/2$. Similarly, the transitions between the LD and MC phases, as well as between the HD and MC phases, are characterized by the conditions $\alpha_\text{eff}=1/2<\beta_\text{eff}$ and $\beta_\text{eff}=1/2<\alpha_\text{eff}$, respectively.

   In the phase diagram with $\mu=0.5$ (cf. Fig.~\ref{phase-diagram-weak-coupling}(a)),  there are only LD and DW phases. Indeed, with such a low value of $\mu$, there are fewer particles and hence the existence of the HD and MC phases  in the TASEP lane is {\em not} possible. We find that this particular phase pattern persists up to $\mu=0.76$ for  $k_{10}=0$ and $k_{20}=0.95$. As $\mu$ exceeds this threshold, the MC and HD phases start to emerge. Consequently, the phase diagrams with $\mu>0.76$ consist of all four phases, namely LD, HD, MC, and DW phase, see the phase diagrams corresponding to $\mu=1$ in Fig.~\ref{phase-diagram-weak-coupling}(b) and $\mu=1.5$ in Fig.~\ref{phase-diagram-weak-coupling}(c). The phase boundary between the LD(HD) and MC phases appears as a straight line parallel to the $\beta$-($\alpha$-)axis in both MFT and MCS studies. Noticeable deviations, however, between MFT and MCS outcomes are observed in obtaining these phase boundaries as $\mu$ grows. %Interestingly, the HD-DW phase boundary as obtained from the MCS studies exhibits very distinct shape from the predictions of MF when $\mu=1$. Instead the MCS boundary between the LD and DW phases shows better agreement with the corresponding MF prediction.
   As $\mu$ increases further, the HD phase expands while the DW region shrinks accordingly, see the phase diagram with $\mu=1.5$ in Fig.~\ref{phase-diagram-weak-coupling}(c). When $\mu$ is raised to a significantly large value, such as $\mu=100$, only three distinct phases, namely LD, HD, and MC, are visible in the phase space, see Fig.~\ref{phase-diagram-weak-coupling}(d). The DW phase region now gets contracted into an inclined line extending from (0,0) to  (1/2,1/2). The domain wall spans the whole of the TASEP lane without any specific or unique position, and hence is a  delocalized domain wall (DDW). %This phenomenon mimics the phase diagrams observed in an open boundary TASEP.

In the following sections, we calculate the mean-field steady-state density profiles in different phases and the boundaries separating them.

 %\vspace{2mm}

 \section{STEADY-STATE DENSITIES AND PHASE BOUNDARIES}
 \label{ssd and pb}

 {
In this section, we  solve the MFT equations to obtain the stationary densities  in the LD, HD, MC, and DW phases and the conditions for their existence. We have further calculated the phase boundaries between these phases. These considerations have led to the phase diagrams in Fig.~\ref{phase-diagram-weak-coupling}.  We have also calculated the domain wall heights and positions as functions of the model parameters. }

 \subsection{ Low-density phase}
 \label{LD Phase model 2 weak}

 %\noindent
 In the LD phase, steady-state bulk density is given by
 \begin{equation}
 \label{LD model 2}
 %\begin{split}
 \rho_\text{LD} = \alpha_\text{eff} = \alpha \frac{N_{1}}{N_{0}}< \frac{1}{2}.
 %\end{split}
 \end{equation}
 Substituting $N_{1}$ from Eq.~(\ref{N1}) in Eq.~(\ref{LD model 2}) and using the steady-state particle current in the LD phase,  $J_\text{LD}=\rho_\text{LD}(1-\rho_\text{LD})$, we obtain a quadratic equation in $\rho_\text{LD}$:
 \begin{equation}
 \label{rhold quadratic}
  \rho_\text{LD}=\alpha\bigg[\frac{k_{2}}{k_{1}+k_{2}}\bigg(1-\frac{\rho_\text{LD}}{\mu}\bigg)-\frac{\rho_\text{LD}(1-\rho_\text{LD})}{N_{0}(k_{1}+k_{2})}\bigg].
 \end{equation}
 Now in the weak coupling limit, particle exchange rates $k_{1(2)}=k_{10(20)}/L$. Substituting that in Eq.~(\ref{rhold quadratic}) and recalling $N_{0}=\mu L$, we get
 \begin{equation}
 \label{rhold quadratic weak}
  \rho_\text{LD}=\alpha\bigg[\frac{k_{20}}{k_{10}+k_{20}}\bigg(1-\frac{\rho_\text{LD}}{\mu}\bigg)-\frac{\rho_\text{LD}(1-\rho_\text{LD})}{\mu(k_{10}+k_{20})}\bigg].
  \end{equation}
  Solving Eq.~(\ref{rhold quadratic weak}) for $\rho_\text{LD}$, we get two solutions:
 \begin{align}
 \label{ld density model 2 weak}
 \begin{split}
  \rho_\text{LD}^{\pm} &= \bigg(\frac{1+k_{20}}{2}+\frac{\mu(k_{10}+k_{20})}{2\alpha}\bigg) \\
  &\quad \pm \bigg[\bigg(\frac{1+k_{20}}{2}+\frac{\mu(k_{10}+k_{20})}{2\alpha}\bigg)^{2}-\mu k_{20}\bigg]^{\frac{1}{2}}.
 \end{split}
 \end{align}
 The condition that $\rho_\text{LD}\rightarrow 0$ when $\mu\rightarrow 0$ establishes the following as the physically acceptable solution for the LD phase density:
 \begin{align}
 \label{ld density model 2 weak acc}
 \begin{split}
  \rho_\text{LD} &= \rho_\text{LD}^{-} \\
  &= \bigg(\frac{1+k_{20}}{2}+\frac{\mu(k_{10}+k_{20})}{2\alpha}\bigg)\\
  &\quad -
  \bigg[\bigg(\frac{1+k_{20}}{2}+\frac{\mu(k_{10}+k_{20})}{2\alpha}\bigg)^{2}-\mu k_{20}\bigg]^{\frac{1}{2}}.
 \end{split}
 \end{align}
 We thus find that $\rho_\text{LD}$ is independent of the exit rate parameter $\beta$. Equation~(\ref{LD model 2}) relates $N_{1}$ with $\rho_\text{LD}$, which has already been expressed in terms of the control parameters $\alpha$, $k_{10}$, $k_{20}$, and $\mu$ of the model in Eq.~(\ref{ld density model 2 weak acc}). Particle number conservation, which states $N_{0}=N_{1}+N_{2}+L\rho_\text{LD}$ in the LD phase, can be used to find $N_{2}$ once $N_{1}$ has been obtained. We find for $N_{1}$ and $N_{2}$
 \begin{eqnarray}
  &&N_1=L\frac{\mu}{\alpha}\rho_\text{LD}, \label{n1-ld-wk}\\
  &&N_2=L\bigg(\mu-\frac{\mu+\alpha}{\alpha}\rho_\text{LD}\bigg). \label{n2-ld-wk}
 \end{eqnarray}

\begin{figure*}[htb]
 \includegraphics[width=\linewidth]{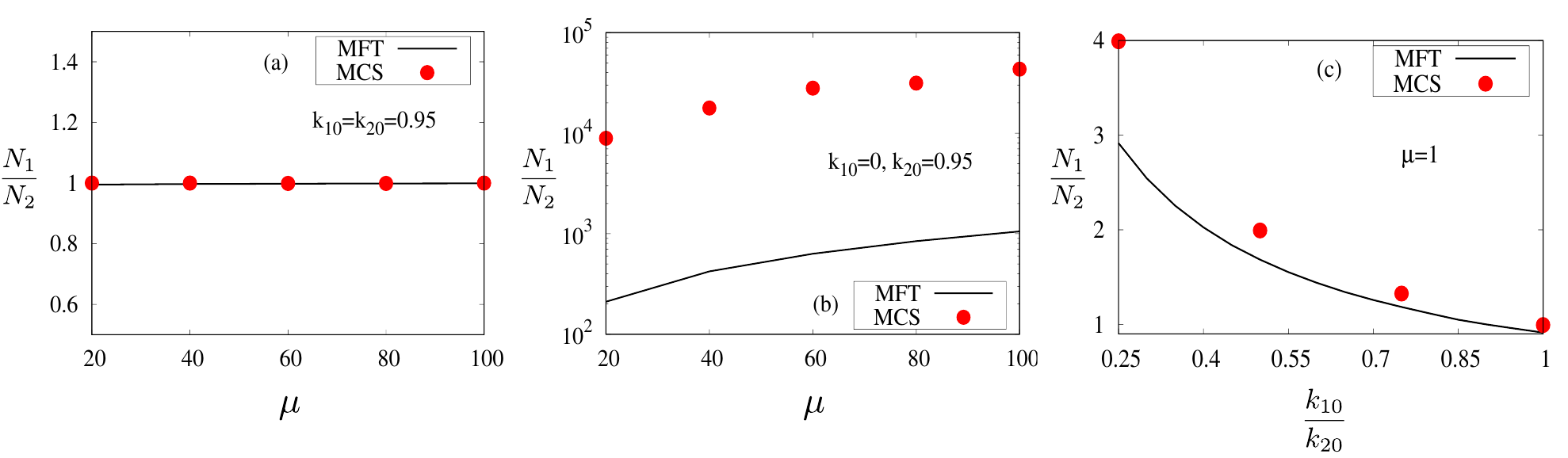}
 \\
\caption{Plot of the reservoir population ratio $N_{1}/N_{2}$ vs $\mu$ in the weak coupling limit of the model for fixed  (a) $k_{10}=k_{20}=0.95$ with good agreement between MFT and MCS results, (b) $k_{10}=0, k_{20}=0.95$ with discrepancies between MFT and MCS results. (c) Plot of $N_{1}/N_{2}$ vs $k_{10}/k_{20}$ for $\mu=1$ in the weak coupling limit. In these plots, the entry and exit rates are carefully selected to ensure that the TASEP lane remains only in one phase (which is here the LD phase).  The values of $\alpha=0.1$ and $\beta=1$ are used. Notice the mismatches between MFT and MCS results increasing with decreasing $k_{10}/k_{20}$. See Eqs.~(\ref{ld density model 2 weak acc})-(\ref{n2-ld-wk}) for the mean-field relation between $N_{1}/N_{2}$, $\mu$, and $k_{10}/k_{20}$.}
\label{n1/n2-vs-mu-and-k10/k20}
\end{figure*}

{We show the variation of the reservoir population ratio ($N_{1}/N_{2}$) with respect to either $\mu$ with constant exchange rates (see Figs.~\ref{n1/n2-vs-mu-and-k10/k20}(a) and \ref{n1/n2-vs-mu-and-k10/k20}(b)) or the ratio of exchange rates ($k_{10}/k_{20}$) with a fixed $\mu$ (see Fig.~\ref{n1/n2-vs-mu-and-k10/k20}(c)) in the weak coupling limit in the LD phase, see Eqs.~(\ref{ld density model 2 weak acc}), (\ref{n1-ld-wk}), and (\ref{n2-ld-wk}). Discrepancies between MFT and MCS results are clearly visible in some cases.}

 %\noindent
 From Eq.~(\ref{ld density model 2 weak acc}), it can be seen that
 \begin{equation}
  \label{rhold-mu-infty-wk}
  \rho_\text{LD} \xrightarrow{\mu \rightarrow \infty} \frac{\alpha k_{20}}{k_{10}+k_{20}}.
 \end{equation}
%{\bf Is it easy to physically understand the above result?}

 We now investigate two special cases: one with equal particle exchange rates, i.e., $k_{10}=k_{20}$, and another when $k_{10}=0, k_{20}\ne 0$. For the first case,  $\rho_\text{LD}$ goes to $\alpha/2$ as $\mu$ approaches $\infty$, see Eq.~(\ref{rhold-mu-infty-wk}). Substituting $\rho_\text{LD}=\alpha/2$ in the infinite-$\mu$ limit in Eqs.~(\ref{n1-ld-wk}) and (\ref{n2-ld-wk}), we find $N_{1}\approx N_{0}/2$ and $N_{2}\approx N_{0}/2$. This implies, for infinitely large value of $\mu$ with equal particle exchange rates, majority of particles are accumulated in the reservoirs with almost equal proportion. For the second case when $k_{10}=0$, $\rho_\text{LD}$ approaches $\alpha$ as $\mu$ tends to $\infty$, see Eq.~(\ref{rhold-mu-infty-wk}). When substituted in Eqs.~(\ref{n1-ld-wk}) and (\ref{n2-ld-wk}), $\rho_\text{LD}=\alpha$ gives $N_{1}\approx N_{0}$ and $N_{2} \approx 0$ in the infinite-$\mu$ limit. Thus, particles are mostly in reservoir $R_{1}$ with reservoir $R_{2}$ (and also the TASEP lane) containing far fewer particles. The phase diagram of the model with $\mu\rightarrow \infty$ in the $\alpha-\beta$ plane is topologically identical to that of an open TASEP for all nonzero values of $k_{10},k_{20}$.

 %\noindent
 %Thus, when particle exchange rates are equal ($k_{10}=k_{20}$), $\rho_\text{LD}$ goes to $\alpha/2$ as $\mu$ approaches $\infty$. To substantiate this, we note as $\mu$ increases, the total particle number in the system ($N_{0}$) becomes significantly larger than the number of sites ($L$) in the TASEP lane. Consequently, majority of particles are allocated in the reservoirs, while the TASEP lane remains sparsely populated. Due to the symmetric exchange rates, the particles are evenly distributed between the two reservoirs, leading to a nearly equal partitioning with $N_{1} \approx N_{2} \approx N_{0}/2$. Hence, the LD phase density, calculated using $\rho_\text{LD}=\alpha_\text{eff}=\alpha N_{1}/N_{0}$, approximates $\alpha/2$. When $k_{10}=0$, on the other hand, $\rho_\text{LD}$ approaches $\alpha$ as $\mu$ tends to $\infty$. In this case, particles can reach reservoir $R_{2}$ from $R_{1}$ only by hopping through $T$ and eventually return to $R_{1}$ being diffused from $R_{2}$. Therefore, as $\mu$ increases, majority of particles accumulate in $R_{1}$, while $R_{2}$ remains nearly empty, i.e., $N_{1}=N_{0}$ and $N_{2}=0$. This leads $\rho_\text{LD}$ approaching $\alpha$. These outcomes, characterized by $\rho_\text{LD} \rightarrow \alpha/2$ for equal exchange rates and $\rho_\text{LD} \rightarrow \alpha$ when $k_{10}=0$, exhibit similarities to the behavior observed in the standard TASEP.

%\noindent

{ Figure~\ref{phase-diagram-weak-coupling} gives the phase diagram of the  present model for different values of $\mu$ with particle exchange rates $k_{10}=0$ and $k_{20}=0.95$. The LD phase is observed for all the values of $\mu$, including large values such as $\mu=100$ in Fig.~\ref{phase-diagram-weak-coupling}(d).}

We now determine the range of $\mu$ over which the LD phase can exist. For sufficiently small values of $\mu$, there will certainly be an LD phase, the density of which will gradually rise with an increase in $\mu$, as shown in Eq.~(\ref{ld density model 2 weak acc}). Eventually, $\rho_\text{LD}$ approaches $\alpha k_{20}/(k_{10}+k_{20})$ as $\mu \rightarrow \infty$, as indicated in (\ref{rhold-mu-infty-wk}). Thus, the LD phase can be found for any value of $\mu$ in our model $0 \le \mu <\infty$.
This can be seen in the phase diagrams of Fig.~\ref{phase-diagram-weak-coupling}, where LD phase appears for small values of $\mu$ ($\mu=0.5$) as well as large values ($\mu=100$).
Furthermore, the condition $\rho_\text{LD}<1/2$ when imposed on Eq.~(\ref{ld density model 2 weak acc}) gives
\begin{equation}
  \label{alpha-less-than-in-weak}
  \alpha < \bigg(\frac{k_{10}+k_{20}}{2k_{20}-\frac{\frac{1}{2}+k_{20}}{\mu}}\bigg).
 \end{equation}
 This implies that, for a given set of values for $k_{10}$, $k_{20}$, and $\mu$, the LD phase can only be obtained for $\alpha$ values satisfying the condition mentioned above. This is illustrated in the phase diagrams of Fig.~\ref{phase-diagram-weak-coupling}. Consider, for example, the case with $k_{10}=0$, $k_{20}=0.95$, and $\mu=1$ Fig.~\ref{phase-diagram-weak-coupling}(b), where the condition (\ref{alpha-less-than-in-weak}) indicates that the LD phase can exist only for $\alpha < 2.11$ according to MFT.

 \subsection{High-density phase}
 \label{HD Phase model 2 weak}

 %\noindent
 The steady-state bulk density in HD phase is given by
 \begin{equation}
  \label{hd-den-n0}
  \rho_\text{HD} = 1-\beta_\text{eff}=1-\beta\bigg(1-\frac{N_{2}}{N_{0}}\bigg)>\frac{1}{2}.
 \end{equation}
 Plugging $N_2$ obtained above in Eq.~(\ref{N2}) into Eq.~(\ref{hd-den-n0}) and using the steady-state current in the HD phase, $J_\text{HD}=\rho_\text{HD}(1-\rho_\text{HD})$, we obtain a quadratic equation in $\rho_\text{HD}$:
 \begin{equation}
  \label{hd-quad}
  \rho_\text{HD}=1-\beta+\beta\bigg[\frac{k_{1}}{k_{1}+k_{2}}\bigg(1-\frac{\rho_\text{HD}}{\mu}\bigg)+\frac{\rho_\text{HD}(1-\rho_\text{HD})}{N_{0}(k_{1}+k_{2})}\bigg].
 \end{equation}
 With $k_{1(2)}=k_{10(20)}/L$ in the weak coupling case and $N_{0}=\mu L$, Eq.~(\ref{hd-quad}) translates into
 \begin{equation}
  \label{hd-quad-weak}
  \rho_\text{HD}=1-\beta+\beta\bigg[\frac{k_{10}}{k_{10}+k_{20}}\bigg(1-\frac{\rho_\text{HD}}{\mu}\bigg)+\frac{\rho_\text{HD}(1-\rho_\text{HD})}{\mu(k_{10}+k_{20})}\bigg],
 \end{equation}
 solving which we get the following two solutions for $\rho_\text{HD}$:
 \begin{align}
\label{hd-den-sol-weak}
\begin{split}
\rho_\text{HD}^{\pm}&=\bigg(\frac{1-k_{10}}{2}-\frac{\mu(k_{10}+k_{20})}{2\beta}\bigg) \pm
\bigg[\bigg(\frac{1-k_{10}}{2}\\
&\quad -\frac{\mu(k_{10}+k_{20})}{2\beta}\bigg)^{2}+
\frac{\mu k_{10}}{\beta}-\mu \bigg(1-\frac{1}{\beta}\bigg)k_{20}\bigg]^{\frac{1}{2}}.
\end{split}
\end{align}
This has two solutions. Physically, in the limit when $N_2\rightarrow 0$, any reservoir crowding effect ceases to operate and as a result $\beta_\text{eff}\rightarrow \beta$. Furthermore, in that case if there are infinite resources, i.e., $\mu\rightarrow \infty$, most of the particles are to be found in $R_1$, giving $N_1\approx N_0$ and $\alpha_\text{eff}\approx \alpha$. Such a situation can be achieved when $k_{10}=0$ and $k_{20}\ne 0$ along with $\mu\rightarrow \infty$. We use these considerations to determine which of the two solutions in (\ref{hd-den-sol-weak}) is physically acceptable.
Indeed, we find that between the two solutions of HD phase density in Eq.~(\ref{hd-den-sol-weak}),  the one with positive discriminant gives
\begin{equation}
 \label{hd-mu-infty-wk}
 \rho_\text{HD} \xrightarrow{\mu \rightarrow \infty} \bigg(1-\frac{\beta k_{20}}{k_{10}+k_{20}}\bigg).
\end{equation}
Thus, when $k_{10}=0$, $\rho_\text{HD}$ approaches ($1-\beta$) in the limit $\mu \rightarrow \infty$. This suggests the following as the HD phase density in the MFT:
 \begin{align}
\label{hd-den-sol-weak-acc}
\begin{split}
\rho_\text{HD}&=\rho_\text{HD}^{+}\\
&=\bigg(\frac{1-k_{10}}{2}-\frac{\mu(k_{10}+k_{20})}{2\beta}\bigg)+
\bigg[\bigg(\frac{1-k_{10}}{2}\\
&\quad -\frac{\mu(k_{10}+k_{20})}{2\beta}\bigg)^{2}+
\frac{\mu k_{10}}{\beta}-\mu \bigg(1-\frac{1}{\beta}\bigg)k_{20}\bigg]^{\frac{1}{2}}.
\end{split}
\end{align}
 Unsurprisingly, $\rho_\text{HD}$ is independent of $\alpha$. To find the reservoir populations in the HD phase, we first obtain $N_{2}$ from Eq.~(\ref{hd-den-n0}) in terms of $\rho_\text{HD}$ (expression of which is obtained in Eq.~(\ref{hd-den-sol-weak-acc}) in terms of model parameters), after which particle number conservation $N_{0}=N_{1}+N_{2}+L\rho_\text{HD}$ in the HD phase can be used to find $N_{1}$. We thus find
 \begin{eqnarray}
  &&N_1=L\bigg[\frac{\mu(1-\rho_\text{HD})}{\beta}-\rho_\text{HD}\bigg], \label{hd-n1-hd-wk}\\
  &&N_2=L\bigg[\mu-\frac{\mu(1-\rho_\text{HD})}{\beta}\bigg]. \label{hd-n2-hd-wk}
 \end{eqnarray}

 %Eq. (\ref{hd-den-sol-weak}) reveals $\rho_\text{HD} \rightarrow 1-[\beta k_{20}/(k_{10}+k_{20})]$ as $\mu \rightarrow \infty$. We now consider two cases as earlier: (i) When $k_{10}=k_{20}$, $\rho_\text{HD} \rightarrow 1-(\beta/2)$ and (ii) when $k_{10}=0$, $\rho_\text{HD} \rightarrow 1-\beta$. Clearly, in the infinite-$\mu$ limit $\rho_\text{HD}$ resembles to the open TASEP. This can be interpreted using the same line of reasoning as in the LD phase. For case (i) in the limit $\mu \rightarrow \infty$, almost all particles are situated in the reservoirs with equal amount $N_{1} \approx N_{2} \approx N_{0}/2$ and we get $\rho_\text{HD} = 1-\beta_\text{eff} \approx 1-(\beta/2)$. On the other hand, for case (ii) nearly all particles are located in $R_{1}$, i.e., $N_{1}\approx N_{0}$ and $N_{2} \approx 0$. This results into $\rho_\text{HD} = 1-\beta_\text{eff} \approx 1-\beta$. \par

  As considered in our MFT for  the LD phase above, we again analyse with two specific cases, $k_{10}=k_{20}$ and $k_{10}=0, k_{20}\ne 0$, in the HD phase. For the first case, i.e., when $k_{10}=k_{20}$, Eq.~(\ref{hd-mu-infty-wk}) indicates that $\rho_\text{HD}$ approaches ($1-\beta/2$) when $\mu\rightarrow\infty$. Consequently, substituting $\rho_\text{HD} = (1-\beta/2)$ in Eqs.~(\ref{hd-n1-hd-wk}) and (\ref{hd-n2-hd-wk}) when $\mu \rightarrow \infty$ yields $N_{1}\approx N_{0}/2$ and $N_{2}=N_{0}/2$. For the second case, i.e., when $k_{10}=0, k_{20}\ne 0$, Eq.~(\ref{hd-mu-infty-wk}) gives $\rho_\text{HD}$ approaching ($1-\beta$) in the limit of infinite-$\mu$. Thus, substituting $\rho_\text{HD}=(1-\beta)$ in Eqs.~(\ref{hd-n1-hd-wk}) and (\ref{hd-n2-hd-wk}) when $\mu \rightarrow \infty$ results into $N_{1}\approx N_{0}$ and $N_{2}=0$.

 %As $\mu$ approaches infinity, Eq.~(\ref{hd-den-sol-weak-acc}) shows $\rho_\text{HD}$ converges to a limiting value determined by the expression $1-[\beta k_{20}/(k_{10}+k_{20})]$. Depending on the specific choice of exchange rates, we discuss two cases here: i) When $k_{10}=k_{20}$, $\rho_\text{HD}$ approaches $1-(\beta/2)$.
%(ii) When $k_{10}\rightarrow0$, $\rho_\text{HD}$ approaches $1-\beta$. These limiting behaviors resemble those observed in the standard TASEP. In case (i), particles are equally distributed between the reservoirs, i.e. $N_{1}=N_{2}\approx N_{0}/2$. Hence, $\rho_\text{HD}=1-\beta(1-N_{2}/N_{0}) \approx 1-(\beta/2)$. In case (ii), where almost all particles accumulate in $R_{1}$, i.e. $N_{1}\approx N_{0}, N_{2}\approx 0$, we obtain $\rho_\text{HD} \approx 1-\beta$.

 { The occurrence of HD phase is illustrated for $\mu=1$ in Fig.~\ref{phase-diagram-weak-coupling}(b), $\mu=1.5$ in Fig.~\ref{phase-diagram-weak-coupling}(c), and $\mu=100$ in Fig.~\ref{phase-diagram-weak-coupling}(d). However, the phase diagram for $\mu=0.5$ in Fig.~\ref{phase-diagram-weak-coupling}(a), includes only the LD and DW phases.}
{Indeed, when $\mu$ is reduced, HD phase is less likely to occur due to insufficient supply of particles. Eventually, below a certain value of $\mu$, HD phase will disappear.}
%\begin{equation}
 % \label{hd-wk-mu-range-lower}
  %\mu>\bigg(\frac{k_{10}-\frac{1}{2}}{\frac{k_{10}+k_{20}}{\beta}-2k_{20}}\bigg),
  %\end{equation}
  {As $\mu \rightarrow \infty$, our model reduces to an open TASEP with $\rho_\text{HD}=1-\beta$(for $k_{10}=0,k_{20}\ne0$), see Eq.~(\ref{hd-mu-infty-wk}). Thus, there will be no upper limit of $\mu$ for HD phase to occur. Using the fact $\rho_\text{HD}>1/2$, one gets the following condition for HD phase existence by using Eq.~(\ref{hd-den-sol-weak-acc}):}
  \begin{equation}
   \label{beta-less-than-hd-weak}
   \beta < \bigg(\frac{k_{10}+k_{20}}{2k_{20}+\frac{k_{10}-\frac{1}{2}}{\mu}}\bigg),
  \end{equation}
{which implies that, for specific values of $k_{10}$, $k_{20}$, and $\mu$, HD phase can occur when the condition mentioned above is satisfied. This is shown in the phase diagrams of Fig.~\ref{phase-diagram-weak-coupling}. For instance, consider the case with $k_{10}=0$, $k_{20}=0.95$, and $\mu=1$ in Fig.~\ref{phase-diagram-weak-coupling}(b), where (\ref{beta-less-than-hd-weak}) suggests that the HD phase will exist for $\beta<0.68$.}

 \subsection{Maximal current phase}
 \label{MC Phase model 2 weak}

 %\noindent
 The MC phase is characterized by $J_\text{max}=1/4$ which is the maximum possible steady-state particle current, and is obtained when the particle density $\rho=1/2$ throughout the bulk of $T$~\cite{blythe}. Thus, in this phase, TASEP lane is half-filled and the particle number conservation reads  $N_{0}=N_{1}+N_{2}+L/2$.

 Analogous to an open TASEP, conditions for the existence of the MC phase in the present model are $\alpha_\text{eff}>1/2$ and $\beta_\text{eff}>1/2$. We determine the boundaries between the LD and MC phases, and between the HD and MC phases by setting $\rho_\text{LD}=1/2$ and $\rho_\text{HD}=1/2$ in Eqs.~(\ref{ld density model 2 weak acc}) and (\ref{hd-den-sol-weak-acc}) respectively, which yields the following as LD-MC and HD-MC phase boundaries, respectively:
\begin{eqnarray}
  &&\alpha=\frac{\mu(k_{10}+k_{20})}{(2\mu-1)k_{20}-\frac{1}{2}}, \label{ld-mc boundary model 2 weak}\\
  &&\beta=\frac{\mu(k_{10}+k_{20})}{k_{10}+2\mu k_{20}-\frac{1}{2}}. \label{hd-mc boundary model 2 weak}
 \end{eqnarray}
 Considering that $\alpha$ and $\beta$ are non-negative real numbers in Eq.~(\ref{ld-mc boundary model 2 weak}) and Eq.~(\ref{hd-mc boundary model 2 weak}), we obtain
  \begin{eqnarray}
  &&\mu>\mu_{1}=\bigg(\frac{1}{2}+\frac{1}{4k_{20}}\bigg), \label{mu-cnd-1}\\
  &&\mu>\mu_{2}=\bigg(\frac{1}{4k_{20}}-\frac{k_{10}}{2k_{20}}\bigg). \label{mu-cnd-2}
 \end{eqnarray}
  Since $\mu_{1}>\mu_{2}$ for any positive value of $k_{10}$ and $k_{20}$, meeting (\ref{mu-cnd-1}) automatically infers fulfillment of (\ref{mu-cnd-2}). Thus the lower threshold of $\mu$ for MC phase existence is
  \begin{equation}
  \label{low-mu-mc}
  \mu_{l}^\text{MC}=\bigg(\frac{1}{2}+\frac{1}{4k_{20}}\bigg).
 \end{equation}
 This is also the lower threshold for HD phase existence, since if $\mu < \mu_{l}^\text{MC}$, the LD-MC phase boundary in Eq.~(\ref{ld-mc boundary model 2 weak}) yields $\alpha < 0$, which is unphysical, thus ruling out the emergence of the HD phase in the phase diagram.
 This is shown in Fig.~\ref{phase-diagram-weak-coupling}(a) for $\mu=0.5$, where the exchange rate $k_{20}=0.95$ yielding the lower threshold of $\mu$ for the existence of the MC and HD phases to be $0.76$. As seen before, the present model reduces to an open TASEP in the limit $\mu \rightarrow \infty$. Thus, MC phase can be present for arbitrarily large values of $\mu$ as well. Taken together, following is the range within which MC and HD phases appear:
 \begin{equation}
  \label{mc-range-mu-weak}
  \bigg(\frac{1}{2}+\frac{1}{4k_{20}}\bigg)<\mu<\infty.
 \end{equation}

 %\noindent
 %and the non-negativity of $\beta$ in (\ref{hd-mc boundary model 2 weak early}) infers the condition

 %\begin{equation}
 %\label{n0-wk-mc-mu-range}
 % \mu>\bigg(\frac{1}{4k_{20}}-\frac{k_{10}}{2k_{20}}\bigg).
 %\end{equation}

\begin{figure*}[htb]
 \includegraphics[width=\linewidth]{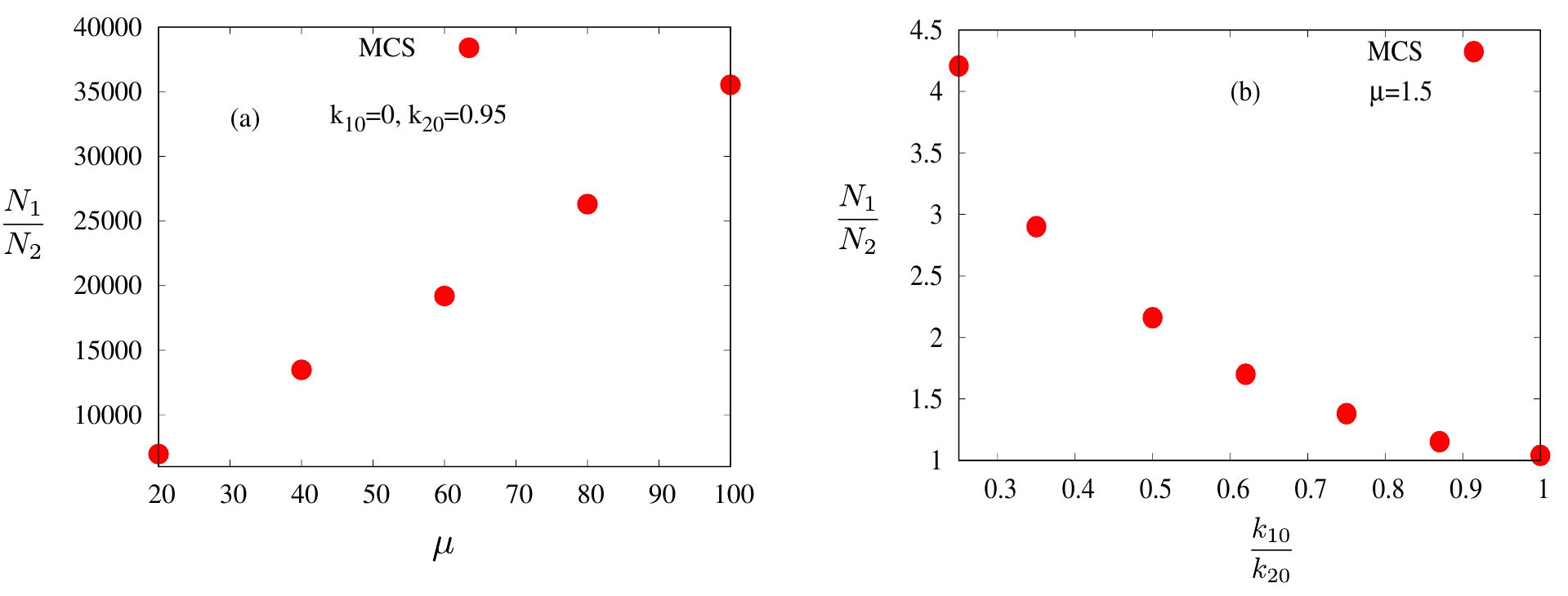}
 \\
\caption{{(a) Plot of the reservoir population ratio $N_{1}/N_{2}$ vs $\mu$ in the weak coupling limit of the model for fixed exchange rates $k_{10}=0, k_{20}=0.95$. Entry and exit rates are $\alpha=\beta=1.3$ for which the system is found to be in MC phase.  (b) Plot of $N_{1}/N_{2}$ vs $k_{10}/k_{20}$ for fixed $\mu=1.5$ in the weak coupling limit. $\alpha=5$, $\beta=3$ for which the system is found to be in MC phase.}}
\label{reser_pop_ratio_mc}
\end{figure*}

 The phase diagrams of the model obtained by MFT and MCS with exchange rates $k_{10}=0$ and $k_{20}=0.95$ are shown in Fig.~\ref{phase-diagram-weak-coupling}. The lower threshold for the existence of the MC phase is given by the condition (\ref{mc-range-mu-weak}) as $\mu=0.76$ for these specific exchange rates, according to MFT. This prediction is confirmed by MCS simulations, where the MC phase does not exist for values of $\mu$ less than $0.76$, such as $\mu=0.5$. The dependence of $N_{1}/N_{2}$ on either $\mu$ for fixed exchange rates $k_{10}=0,k_{20}=0.95$ or on $k_{10}/k_{20}$ for fixed $\mu=1.5$ are shown respectively in Figs.~\ref{reser_pop_ratio_mc}(a) and \ref{reser_pop_ratio_mc}(b).

 \subsection{Domain wall phase}
 \label{SP Phase model 2 weak}

 %\begin{figure}[!h]
 %\centering
 %\includegraphics[width=0.55\textwidth]{n0_wk_dw_new.pdf}
 %\caption{Steady state density profiles in the DW phase, under weak coupling conditions of the model. Black solid lines indicate the MF predicted domain walls, wherein colored discrete points are the MCS results. Filling factor and entrance rate are fixed to the values $\mu=3/2$ and $\alpha=0.2$. The plots compare two scenarios: one with exchange rates $k_{10}=k_{20}=0.7$ and $\beta=0.1$ (red plot), another with $k_{10}=0$ and $k_{20}=0.7$ and $\beta=0.12$ (blue plot). Clearly, a localized domain wall (LDW) is formed for these specific choices of control paramters.}
 %\label{ldw-wk-n0}
 %\end{figure}

 %\begin{figure}[!h]
 %\centering
 %\includegraphics[width=0.55\textwidth]{n0_ddw_wk_c1_new.pdf}
 %\caption{This plot shows a Monte Carlo simulation of the steady-state density profile in the DW phase of the model under weak coupling conditions, with filling factor $\mu=1000$ and exchange rates $k_{10}=k_{20}=0.7$. As $\mu$ increases, the domain wall becomes delocalized in space and forms a DDW pattern, resembling the behavior of the standard open boundary TASEP.}
 %\label{ddw-wk-n0}
 %\end{figure}

 %\noindent
 In the DW phase, the density profile is piecewise continuous forming a DW that connects the LD and HD regions of a stationary density profile. Analogous to an open TASEP one has $\rho_\text{LD}=1-\rho_\text{HD}<1/2$ in the DW phase. However, unlike in an open TASEP, the precise location of the DW in our model can be obtained by using particle number conservation. Therefore, we get a localized domain wall (LDW), which is formed, say at a position $x_{w}$, in the bulk of the lane. The nonuniform bulk density in the DW phase can be represented in general as
\begin{equation}
\label{rho-sp-theta}
\rho(x)=\rho_\text{LD}+\Theta(x-x_{w})(\rho_\text{HD}-\rho_\text{LD}),
\end{equation}
where $\rho_\text{LD}$ and $\rho_\text{HD}$ are the densities respectively in LD and HD domain of the DW and $\Theta$ is the Heaviside step function, defined as $\Theta(x)=1$ for $x>0$ and $\Theta(x)=0$ for $x<0$.

Particle number in $T$ is given by %calculated as the integral of the density function $\rho(x)$ over the interval [0, 1]:
 \begin{equation}
 \label{nt-sp}
 N_{T} = \sum_{i=1}^{L}\rho_{i} = L \int_{0}^{1} \rho(x) dx.
 \end{equation}
% where a multiplicative factor $L$ is introduced in the right-hand side to rescale the integration limit of the position variable $x$.
Plugging $\rho(x)$ from Eq.~(\ref{rho-sp-theta}) into Eq.~(\ref{nt-sp}), we get after simplifying
 \begin{equation}
 \label{NT}
 N_{T} = L\bigg[ \alpha \frac{N_{1}}{N_{0}} (2x_{w}-1)+1-x_{w}\bigg].
 \end{equation}
 Substituting $N_{T}$ from Eq.~(\ref{NT}) into the particle number conservation, and the expression for the steady-state current $J_\text{DW}=\rho_\text{LD}(1-\rho_\text{LD})$ or $J_\text{DW}=\rho_\text{HD}(1-\rho_\text{HD})$ in the DW phase into Eq.~(\ref{N1}), we obtain two equations coupled by $N_{1}/N_{0}$ and $x_{w}$:
 \begin{eqnarray}
  &&\frac{N_{1}}{N_{0}} \bigg[\mu\bigg(1-\frac{\alpha}{\beta}\bigg) + \alpha(2x_{w}-1)\bigg]-x_{w}+1=0, \label{First coupled equation model 2}\\
  &&\begin{split}
  \label{Second coupled equation model 2}
  \frac{N_{1}}{N_{0}}&=\frac{k_{2}}{k_{1}+k_{2}}\bigg[1-\frac{1}{\mu}\bigg\{\alpha \frac{N_{1}}{N_{0}} (2x_{w}-1)+1-x_{w}\bigg\}\bigg]\\
  &\quad -\frac{1}{N_{0}(k_{1}+k_{2})}\alpha\frac{N_{1}}{N_{0}}\bigg(1-\alpha\frac{N_{1}}{N_{0}}\bigg).
 \end{split}
 \end{eqnarray}
 These equations provide two solutions for $N_{1}/N_{0}$ as follows:
\begin{align}
  \label{n1/n0-weak-pm-pos}
  \begin{split}
  \bigg(\frac{N_{1}}{N_{0}}\bigg)^{\pm}&=\bigg(\frac{1}{2\alpha}+\frac{\mu k_{10}}{2\alpha^{2}}+\frac{\mu k_{20}}{2\alpha\beta}\bigg)\\
  &\quad \pm \bigg[\bigg(\frac{1}{2\alpha}+\frac{\mu k_{10}}{2\alpha^{2}}+\frac{\mu k_{20}}{2\alpha\beta}\bigg)^{2}-\frac{\mu k_{20}}{\alpha^{2}}\bigg]^{\frac{1}{2}},
  \end{split}
 \end{align}
 %\noindent
%\textit{Second solution}:
%\begin{align}
  %\label{n1/n0-weak-second-neg}
  %\begin{split}
  %\bigg(\frac{N_{1}}{N_{0}}\bigg)^{-}&=\bigg(\frac{1}{2\alpha}+\frac{\mu k_{10}}{2\alpha^{2}}+\frac{\mu k_{20}}{2\alpha\beta}\bigg) \\
  %&\quad-\bigg[\bigg(\frac{1}{2\alpha}+\frac{\mu k_{10}}{2\alpha^{2}}+\frac{\mu k_{20}}{2\alpha\beta}\bigg)^{2}-\frac{\mu k_{20}}{\alpha^{2}}\bigg]^{\frac{1}{2}}.
  %\end{split}
 %\end{align}
  With these two possible solutions in Eq.~(\ref{n1/n0-weak-pm-pos}), the density in the LD region of the domain wall becomes
 \begin{align}
  \label{rhold-weak}
  \begin{split}
  \rho_\text{LD}&=\alpha\bigg(\frac{N_{1}}{N_{0}}\bigg)^{\pm}\\
  &=\bigg(\frac{1}{2}+\frac{\mu k_{10}}{2\alpha}+\frac{\mu k_{20}}{2\beta}\bigg) \\
  &\quad \pm \bigg[\bigg(\frac{1}{2}+\frac{\mu k_{10}}{2\alpha}+\frac{\mu k_{20}}{2\beta}\bigg)^{2}-\mu k_{20}\bigg]^{\frac{1}{2}}.
  \end{split}
 \end{align}
 At the LD-DW phase boundary, the density corresponding to the LD phase (Eq.~(\ref{ld density model 2 weak acc})) must be identical to the acceptable density in the LD part of domain wall (Eq.~(\ref{rhold-weak})). Considering the solution of Eq.~(\ref{rhold-weak}) with negative discriminant, we found that it is identical to the LD phase density given in Eq.~(\ref{ld density model 2 weak acc}) under the following condition:
 \begin{equation}
  \label{lddw-bound-weak-1}
  \mu\bigg(1-\frac{\alpha}{\beta}\bigg)+\alpha=0.
 \end{equation}
 Equation~(\ref{lddw-bound-weak-1}) is exactly the same equation that defines the boundary between LD and DW phases, which we will obtain later in Eq.~(\ref{ld-sp boundary model 2 weak}) {by considering the domain wall position at the extreme right end or exit end of the TASEP lane}. Thus the acceptable solution for $N_{1}/N_{0}$ is
 \begin{align}
  \label{n1/n0-weak-acc}
  \begin{split}
  \frac{N_{1}}{N_{0}}&=\bigg(\frac{1}{2\alpha}+\frac{\mu k_{10}}{2\alpha^{2}}+\frac{\mu k_{20}}{2\alpha\beta}\bigg)\\
  &\quad - \bigg[\bigg(\frac{1}{2\alpha}+\frac{\mu k_{10}}{2\alpha^{2}}+\frac{\mu k_{20}}{2\alpha\beta}\bigg)^{2}-\frac{\mu k_{20}}{\alpha^{2}}\bigg]^{\frac{1}{2}}.
  \end{split}
 \end{align}
 In the DW phase
 \begin{align}
  &\rho_\text{LD}+\rho_\text{HD}=1, \label{dw1} \\
  \implies & \alpha_\text{eff}=\beta_\text{eff}, \label{dw2} \\
  \implies & \alpha\frac{N_{1}}{N_{0}}=\beta\bigg(1-\frac{N_{2}}{N_{0}}\bigg). \label{dw3}
 \end{align}
One then gets the expression of $N_{2}$ from that of $N_{1}$:
  \begin{align}
  \label{n2/n0-wk}
  \begin{split}
  \frac{N_{2}}{N_{0}}&=1-\frac{\alpha}{\beta}\frac{N_{1}}{N_{0}}\\
  &=1-\bigg(\frac{1}{2\beta}+\frac{\mu k_{10}}{2\alpha\beta}+\frac{\mu k_{20}}{2\beta^{2}}\bigg) \\
  &\quad +\bigg[\bigg(\frac{1}{2\beta}+\frac{\mu k_{10}}{2\alpha\beta}+\frac{\mu k_{20}}{2\beta^{2}}\bigg)^{2}-\frac{\mu k_{20}}{\beta^{2}}\bigg]^{\frac{1}{2}}.
  \end{split}
  \end{align}
 Consequently, the DW position ($x_{w}$) is obtained by solving Eq.~(\ref{First coupled equation model 2}). We obtain
\begin{equation}
\label{xw model N0 weak coupling}
\begin{split}
x_{w}=\frac{1-\frac{N_{1}}{N_{0}}\bigg[\mu\bigg(\frac{\alpha}{\beta}-1\bigg)+\alpha\bigg]}{1-2\alpha\frac{N_{1}}{N_{0}}}.
\end{split}
\end{equation}
Substituting $N_{1}/N_{0}$ from Eq.~(\ref{n1/n0-weak-acc}) into Eq.~(\ref{xw model N0 weak coupling}), one finds $x_w$ in terms of the control parameters, $\alpha$, $\beta$, $\mu$, $k_{10}$, and $k_{20}$, of the model. Furthermore, densities corresponding to the LD and HD regions of the DW phase can be calculated as follows:
\begin{align}
  \rho_\text{LD} &= \alpha\frac{N_{1}}{N_{0}} \nonumber\\
  &=\bigg(\frac{1}{2}+\frac{\mu k_{10}}{2\alpha}+\frac{\mu k_{20}}{2\beta}\bigg) \nonumber\\
  &\quad - \bigg[\bigg(\frac{1}{2}+\frac{\mu k_{10}}{2\alpha}+\frac{\mu k_{20}}{2\beta}\bigg)^{2}-\mu k_{20}\bigg]^{\frac{1}{2}}, \label{ld-den-dw-wk}
  %\rho_\text{HD} &= 1-\rho_\text{LD}. \label{hd-den-dw-wk}
\end{align}
with $\rho_\text{HD} = 1-\rho_\text{LD}$. The LDW height $\Delta$ is then
\begin{align}
  \Delta &= \rho_\text{HD}-\rho_\text{LD} \nonumber\\
  &= \bigg[\bigg(1+\frac{\mu k_{10}}{\alpha}+\frac{\mu k_{20}}{\beta}\bigg)^{2}-4\mu k_{20}\bigg]^\frac{1}{2}-\bigg(\frac{\mu k_{10}}{\alpha}+\frac{\mu k_{20}}{\beta}\bigg). \label{delta model L weak coupling}
 \end{align}

%\begin{widetext}

 \begin{figure}[!h]
 \centering
\includegraphics[width=\linewidth]{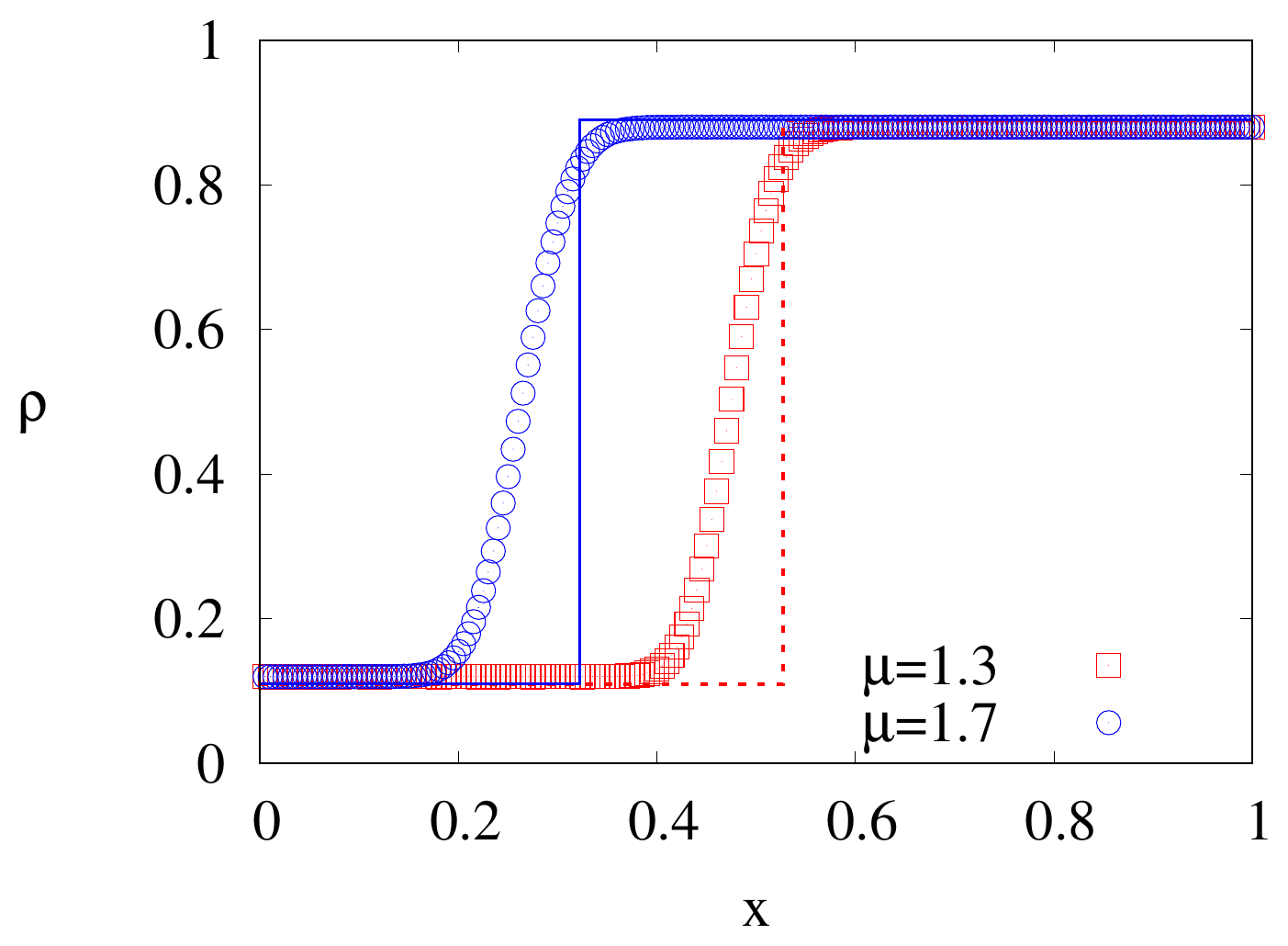}
\caption{Plot of an LDW in the weak coupling limit of the model. System size is $L=1000$ and $2 \times 10^{9}$ Monte Carlo steps are performed to get the density profiles. Lines (solid and dotted) represent the MFT predictions (see Eqs.~(\ref{n1/n0-weak-acc}), (\ref{xw model N0 weak coupling}), and (\ref{ld-den-dw-wk})), whereas discrete points are the MCS results. For smaller values of $\mu$ we get an LDW. Two different sets of parameters are chosen: $\alpha=0.2$, $\beta=0.12$, $\mu=1.3$, $k_{10}=0$, and $k_{20}=0.95$ (red squares); $\alpha=0.2$, $\beta=0.12$, $\mu=1.7$, $k_{10}=0$, and $k_{20}=0.95$ (blue circles). Clearly, increasing the value of $\mu$ shifts the LDW position to the left or entry-end of TASEP, according to MFT Eq.~(\ref{xw model N0 weak coupling}).}
\label{ldw-ddw-wk}
\end{figure}

%\end{widetext}

 %\begin{figure*}[htb]
 %\includegraphics[width=\columnwidth]{n0_wk_ldw_a0.2_b0.12_m1.3_1.7_crop_new.pdf}
 %\hfill
 %\includegraphics[width=\columnwidth]{n0_wk_ddw_ab0.25_m1000_crop.pdf}
 %\hfill
 %\includegraphics[width=\columnwidth]{n0_wk_k10_0.95_k20_0.01_ldw_new.pdf}
 %\\
%\caption{Localised and delocalized domain walls (LDW and DDW) in the weak coupling limit of the model. Black solid lines represent the MF predictions, wherein colored discrete points are the MCS results. \textbf{(Left)} For smaller values of $\mu$ domain walls are localized. Two different sets of parameters are chosen: $\alpha=0.2$, $\beta=0.12$, $\mu=1.3$, $k_{10}=0$, and $k_{20}=0.95$ (red); $\alpha=0.2$, $\beta=0.12$, $\mu=1.7$, $k_{10}=0$, and $k_{20}=0.95$ (blue). Clearly, increasing the value of $\mu$ shifts the domain wall to the left. MF predicts LDW densities closer to MCS values, whereas LDW position has slight deviations. \textbf{(Right)} When $\mu$ is increased to a sufficiently large value, domain wall gets delocalized. Parameter values are: $\alpha=\beta=0.25$, $\mu=1000$, $k_{10}=0$, and $k_{20}=0.95$. \textbf{(3)} Parameter values: $\alpha=10$, $\beta=7$, $\mu=100$, $k_{10}=0.95$, and $k_{20}=0.01$ (red); $\alpha=23$, $\beta=22$, $\mu=1000$, $k_{10}=0.95$, and $k_{20}=0.01$ (blue). With these sets of parameter values where $k_{10}>>k_{20}$, only localized domain walls are formed even with very large values of $\mu$.}
%\label{ldw-ddw-wk}
%\end{figure*}

 To determine the boundaries between the LD and the HD phases with the DW phase, we make use of the fact that the domain wall is situated at the far right and left ends of the TASEP lane at the point of LD-DW and HD-DW transitions, respectively. By setting $x_{w}=1$ and $x_{w}=0$ in Eq.~(\ref{xw model N0 weak coupling}), we obtain the following equations that define the LD-DW and HD-DW phase boundaries respectively:
\begin{eqnarray}
  &&\mu\bigg(1-\frac{\alpha}{\beta}\bigg)+\alpha=0, \label{ld-sp boundary model 2 weak}\\
  &&\frac{N_{1}}{N_{0}}\bigg[\mu\bigg(1-\frac{\alpha}{\beta}\bigg)-\alpha\bigg]+1=0, \label{hd-sp boundary model 2 weak}
 \end{eqnarray}
 with $N_{1}/N_{0}$ given in Eq.~(\ref{n1/n0-weak-acc}). The LDW position $x_w$ obtained in Eq.~(\ref{xw model N0 weak coupling}) spans from 0 to 1 with $x_{w}=0$ and $x_{w}=1$ corresponding to the HD-DW and LD-DW phase boundaries, respectively. Analogous to an open TASEP, the DW phase in the present model is characterized by $\alpha_\text{eff}=\beta_\text{eff}<1/2$. Since $\alpha_\text{eff}=\alpha N_{1}/N_{0}<1/2$, the denominator in Eq.~(\ref{xw model N0 weak coupling}) is always positive. The condition $x_{w}>0$ then is met only if the numerator in Eq.~(\ref{xw model N0 weak coupling}) is positive, which leads to
  \begin{align}
    \mu\bigg(1-\frac{\alpha}{\beta}\bigg)>\alpha-\frac{N_{0}}{N_{1}}. \label{numerator-positive}
  \end{align}
   Also, the condition $x_w<1$ when applied in Eq.~(\ref{xw model N0 weak coupling}) yields the following:
\begin{equation}
 \label{xw-less-1}
 \mu\bigg(1-\frac{\alpha}{\beta}\bigg) < -\alpha.
\end{equation}
Within the DW phase, the right-hand side of (\ref{numerator-positive}), $\alpha-N_{0}/N_{1}$, is negative. Moreover, the right-hand side of (\ref{xw-less-1}) is clearly negative ($\alpha$ being positive always) indicating the left-hand side, $\mu(1-\alpha/\beta)$, to be negative. Given $\mu>0$, this implies $(1-\alpha/\beta)$ to be negative or $\alpha>\beta$ over the DW region. We recast these inequalities (\ref{numerator-positive}) and (\ref{xw-less-1}) as
 \begin{eqnarray}
    &&\mu\bigg(\frac{\alpha}{\beta}-1\bigg)<\frac{N_{0}}{N_{1}}-\alpha, \label{numerator-positive-1}\\
    &&\mu\bigg(\frac{\alpha}{\beta}-1\bigg) > \alpha. \label{xw-less-1-1}
  \end{eqnarray}
Taken together, (\ref{numerator-positive-1}) and (\ref{xw-less-1-1}) delineate the range of $\mu$ within which the DW phase exists:
\begin{equation}
 \label{range-of-mu-dw}
 \bigg(\frac{\alpha}{\frac{\alpha}{\beta}-1}\bigg)<\mu<\bigg(\frac{\frac{N_{0}}{N_{1}}-\alpha}{\frac{\alpha}{\beta}-1}\bigg).
\end{equation}
The thresholds of $\mu$ for DW phase existence obtained in (\ref{range-of-mu-dw}) are not fixed but can be varied with the control parameters involved. Only the region in the $\alpha-\beta$ plane for a given $\mu$ that satisfies {\em both} (\ref{numerator-positive-1}) and (\ref{xw-less-1-1}) simultaneously admits an LDW. Notice that the line $\alpha=\beta$ satisfies (\ref{numerator-positive-1}) but not (\ref{xw-less-1-1}), meaning the line $\alpha=\beta$ falls {\em outside} the DW phase region in the phase diagram. From Eqs.~(\ref{numerator-positive-1}) and (\ref{xw-less-1-1}), we find that as $\mu$ approaches infinity, both LD-DW and HD-DW boundaries converge to the $\alpha=\beta$ line, which is below them for any finite $\mu$; see the phase diagrams in Fig.~\ref{phase-diagram-weak-coupling}.

%Remarkably, in the infinite-$\mu$ limit, we get $\alpha=\beta$ in the DW phase. The phase diagrams depicted for the values of $\mu$ at 0.5, 1, and 1.5 in Fig.~\ref{phase-diagram-weak-coupling} clearly illustrate that within the DW phase region $\alpha$ is greater than $\beta$. Additionally, for a sufficiently large value of $\mu$ (such as $\mu=100$), it is observed that the DW phase region converges into the $\alpha=\beta$ line, alike the coesistence line in an open TASEP.

See Fig.~\ref{ldw-ddw-wk} for stationary densities in the form of LDWs. In MFT, particle number conservation gives a solution for the position $x_w$ of the DW, corresponding to an LDW. However, our results from the MCS studies reveal that for smaller values of $\mu$, an LDW is obtained, whereas for  larger values of $\mu$, the DW gets delocalized. We attribute this to fluctuations; see Sec.~\ref{deloc} below for additional results and discussions.

 %Fig.~\ref{n0_wk_dw_new2} presents the localized domain wall (LDW) in the weak coupling limit of the present model at half-filling limit, with a filling factor of $\mu=3/2$. Two cases are considered: one where the exchange rates $k_{10}$ and $k_{20}$ are both set to 0.7, and another where $k_{10}=0$ and $k_{20}=0.7$. Position and height of the LDW is determined using both mean-field theory (MF) and Monte Carlo simulations (MCS), and a comparison is made between the obtained results. Upon increasing the filling fraction to a significantly larger value, such as $\mu=1000$, the domain wall gets delocalized, as illustrated in Fig.~\ref{ddw-wk-n0}. The DDW spans the entire TASEP lane, with a mean position denoted by $\langle x_{w} \rangle=1/2$, exhibiting striking similarities to the open TASEP configuration. In this large-$\mu$ regime, the conservation of particle number becomes inconsequential in determining the DW location uniquely.

 \begin{figure*}[htb]
 \includegraphics[width=\columnwidth]{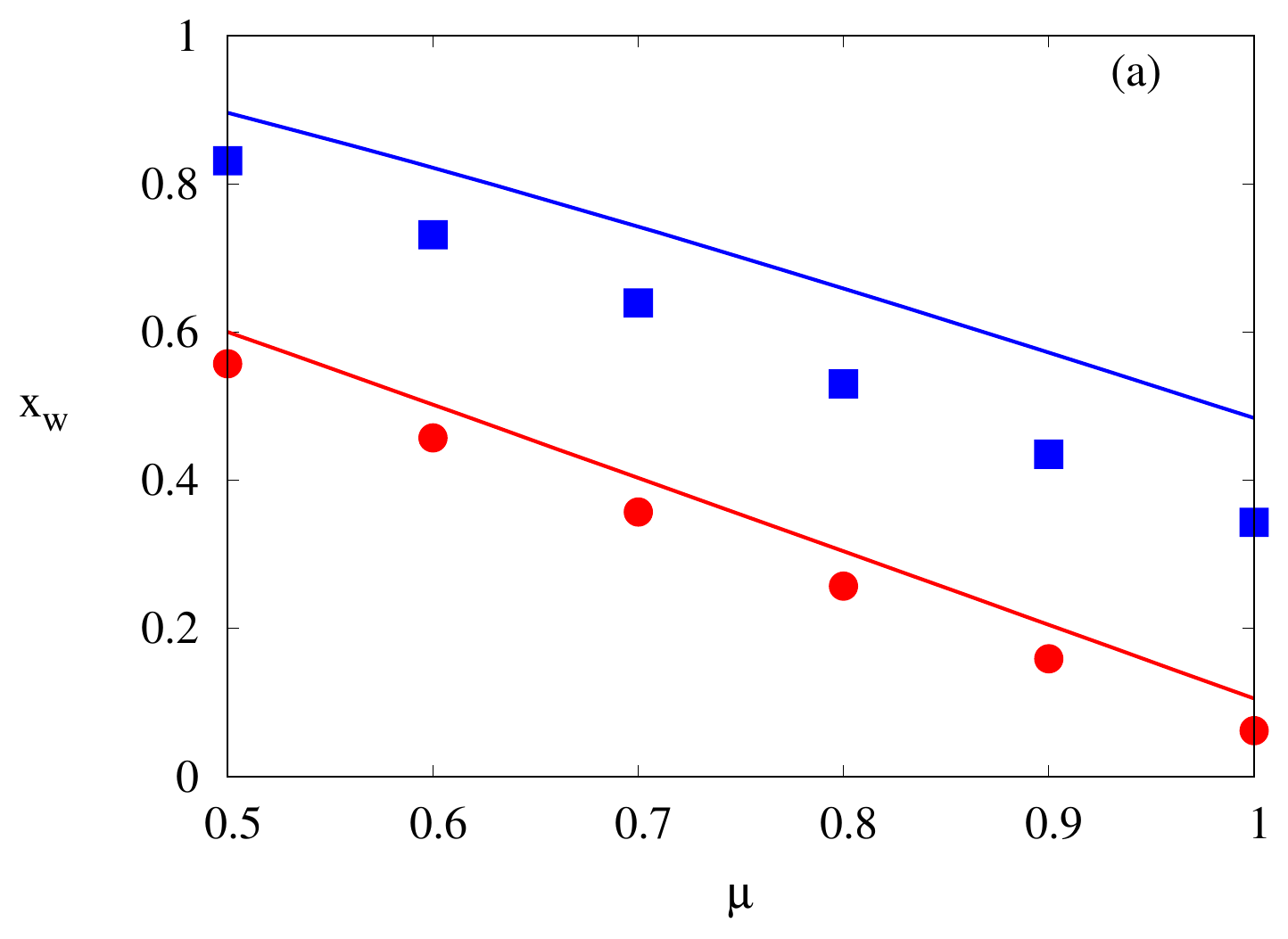}
 \hfill
 \includegraphics[width=\columnwidth]{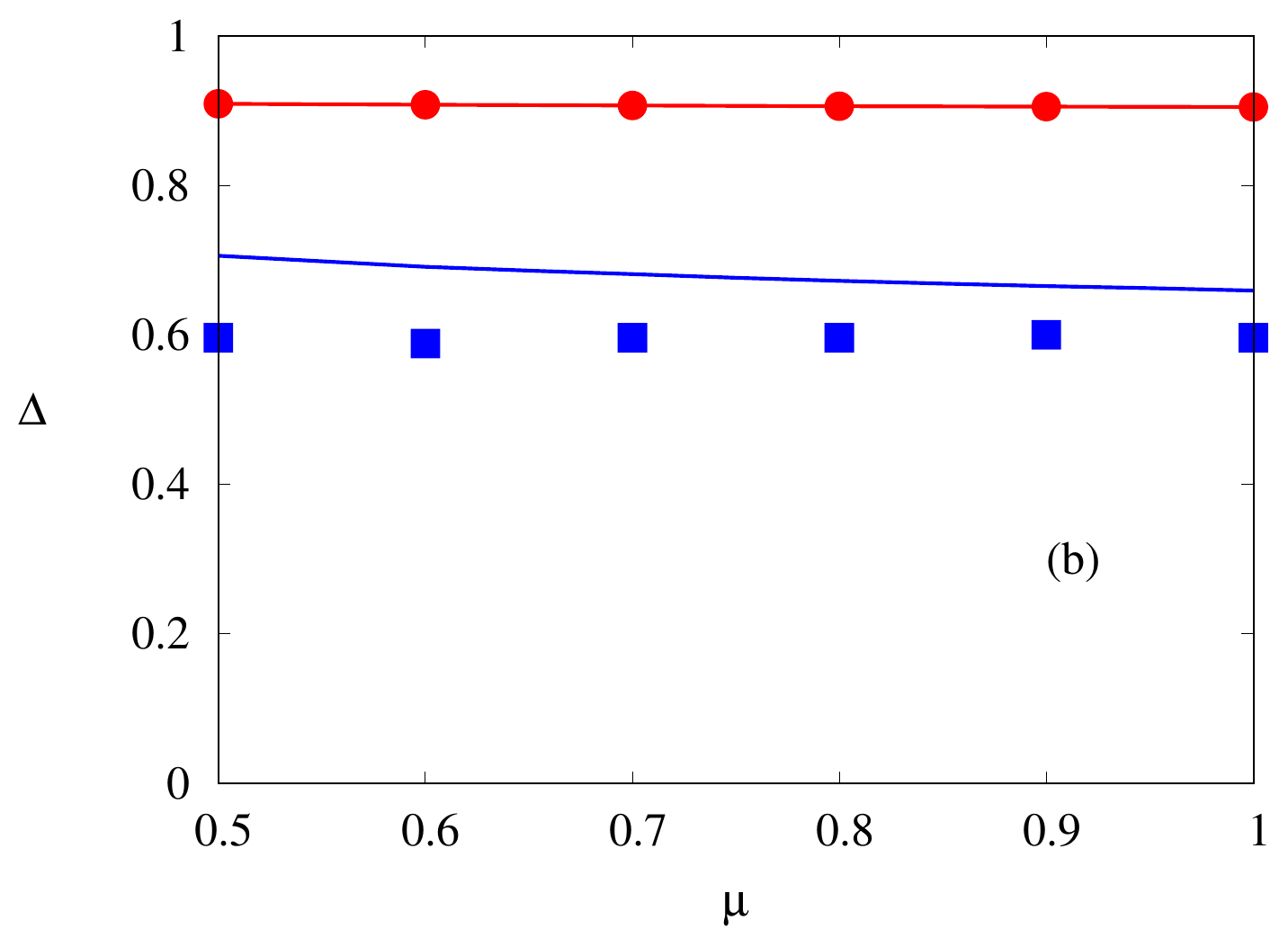}
 \\
\caption{(a) Plots of DW position $x_{w}$ vs $\mu$ in the weak coupling limit, with parameter values: $\alpha=0.5$, $\beta=0.05$, $k_{10}=0$, and $k_{20}=0.95$ (red circles); $\alpha=0.5$, $\beta=0.2$, $k_{10}=0$, and $k_{20}=0.95$ (blue squares). Solid lines depict MF results, while discrete points show MCS results. MF and MCS results agree better for lower parameter values, as seen for the red set.
(b) Plots of DW height $\Delta$ vs $\mu$ in the weak coupling limit with parameter values same as in Fig.~\ref{xw-del-vs-mu-weak}(a). These plots demonstrate that $x_{w}$ diminishes as $\mu$ increases according to Eq.~(\ref{xw model N0 weak coupling}), whereas $\Delta$ remains relatively constant with changes in $\mu$, see Eq.~(\ref{delta model L weak coupling}).}
\label{xw-del-vs-mu-weak}
\end{figure*}

\begin{figure*}[htb]
 \includegraphics[width=\columnwidth]{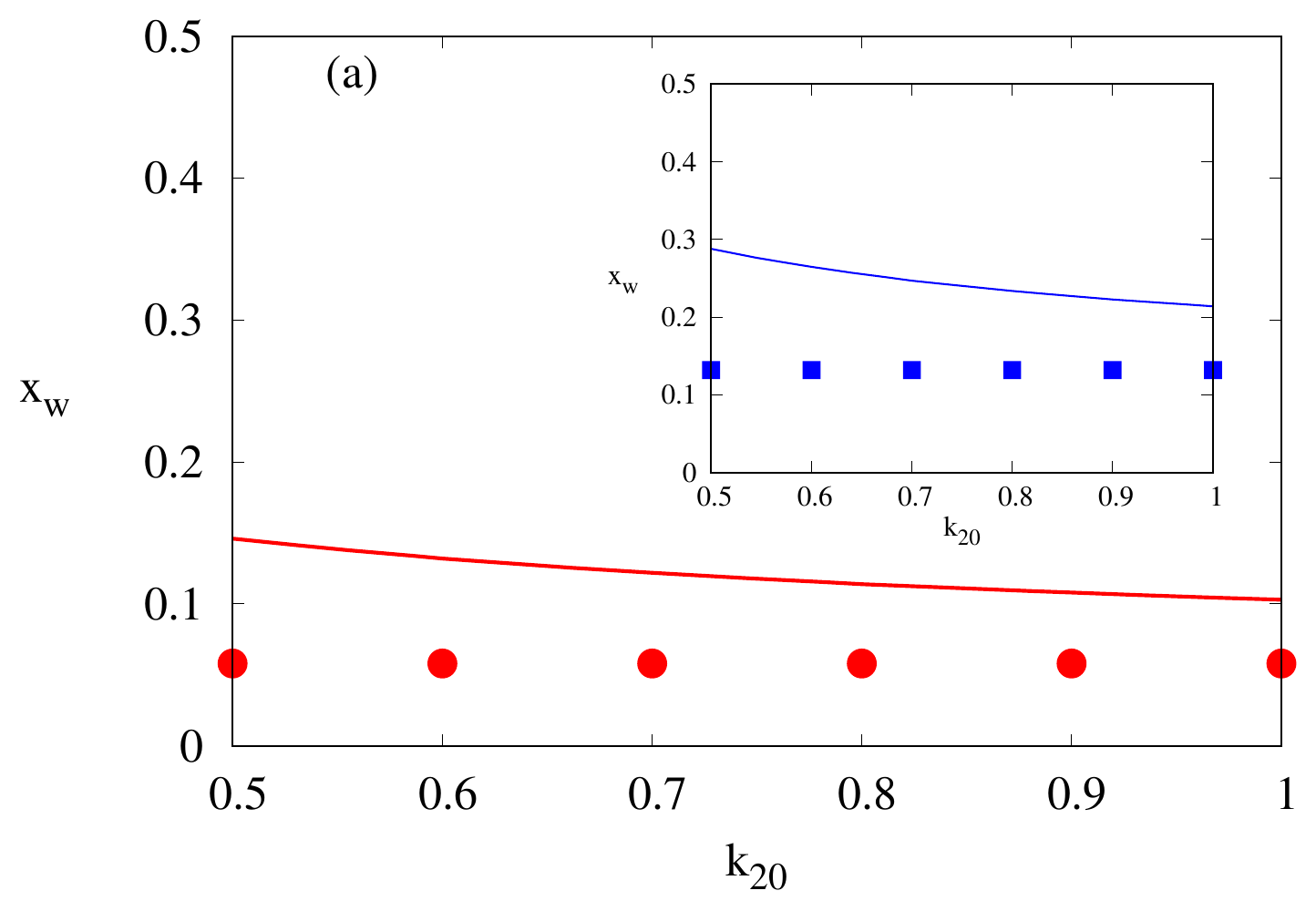}
 \hfill
 \includegraphics[width=\columnwidth]{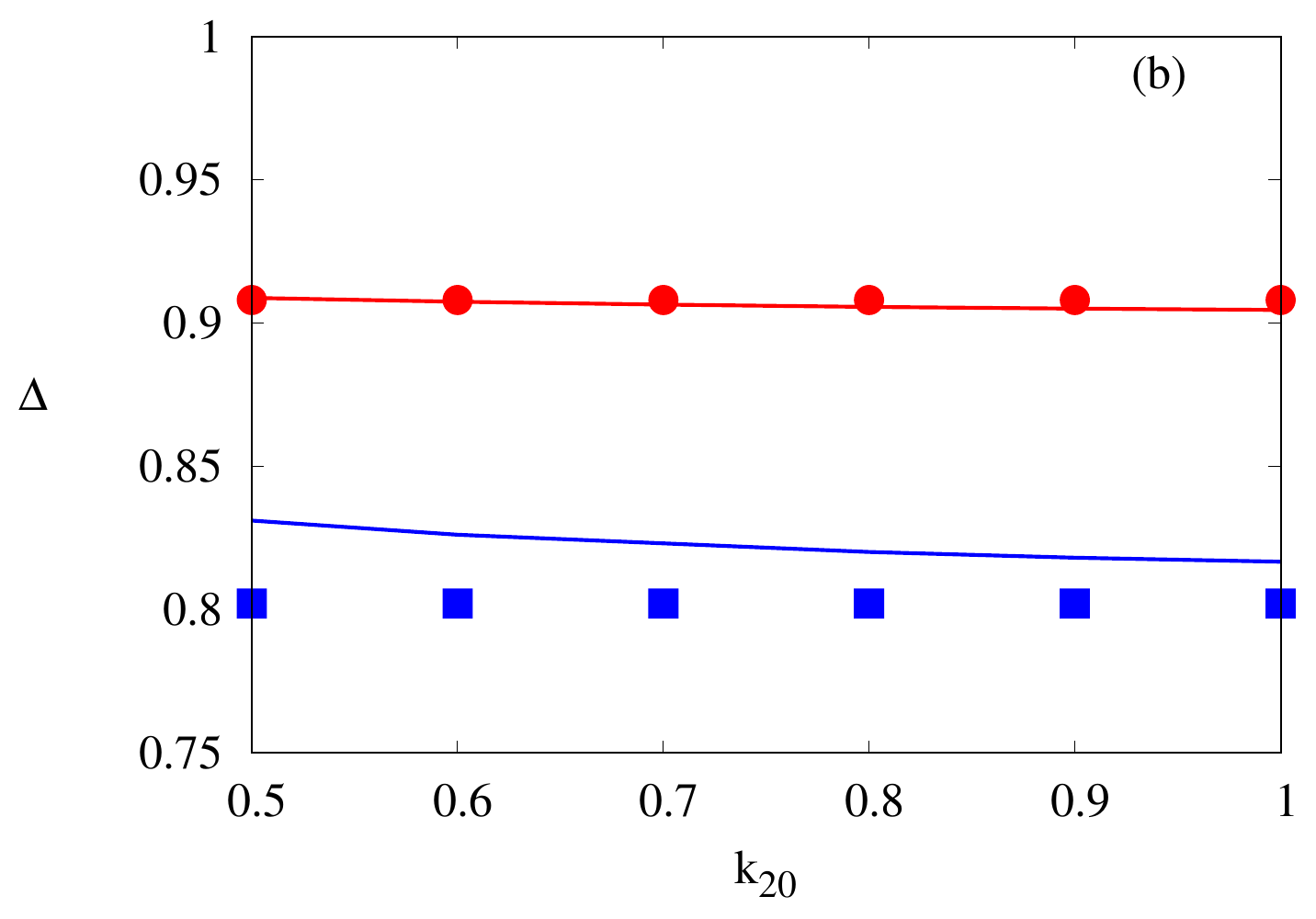}
 \\
\caption{(a) Plots of DW position $x_w$ vs $k_{20}$ in the weak coupling limit. Parameter values are: $\alpha=0.5$, $\beta=0.05$, $\mu=1$, and $k_{10}=0$ for the main plot (red circles); $\alpha=0.5$, $\beta=0.1$, $\mu=1$, and $k_{10}=0$ for the inset plot (blue squares).
(b) Plots of DW height $\Delta$ vs $k_{20}$ in the weak coupling limit. Parameter values are same as in Fig.~\ref{xw-del-vs-k20-weak}(a). MFT results obtained in Eqs.~(\ref{xw model N0 weak coupling}) and (\ref{delta model L weak coupling}) for these parameter values are shown by solid lines, while MCS outcomes are depicted as discrete points.}
\label{xw-del-vs-k20-weak}
\end{figure*}

Our MCS studies reveal that increasing the particle number causes them being accumulated near the entrance end of $T$, thus positioning the DW towards left side. This is illustrated in Fig.~\ref{xw-del-vs-mu-weak}(a), where $x_{w}$ is observed to decrease with increasing $\mu$. The DW height $\Delta$, however, remains nearly unchanged with varying $\mu$, see Fig.~\ref{xw-del-vs-mu-weak}(b). In Fig.~\ref{xw-del-vs-k20-weak}, the variation of $x_{w}$ and $\Delta$ with $k_{20}$ for a fixed value of $\mu$  is shown.

\subsection{{Phase boundaries meet at a common point}}\label{phase-meet-weak}

Within a specific range of $\mu$ where all four phases can be present in the $\alpha-\beta$ plane. In that case, there is a common point known as the (four-phase) \textit{multicritical point}. Coordinate $(\alpha_{c}, \beta_{c})$ of this point can be obtained where the LD-MC, HD-MC, LD-DW, and HD-DW phase boundaries obtained in Eqs.~(\ref{ld-mc boundary model 2 weak}), (\ref{hd-mc boundary model 2 weak}), (\ref{ld-sp boundary model 2 weak}), and (\ref{hd-sp boundary model 2 weak}) intersect:
\begin{equation}
 \label{mult-point}
 (\alpha_{c}, \beta_{c})=\bigg(\frac{\mu(k_{10}+k_{20})}{(2\mu-1)k_{20}-\frac{1}{2}}, \frac{\mu(k_{10}+k_{20})}{k_{10}+2\mu k_{20}-\frac{1}{2}}\bigg).
\end{equation}
Thus $(\alpha_c, \beta_c)$ depends explicitly on $k_{10}$, $k_{20}$, and $\mu$. In the limit $\mu \rightarrow \infty$, the multicritical point approaches

\begin{equation}
\label{mult_infty_mu}
 (\alpha_{c}, \beta_{c})_{\mu \rightarrow \infty}=\bigg(\frac{k_{10}+k_{20}}{2k_{20}}, \frac{k_{10}+k_{20}}{2k_{20}}\bigg).
\end{equation}
Thus, for $k_{10}=k_{20}$, $(\alpha_{c}, \beta_{c})_{\mu \rightarrow \infty}=(1,1)$ and for $k_{10}=0,k_{20} \ne 0$ we have $(\alpha_{c}, \beta_{c})_{\mu \rightarrow \infty}=(1/2,1/2)$. We find from (\ref{mult-point}) that this multicritical point exists for
\begin{equation}
 \label{mu-limit-mult}
 \mu>\bigg(\frac{1}{2}+\frac{1}{4k_{20}}\bigg),
\end{equation}
which is the lower threshold of $\mu$ for MC phase existence; see also (\ref{mc-range-mu-weak}).

The distance $d$ between the origin (0,0) and the multicritical point $(\alpha_{c}, \beta_{c})$ is
\begin{equation}
 \label{d-weak}
 d=\sqrt{\bigg(\frac{\mu(k_{10}+k_{20})}{(2\mu-1)k_{20}-\frac{1}{2}}\bigg)^{2} + \bigg(\frac{\mu(k_{10}+k_{20})}{k_{10}+2\mu k_{20}-\frac{1}{2}}\bigg)^{2}}.
\end{equation}
This implies that the distance $d$ diverges as $\mu$ approaches the threshold $(1/2+1/4k_{20})$ from above. We get
\begin{equation}
 \label{d-weak-mu-infty}
 d\xrightarrow{\mu \rightarrow \infty}\frac{k_{10}+k_{20}}{\sqrt{2}k_{20}}.
\end{equation}
Consequently for $k_{10}=k_{20}$ one gets $d \rightarrow \sqrt{2}$, while $d \rightarrow 1/\sqrt{2}$ when $k_{10}=0,k_{20} \ne 0$.

  \section{NATURE OF THE PHASE TRANSITIONS}
 \label{dw and pt}

 %To provide context, let us recall the phase diagrams of the standard open TASEP where the LD, HD, and MC phases appear with a domain wall delocalized in space due to particle number non-conservation. In this model, steady-state bulk density $\rho_\text{bulk}$ serves as the order parameter. The transition from LD(HD) to MC phase is identified by a continuous change in $\rho_\text{bulk}$, wherein the LD to HD phase transition is marked by a \textit{sudden} jump in $\rho_\text{bulk}$. Thus, two second-order (LD or HD to MC) and one first-order (LD to HD) phase transitions are involved. All these phases meet at a common point ($1/2,1/2$) in the phase space of control parameters $\alpha$ and $\beta$.

 As shown in Fig.~\ref{phase-diagram-weak-coupling}, the phase diagrams in weak coupling limit of the present model are markedly different from that for an open TASEP for $\mu$ not too large. This is attributed to the global particle number conservation and dynamically controlled effective entry and exit rates for the TASEP channel. Depending on the values of $\mu$, there are either two or four phases in the $\alpha-\beta$ plane; see Fig.~\ref{phase-diagram-weak-coupling}. We now briefly discuss the nature of the phase transitions. As in an open TASEP, we consider the steady-state bulk density ($\rho$) as the order parameter. We recall that in open TASEP, the transition between LD and HD phases is marked by a sudden jump in $\rho$, thus implying it a first-order phase transition. On the other hand, the transitions between either LD or HD phases and the MC phase involve continuous changes in $\rho$, indicative of  second-order phase transitions. In the present model, there are {\em no} phase transitions between the LD and HD phases for any finite $\mu$. Instead, there are transitions from the LD to DW and DW to HD phases, both of which are associated with continuous change in $\rho$ across the corresponding phase boundaries.  Similarly, all the other phase transitions, such as LD-MC and HD-MC transitions in the present model are also second-order in nature characterized by a continuous change in $\rho$. Finally, in the asymptotic limit of $\mu\rightarrow \infty$, the LD-DW and DW-HD phase boundaries merge and overlap with the line $\alpha=\beta$, effectively giving a transition between the LD and the HD phases. This transition is {characterized by a discontinuous change in $\rho$}, thus a first-order transition as in an open TASEP.

\section{ROLE OF FLUCTUATIONS}
\label{role of fluctuations}

%\begin{widetext}

 \begin{figure*}[ht]
 \begin{minipage}{0.32\linewidth}
%\centering  % redundant
\includegraphics[width=\textwidth]{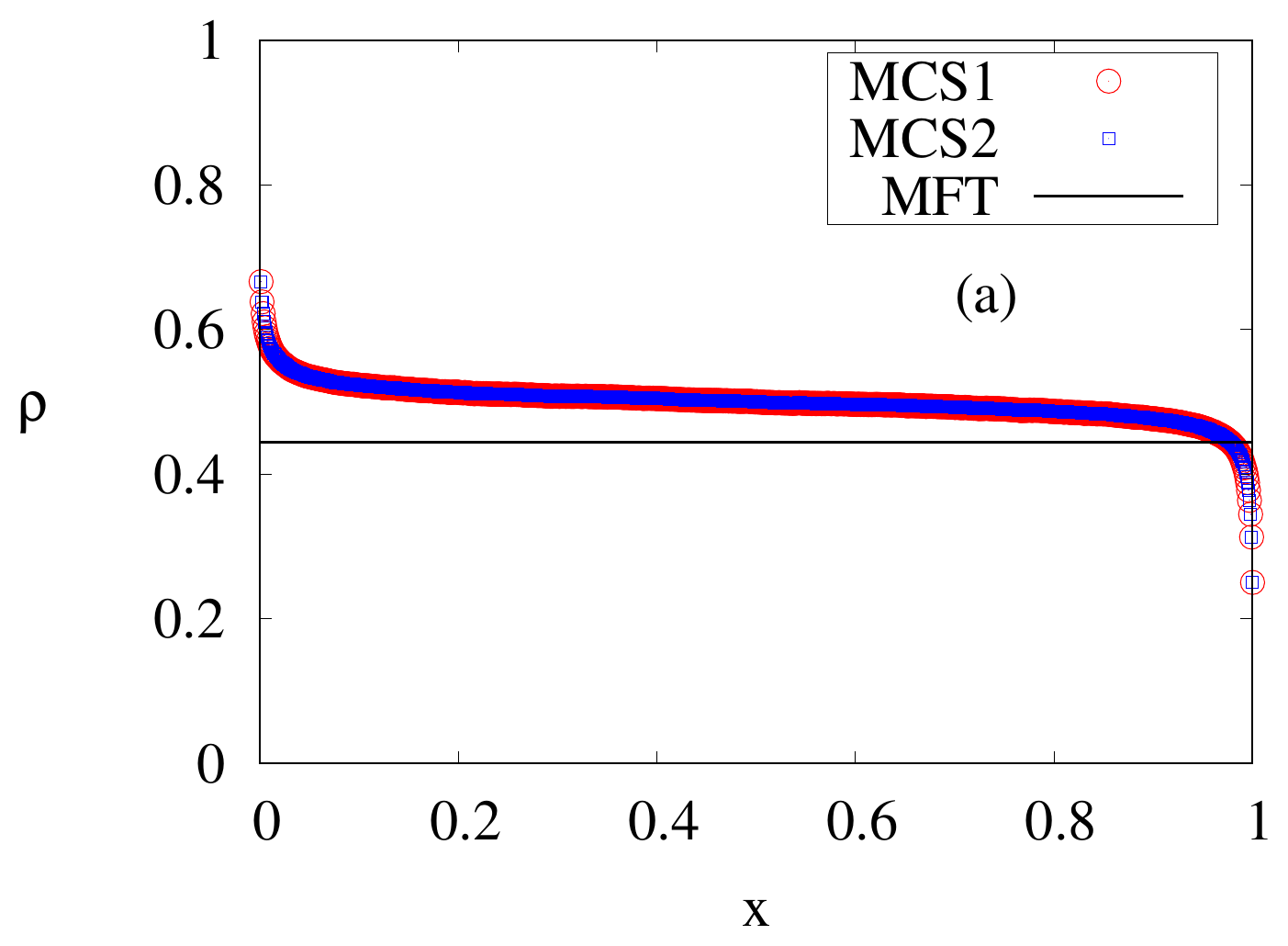}
%\caption{(d)}
\label{subfigure a}
\end{minipage}%
\hfill% not: "\hspace{0.5cm}"
\begin{minipage}{0.32\linewidth}
%\centering  % redundant
\includegraphics[width=\textwidth]{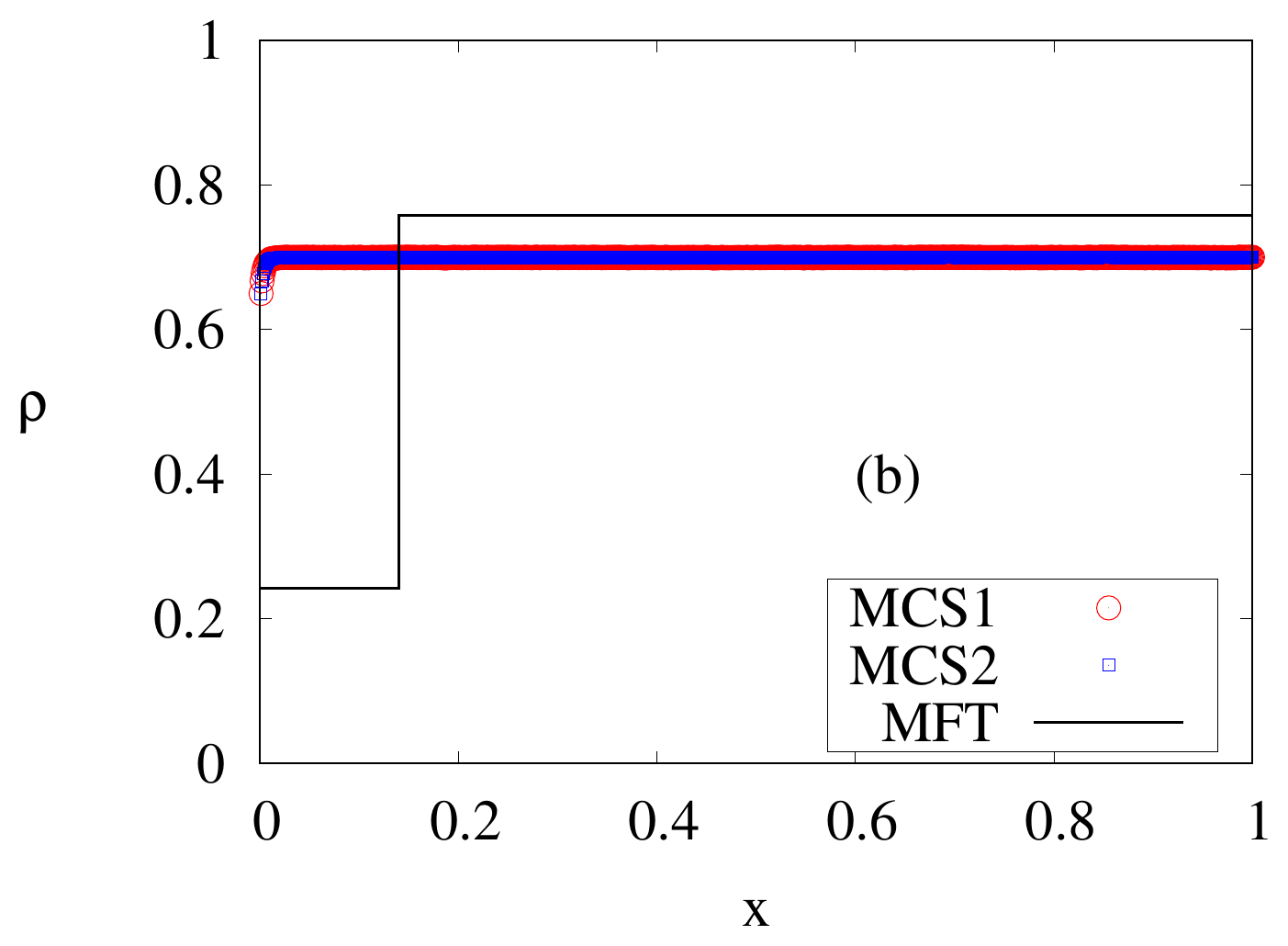}
%\caption{(e)}
\label{subfigure b}
\end{minipage}%
\hfill% not: "\hspace{0.5cm}"
\begin{minipage}{0.32\linewidth}
%\centering  % redundant
\includegraphics[width=\textwidth]{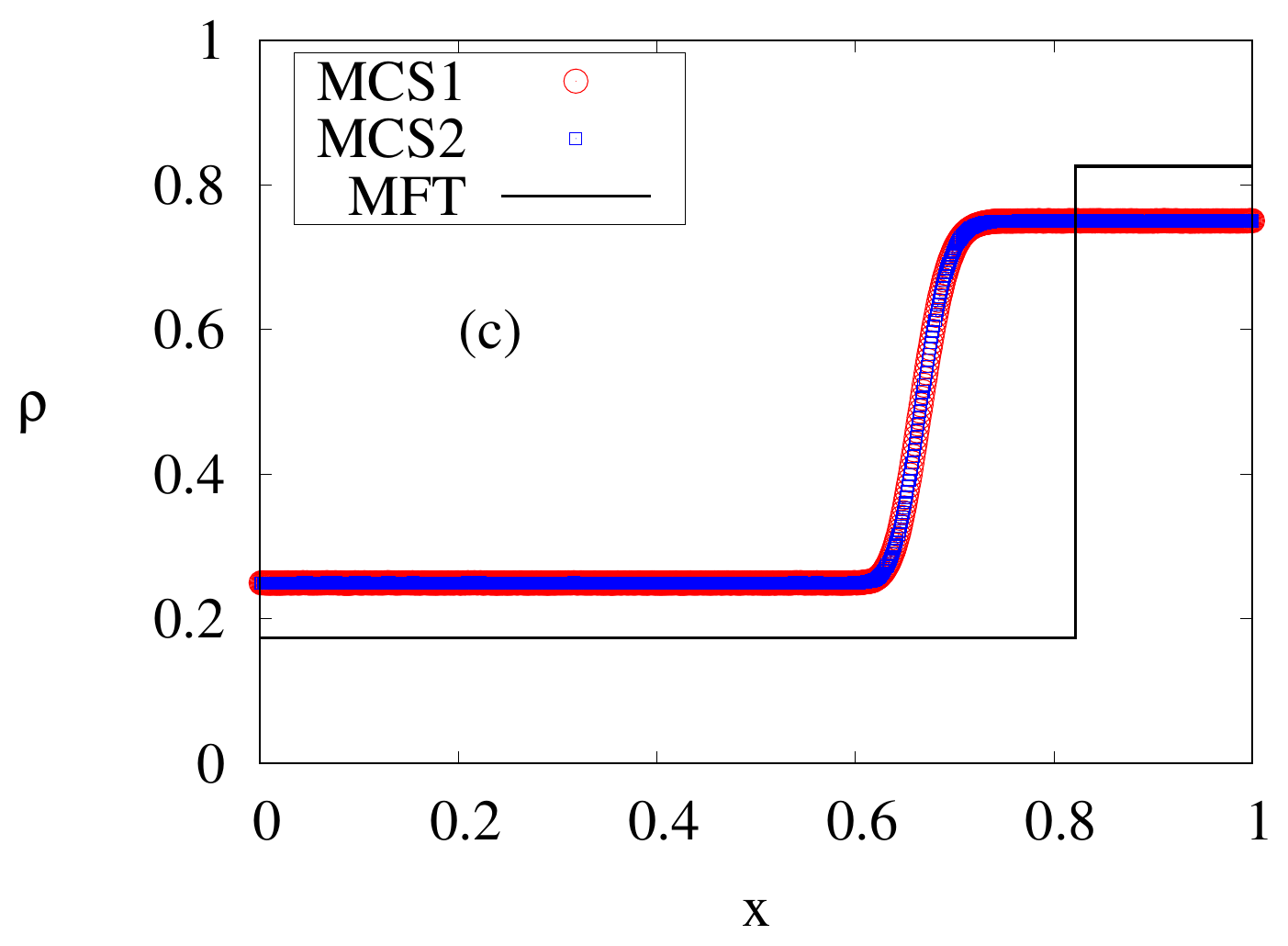}
%\caption{(f)}
\label{subfigure c}
\end{minipage}
\caption{Disagreements in the MFT and MCS density profiles in the weak coupling limit of the model. Solid lines represent MFT densities, while discrete points show MCS densities. Two sets of MCS results, labeled as MCS1 (red circles) and MCS2 (blue squares), correspond to time-averaged densities with $2 \times 10^{9}$ and $2 \times 10^{10}$ runs, respectively. Notably, fluctuations persist even for longer runs. The system size is fixed at $L=1000$. (a) For parameters $\alpha=1.5$, $\beta=1.2$, $\mu=1$, $k_{10}=0$, and $k_{20}=0.95$, MFT (see Eq.~(\ref{ld density model 2 weak acc})) predicts an LD phase with density $\rho_\text{LD}=0.44$, while MCS suggests an MC phase with density $\rho_\text{MC}=0.5$. (b) In the case of $\alpha=2$, $\beta=0.3$, $\mu=1$, $k_{10}=0$, and $k_{20}=0.95$, MFT (see Eq.~(\ref{xw model N0 weak coupling}) and (\ref{ld-den-dw-wk})) and MCS predict DW and HD phases (with density $\rho_\text{HD}=0.69$), respectively. (c) With parameters $\alpha=1.5$, $\beta=0.25$, $\mu=0.5$, $k_{10}=0$, and $k_{20}=0.95$, both MFT (see Eq.~(\ref{xw model N0 weak coupling}) and (\ref{ld-den-dw-wk})) and MCS predict a DW phase. However, MFT suggests an LDW with different values of low and high densities and positions compared to MCS.}
\label{rof}
\end{figure*}

%\end{widetext}

A significant deviation, particularly for small and intermediate values of $\mu$, between our analytical mean-field and simulated Monte Carlo results has been observed in the weak coupling limit of the model; see Figs.~\ref{phase-diagram-weak-coupling}-\ref{xw-del-vs-k20-weak}, and Figs.~\ref{rof}, \ref{DDW-weak}. In contrast, we have observed excellent agreement between MFT and MCS results in the strong coupling case, see Fig.~\ref{pd strong}-\ref{xw-del-vs-k20-strong}, and Fig.~\ref{n1/n2-vs-mu-and-k1/k2} in the Appendix~\ref{mft strong coupling}.  %We 
%also explain the behavior of a domain wall for large $\mu$.
In Fig.~\ref{rof}, discrepancies between MFT predictions and MCS results in steady-state density profiles in the weak coupling limit are shown for a set of control parameters. For the parameters specified in Fig.~\ref{rof}(a), MFT gives an LD phase with density $\rho_\text{LD}=0.44$, while MCS gives an MC phase. In Fig.~\ref{rof}(b), MFT predicts an LDW, whereas MCS yields HD phase with density $\rho_\text{HD}=0.69$. In Fig.~\ref{rof}(c), both MFT and MCS give an LDW, but with different densities and positions. The system size is $L=1000$ and time-average with $2 \times 10^{9}$(red circles) and $2 \times 10^{10}$(blue squares) number of runs are performed. Nonetheless, the disagreements between the MFT and MCS predictions persist, with no significant improvement. In this section, we explain the reason behind such discrepancies; see also Ref.~\cite{sourav-1} for similar discussions.

Below we set up the heuristic arguments to explain the fluctuation-induced disagreements between the MFT and MCS results in the thermodynamic limit by following and extending the arguments developed in Ref.~\cite{sourav-1}.
Fluctuations induce deviations in the effective values of reservoir populations from their respective MFT values. One can thus generalize the relations $\alpha_\text{eff}=\alpha N_1/N_0$ and $\beta_\text{eff}=\beta(1-N_{2}/N_0)$ to
\begin{equation}
\label{eff_alpha_beta}
 \tilde\alpha_\text{eff}=\frac{\alpha}{N_0}N_1^\text{eff}, \hspace{5mm} \tilde\beta_\text{eff}=\beta\bigg(1-\frac{N_{2}^\text{eff}}{N_0}\bigg)
\end{equation}
with
\begin{equation}
 N_1^\text{eff}=N_1 +\delta N_1, \hspace{5mm} N_2^\text{eff}=N_2 +\delta N_2,
\end{equation}
where $N_1$, $N_{2}$ are the solution for the populations of reservoirs $R_1$, $R_{2}$ in MFT, i.e., solutions in Eqs.~(\ref{N1w}) and (\ref{N2w}), $\delta N_1$, $\delta N_2$ are the corresponding deviations due to the fluctuations from their respective mean-field values, and $\tilde\alpha_\text{eff}, \tilde\beta_\text{eff}$ are the effective entry, exit rates in MCS studies. Since $\delta N_1$ and $\delta N_2$ are fluctuations, they can be positive or negative, making $\tilde\alpha_\text{eff}-\alpha_\text{eff}$ and $\beta_\text{eff}-\tilde\beta_\text{eff}$ positive or negative accordingly.

In the weak coupling limit, as shown in Eqs.~(\ref{N1w}) and (\ref{N2w}), $N_{1}$ and $N_{2}$ have dependence on TASEP current $J_{T}$ that scales as $LJ_{T}$. $J_{T}$ being a bounded quantity ($J_{T_\text{max}}=1/4$), fluctuation in $J_{T}$ is $\mathcal{O}(1)$. Hence the reservoir populations $N_{1},N_{2}$ as well as their fluctuations $\delta N_{1},\delta N_{2}$ must be $\mathcal{O}(L)$ quantities. Indeed, the ratio of $N_{1}, N_{2}$ depends explicitly on $J_{T}$. This immediately gives $\tilde\alpha_\text{eff}-\alpha_\text{eff}\sim {\cal O}(1)$ and $\beta_\text{eff}-\tilde\beta_\text{eff}\sim {\cal O}(1)$ even in the thermodynamic limit. Furthermore, the sum of the two reservoir populations $N_R=N_0-N_T$ should scale with $L$ (for a fixed $\mu$) and hence should have typical fluctuations of the size ${\cal O}(\sqrt {N_R})\sim {\cal O}(\sqrt L)\ll$ fluctuations in $N_1$ or $N_2$ in the limit of large $L$. Given that $N_1 + N_2 = N_R$, we must have $\delta N_1\sim -\delta N_2$ to the leading order in $L$ for sufficiently large $L$. This in turn means when $N_1^\text{eff}$ is larger (smaller) than $N_1$, $N_2^\text{eff}$ is smaller (larger) than $N_2$, giving   $\tilde\alpha_\text{eff}>(<) \alpha_\text{eff}$ and $\tilde\beta_\text{eff} >(<) \beta_\text{eff}$. This suggests that in the weak coupling case, fluctuations persist in the thermodynamic limit, causing the transition between LD or HD phases to occur at values of $\alpha$ and $\beta$ distinct from those predicted by MFT. Contrastingly, in the strong coupling regime, $N_1$ and $N_2$ become independent of $J_T$ in the thermodynamic limit and are found to maintain a fixed ratio, $N_{1}/N_{2}=k_{2}/k_{1}$, see Appendix. The particle exchange dynamics between reservoirs effectively resemble equilibrium dynamics, leading to $\delta N_1$ and $\delta N_2$ scaling as ${\cal O}(\sqrt L)$. As the MFT expressions for $N_1$ and $N_2$ scale with $L$, in the thermodynamic limit, $\tilde\alpha_\text{eff} \rightarrow \alpha_\text{eff}$ and $\tilde\beta_\text{eff} \rightarrow\beta_\text{eff}$. This convergence ensures a strong quantitative agreement between MFT and MCS results in the strong coupling case.

Interestingly, the phase diagrams in both weak and strong coupling limit reveals that MFT and MCS results are in excellent agreement for large values of $\mu$, see Fig.~\ref{phase-diagram-weak-coupling}(d) and Fig.~\ref{pd strong}(d). This is argued below. With $N_{0}=\mu L$, Eq.~(\ref{eff_alpha_beta}) leads to
\begin{eqnarray}
 &&\tilde\alpha_\text{eff}=\frac{\alpha(N_{1}+\delta N_{1})}{N_0}=\alpha_\text{eff}+\frac{1}{\mu}\frac{\alpha \delta N_{1}}{L},\label{alpha_fluc}\\
 &&\tilde\beta_\text{eff}=\beta\bigg(1-\frac{N_{2}+\delta N_{2}}{N_0}\bigg)=\beta_\text{eff}-\frac{1}{\mu}\frac{\beta \delta N_{2}}{L}, \label{beta_fluc}
\end{eqnarray}
 where, as above, entry and exit rates \textit{with} and \textit{without} tilde corresponds to MCS and MFT respectively. As $\delta N_{1}, \delta N_{2}$ are $\mathcal{O}(L)$ quantities in the weak coupling limit, fluctuations in the MCS values of entry and exit rates (second part of Eqs. (\ref{alpha_fluc}) and (\ref{beta_fluc})) are clearly $\mathcal{O}(1/\mu)$ quantities that vanishes in the limit $\mu \rightarrow \infty$, thereby giving $\tilde\alpha_\text{eff} \rightarrow \alpha_\text{eff}$ and $\tilde\beta_\text{eff} \rightarrow \beta_\text{eff}$ in the thermodynamic limit.

\section{Delocalization of the Domain Wall}\label{deloc}

\begin{figure}[htb]
 \includegraphics[width=\columnwidth]{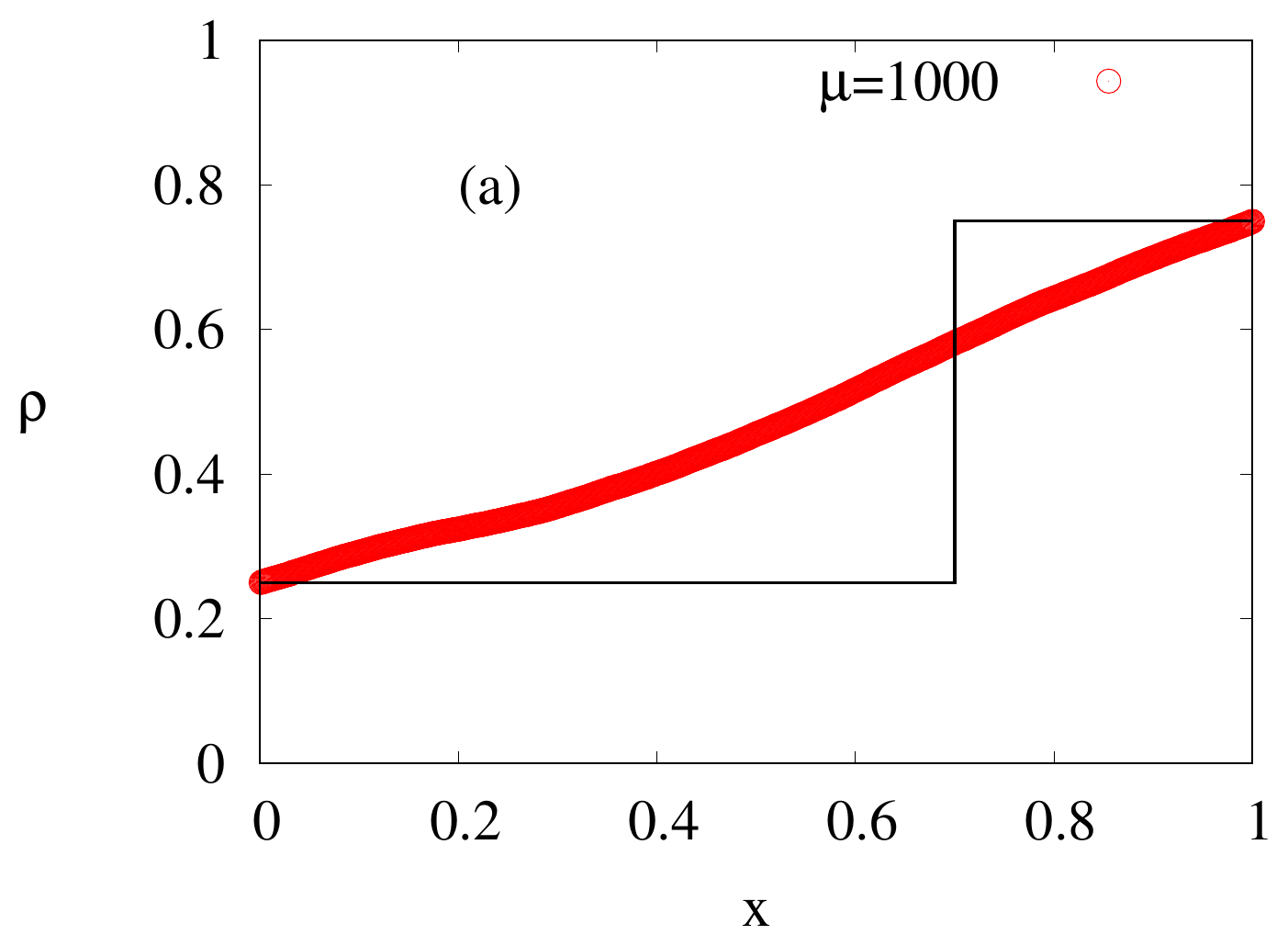}\\
%\caption{(e)}
%\centering  % redundant
\includegraphics[width=\columnwidth]{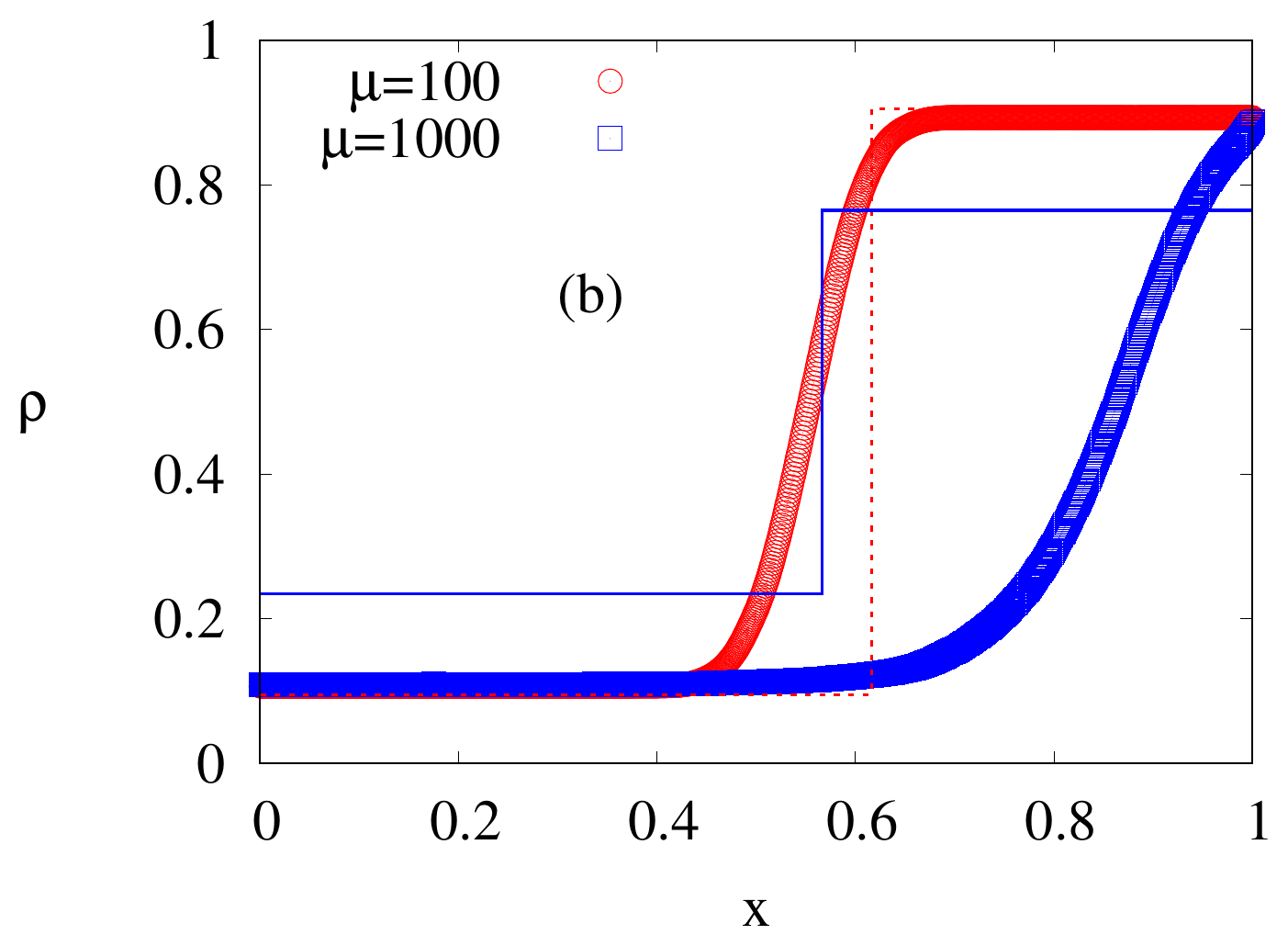}
%\caption{(f)}
\caption{Plots of domain walls in the weak coupling limit of the model with system size $L=1000$. (a) $\alpha=0.25$, $\beta=0.2499$, $\mu=1000$, $k_{10}=0$, and $k_{20}=0.95$. Increasing $\mu$ sufficiently (of the order of $L$), we get DDW. (b) $\alpha=10$, $\beta=7$, $\mu=100$, $k_{10}=0.95$, and $k_{20}=0.01$ (red circles); $\alpha=23$, $\beta=22$, $\mu=1000$, $k_{10}=0.95$, and $k_{20}=0.01$ (blue squares). With these sets of parameter values where $k_{10}>>k_{20}$, only LDWs are formed even with very large values of $\mu$; see text in Section~\ref{deloc}.}
\label{DDW-weak}
\end{figure}

%\begin{minipage}{0.32\linewidth}
%\centering  % redundant
For smaller values of $\mu$, one gets localized DWs, as can be seen in Fig.~\ref{ldw-ddw-wk}, positions $x_{w}$ of which can be obtained from Eq.~(\ref{xw model N0 weak coupling}). However, for large values of $\mu$, the DW clearly gets delocalized; see Fig.~\ref{DDW-weak}(a) and also Fig.~\ref{ddw-scaling}. More intriguingly, for $k_{10}>>k_{20}$, the tendency to delocalize vanishes even with very large values of $\mu$; see Fig.~\ref{DDW-weak}(b). We now explore these systematically below by using our analysis in Section~\ref{role of fluctuations} above.

%In the middle plot when $\mu=1000$, we get a DDW. The system size is $L=1000$. The delocalization of domain wall occurs when $\mu \sim L$ for both weak and strong coupling case, see Fig.~\ref{ddw-scaling}. This is argued below. We know,

%The steady-state densities in the LD and HD phases read $\rho_\text{LD}=\alpha_\text{eff}$ and $\rho_\text{HD}=1-\beta_\text{eff}$. When $\mu \rightarrow \infty$, we get $\alpha_\text{eff} \rightarrow \alpha/2$, $\beta_\text{eff} \rightarrow \beta/2$ when particle exchange rates are equal ($k_{10}=k_{20}$ or $k_{1}=k_{2}$), and $\alpha_\text{eff} \rightarrow \alpha$, $\beta_\text{eff} \rightarrow \beta$ when $k_{10}=0$ or $k_{1}=0$, see Eqs.~(\ref{rhold-mu-infty-wk}), (\ref{hd-mu-infty-wk}), (\ref{ld-st-mu-infty}), and (\ref{hd-st-mu-infty}). Hence for large $\mu$, $N_{1},N_{2} \sim \mathcal{O}(\mu L)$, except for the limiting cases where one or the other may be small.
We have in general, $N_{1},N_{2} \sim \mathcal{O}(\mu L)$.  This means typical fluctuations in $N_{1},N_{2}$ $\sim {\cal O} (\sqrt{\mu L})$. Now, for a single DDW-like domain wall to exist in TASEP lane, domain wall position $x_{w}$ must fluctuate over a scale comparable to $L$, which gives a typical number fluctuation in TASEP as ${\cal O}(L)$. Number conservation implies number fluctuations in the TASEP must be the same as the number fluctuations in $N_1,\,N_2$. This means that fluctuations in $N_{1}$ and $N_{2}$ $\sim \mathcal{O}(L)$, for a DDW to exist, which immediately gives $\mu \sim L$ for DDW to occur~\cite{parna-anjan}. Our MCS results in Fig.~\ref{DDW-weak}(a) satisfies this. However, in Fig.~\ref{DDW-weak}(b) with $k_{10}>>k_{20}$ we do {\em not} get a DDW, in spite of having a large $\mu$. This can be explained by considering the Fokker-Planck equation for the probability $P$ to find the DW at a point $x_w$ in the bulk~\cite{tirtha-qkpz,astik-parna,erwin-tobias}, which has the form
\begin{equation}
 \frac{\partial P}{\partial t}=D\frac{\partial^2 P}{\partial x^2}.
\end{equation}
Here, $D\equiv [\alpha_\text{eff}(1-\alpha_\text{eff}) + \beta_\text{eff}(1-\beta_\text{eff})]$ is the diffusivity. For a DW, $\alpha_\text{eff}=\beta_\text{eff}$, for which $D$ in Fig.~\ref{DDW-weak}(b) becomes very small. This means the typical time-scale $\tau \equiv 1/D$ for the DW to span the entire TASEP channel due to fluctuations becomes very large, which effectively localizes the DW. This explains an LDW in Fig.~\ref{DDW-weak}(b).

 \section{CONCLUSIONS AND OUTLOOK}
 \label{conclusions}

 In this work, we have studied the generic nonequilibrium steady-state density profiles and phase diagrams of a  TASEP lane executing asymmetric exclusion processes and connecting two particle reservoirs $R_{1}$ and $R_{2}$ with unlimited storage capacity and which directly exchange particle. %Particles from the left reservoir $R_{1}$ enter the TASEP lane, followed by a unidirectional (from left to right) hopping throughout the TASEP lane, subject to exclusion and eventually exit to the right reservoir $R_{2}$. To model diffusion in our system, the reservoirs are allowed to exchange particles directly between them. 
 % and uncorrelated. 
 %In this model, there is no upper bound for the overall particle number (including the TASEP and reservoirs) which is held conserved by the dynamics. This 
%sets our model apart from the particle nonconserving dynamics of open TASEP. 
Actual or effective entry and exit rates in our model are controlled by the entry and exit rate parameters, $\alpha$ and $\beta$ respectively, together with the two functions $f$ and $g$. We choose simple forms of $f$ and $g$, which allow unlimited capacity, see Eq.~(\ref{Model 1}).%, where $f$ and $g$ respectively are the monotonically increasing and decreasing functions of $N_{1}$ and $N_{2}$ with values between 0 and 1, $N_{1}$ and $N_{2}$ being the populations of reservoirs $R_{1}$ and $R_{2}$. Taken together, the dynamics of the system is influenced by five control parameters -- one entry and exit rate ($\alpha$ and $\beta$), two particle exchange rates ($k_{1}$ and $k_{2}$), and a filling factor ($\mu$). 
The behavior of the system is particularly interesting when particle hopping throughout the TASEP and direct particle exchange between the reservoirs compete. This corresponds to the \textit{weak coupling limit}, captured by scaling $k_{1}$ and $k_{2}$ with $\mathcal{O}(1/L)$, where $L$ represents the lattice size. We have also studied the \textit{strong coupling limit}, where direct particle exchange between reservoirs dominates particle current through TASEP lane, i.e., when $k_{1},k_{2} \sim \mathcal{O}(1)$. %We have investigated the phase diagrams for two special cases: (i) $k_{1}=k_{2}$ and (ii) $k_{1}=0,k_{2}\ne0$.

 { We have analyzed our model at the mean-field level and corroborated the results with Monte Carlo simulations. For both the weak and strong coupling limits of the model, depending on the value of $\mu$, generally either two or four phases appear simultaneously in the $\alpha-\beta$ phase plane, with continuous transitions between them. 
 Interestingly, in the limit of infinite $\mu$, the phase diagram of our model mimics that of an open TASEP. Indeed, when $\mu \rightarrow \infty$, the phase space of the present model includes three phases -- LD, HD, and MC -- with a DW region in the form of an inclined line connecting LD phase to the HD phase. Manifested as localized domain walls (LDWs) for smaller $\mu$ values, these are now  delocalized domain walls (DDWs) in the limit of $\mu \rightarrow \infty$. While particle number conservation together with MFT still predicts an LDW, large fluctuations around the mean-field position of the DW effectively delocalizes the domain wall, giving a DDW. The weak coupling limit differs from its strong counterpart for moderate $\mu$, when fluctuation effects are important for the reservoir-TASEP couplings, which in turn lead to quantitative failure of the MFT. This has been explained by phenomenological arguments, which clearly distinguishes the weak coupling limit from the strong coupling limit for moderate $\mu$. For very large $\mu$, weak and strong coupling limit results coincide, a fact explained phenomenologically above.}
 
 %Overall particle number, though conserved, no longer precisely determines their location, as the enhanced particle population delocalizes the DW, akin to the conventional DDW in an open TASEP.

{ The model in Ref.~\cite{sourav-1} and the one studied here have similar structures in having one TASEP lane connecting two reservoirs, which in turn can exchange particles directly. These two models nonetheless are clearly distinct, as delineated by their respective reservoir-TASEP couplings and the associated constraints imposed on them, which have strong consequences on their  stationary densities and phase diagrams. First of all, the reservoirs of the model in Ref.~\cite{sourav-1} have finite carrying capacities, as controlled by the specific reservoir-TASEP couplings: the maximum number of particles (or, equivalently an upper limit on $\mu$), or ``maximum resources'' that the model reservoirs in Ref.~\cite{sourav-1} can accommodate is {\em finite}.  This appears as a constraint on the model, in addition to the other constraint of actually available resources, or total number of particles. The presence of these two constraints  allows the model of Ref.~\cite{sourav-1} to study the interplay between the two constraints of finite resources and finite carrying capacities, which ultimately determine the steady-states.  In contrast, the reservoir carrying capacities of the model in the present study is {\em unrestricted} and hence $\mu$ can be freely chosen.  This makes the present model a one constraint model, having the {{finite}} resources as the only constraint acting on it. At a technical level, the model in Ref.~\cite{sourav-1} has the particle-hole symmetry for equal particle exchange rates between the two reservoirs, whereas the model in the present study {\em does not} admit any such symmetry for {\em any } values of the model parameters. This is attributed to the specific reservoir-TASEP couplings used in the two models.

Due to the specific reservoir-TASEP couplings, the model in Ref.~\cite{sourav-1} {\em does not} reduce to an open TASEP for any  value of the filling fraction $\mu$.  In contrast,  the phase diagram of the present model in the plane of the control parameters smoothly reduces to that for an open TASEP as the filling fraction $\mu$ grows bigger without any bound. This holds both in the weak and strong coupling limits of the model; see the phase diagrams in Fig.~\ref{phase-diagram-weak-coupling}(d) and Fig.~\ref{pd strong}(d).  Furthermore, the quantitative differences between the weak and strong coupling limits of the model disappear for a sufficiently large $\mu$, in contrast to Ref.~\cite{sourav-1}. Related to this and in stark contrast to the model in Ref.~\cite{sourav-1}, fluctuation effects are increasingly less important with the mean-field theories becoming quantitatively accurate even in the weak coupling limit of the present model, as $\mu$ grows.  In addition,  a domain wall in the present model can delocalize if sufficiently large number of particles are available, or for sufficiently large resources (or $\mu$) for appropriate choice of the other model parameters.%; see, e.g., the stationary densities in Fig.~\ref{DDW-weak}(a) and Fig.~\ref{ddw-scaling}. 
This is {\em not} possible in the model of Ref.~\cite{sourav-1}, due to the finite carrying capacity there, making a domain wall localized~\cite{sourav-1}.  %This ultimately leads to the significant differences in the phase diagrams in the two models for large ``resources'' or large number of particles, i.e., for large values of $\mu$.  
Lack of one constraint leading to unrestricted carrying capacity thus establishes the present model as belonging to a distinct class, different from the one for the model in Ref.~\cite{sourav-1}. }

{ Our quantitative results here are obtained for the particular  choices of the functions $f$ and $g$, as given in Eq.~(\ref{effective entry and exit rates}), which define the reservoir-TASEP couplings. It then begs the question how ``universal'' are these results, or how much will the results change if the specific forms for $f$ and $g$ are altered. This change in $f$ and $g$ can be made by  retaining their ``basic forms'', i.e., their monotonic dependence on their arguments (as in Eq.~(\ref{effective entry and exit rates})), still allowing for unrestricted carrying capacities. From our calculational scheme above, such modifications in $f$ and $g$ will change the specific locations of the phase boundaries in the $\alpha-\beta$ plane, and the quantitative dependence of the stationary densities on $\alpha,\beta$. However, the general topology of the phase diagrams and the fact that for very high $\mu$, the phase diagrams are identical to that of an open TASEP, should hold. This reveals a type of universality, discussed in Ref.~\cite{tirtha-qkpz} in a partially related context. In contrast, if the modifications in $f$ and $g$ are such that these become nonmonotonic functions of their arguments, or introduce new constraints to the model, e.g., bringing in capacity restrictions, the phase diagrams should change significantly, altering their topology, see, e.g., Ref.~\cite{sourav-1}.}

%We have used analytical MFT, supplemented by extensive MCS studies to obtain the stationary densities and the phase diagrams.  In the weak coupling limit, we find significant disagreements between the MFT and MCS results for intermediate values of $\mu$. Such disagreements disappear in the strong coupling limit of the model. We further find that the domain wall can delocalize for some choices of $k_{10},k_{20}$, but does not delocalize for other choices even for sufficiently larger values of $\mu$. We are able to explain these results in terms of a set of heuristic arguments that show the crucial role of fluctuations that survive in the thermodynamic limit in the weak coupling limit, but not in the strong coupling case. 
 The failure of the MFT in the weak coupling limit suggests using more sophisticated methods that can account for fluctuations, neglected in MFT, in a more quantitative way, giving a firmer basis to our heuristic arguments in Sec.~\ref{role of fluctuations}. A particularly good candidate is macroscopic fluctuation theory~\cite{fluc-th,bal-fluc-th}, that can  provide
a unified macroscopic treatment of such states for driven diffusive systems. %This was subsequently extended to ballistic systems~\cite{bal-fluc-th}.  
A complete treatment of the present model within macroscopic fluctuation theory should require combining and further extending the approaches developed in Refs.~\cite{fluc-th,bal-fluc-th}, since our model has both directed and diffusive (particle exchanges) motion. This remains a theoretically challenging task at present. %We hope our studies here will provide impetus to future research in this line.}

Our studies can be extended in various ways. One can consider several TASEP lanes connecting the two reservoirs. More interestingly, one can consider a ``mixed model'', where one of the reservoirs has the structure of the reservoirs in Ref.~\cite{sourav-1}, and the other has a form similar to the present study. This can lead to possible unexpected competitions between the reservoir structures and the resulting phase diagrams. We hope our studies here will stimulate further research along these lines.

\section{Acknowledgement}

A.B. thanks SERB (DST), India for partial financial support through the CRG scheme [file:
CRG/2021/001875].

 \appendix

 \section{STRONG COUPLING LIMIT}
 \label{mft strong coupling}

 In this section, we focus on the strong coupling limit of the model, and study the phase diagrams and density profiles. The strong coupling limit is characterized by the particle exchange rates independent of system size $L$, i.e., $k_{1},k_{2} \sim \mathcal{O}(1)$, see Sec.~\ref{Model}. %This means the direct exchange of particles between reservoirs dominates over the particle transport via hopping through the TASEP. Therefore, in this limit, the reservoir populations $N_{1}$ and $N_{2}$ maintain a fixed ratio for a given set of particle exchange rates. This considerably simplifies the corresponding MFT calculations, which agrees quantitatively with the MCS results, contrasting the weak coupling limit.

 \subsection{Phase diagrams and steady-state densities in the strong coupling case}
 \label{MF pd strong coupling}

 \begin{figure*}[htb]
 \includegraphics[width=\linewidth]{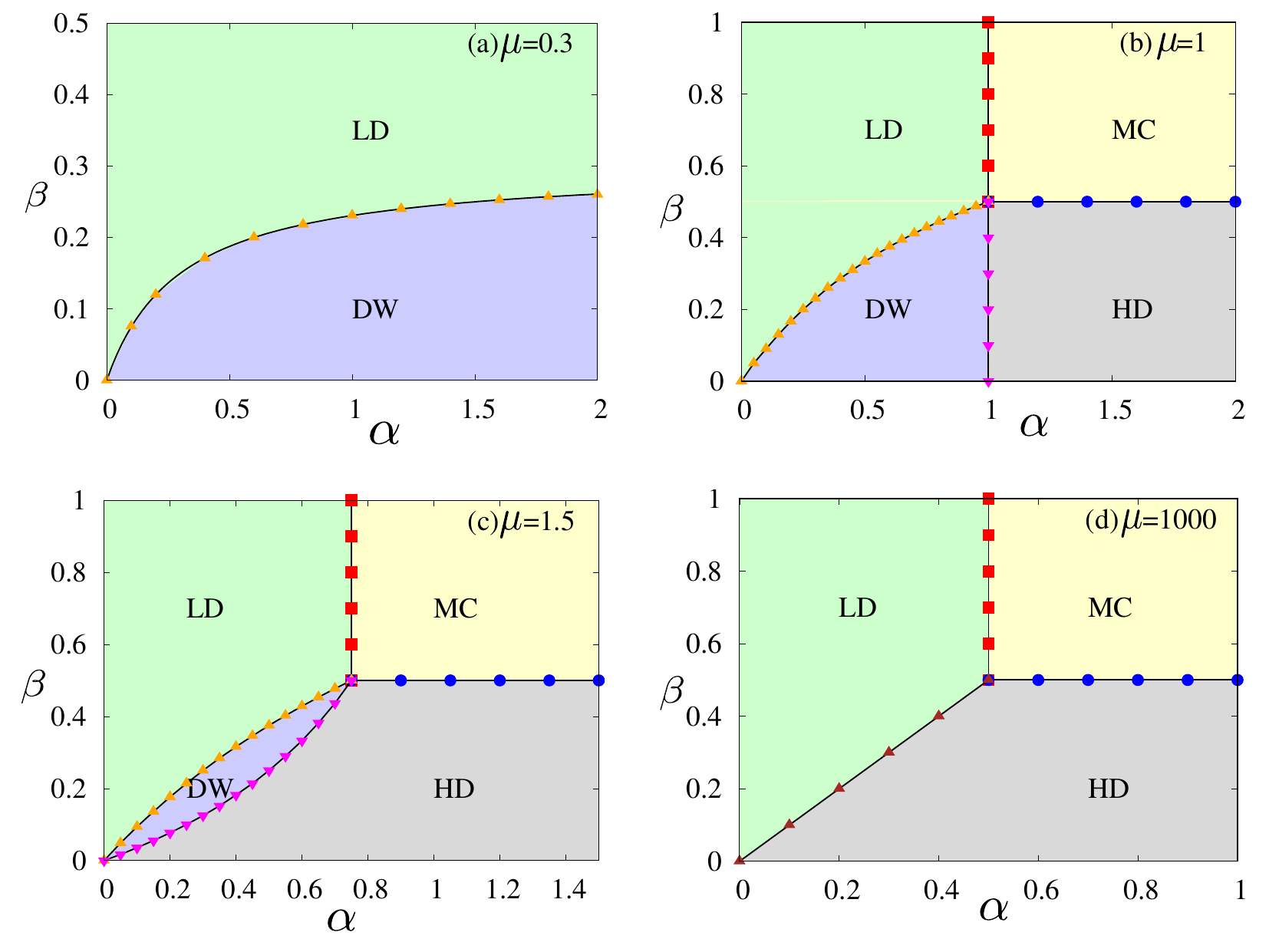}
 \\
\caption{Phase diagrams  in the $\alpha-\beta$ plane in the strong coupling limit of the model for different values of the filling factor $\mu$, (a) $\mu=0.3$, (b) $\mu=1$, (c) $\mu=1.5$, and (d) $\mu=1000$, with  $k_{1}=0$ and $k_{2}=0.95$. MFT predicts four distinct phases (LD, HD, MC, and DW), represented by the green, gray, yellow, and blue regions, respectively, with black solid lines separating them, see Eqs.~(\ref{ldmc strong}), (\ref{hdmc strong}), (\ref{ldsp boundary strong}), and (\ref{hdsp boundary strong}) for MFT phase boundaries. The phase boundaries obtained from MCS are depicted by colored points: red (LD-MC), blue (HD-MC), orange (LD-DW), and magenta (HD-DW). The MFT and MCS results exhibit excellent match. In the limit of large $\mu$, see Fig.~\ref{pd strong}(d), the phase diagram of the present model is identical to the familiar phase diagram of an open TASEP. }
\label{pd strong}
\end{figure*}

 Phase diagrams in the strong coupling limit are shown in Fig.~\ref{pd strong} for a set of representative values of $\mu$ with $k_{1}=0$ and $k_{2}=0.95$. As explained in Sec.~\ref{steady-state density model N0} above, in the strong coupling limit
\begin{equation}
 \label{n1byn2-st}
 \frac{N_{1}}{N_{2}}=\frac{k_{2}}{k_{1}}
\end{equation}
in the thermodynamic limit ($L \rightarrow \infty$). To calculate the steady-state densities in the different phases in the strong coupling limit, we follow the logic and calculational scheme outlined above for the weak coupling limit of the model.

 \subsubsection{ Low-density phase}
 \label{ld phase strong}
  We start by using Eq.~(\ref{rhold quadratic}). %Last part in the right-hand side of Eq.~(\ref{rhold quadratic}) is $\mathcal{O}(1/L)$ and hence can be neglected in the thermodynamic limit ($L\rightarrow \infty$). 
  When $L\rightarrow \infty$, the quadratic equation in (\ref{rhold quadratic}) becomes linear in $\rho_\text{LD}$, solving which we obtain
\begin{equation}
 \label{ld density strong}
  \rho_\text{LD}=\frac{\alpha k_{2}}{k_{1}+\bigg(1+\frac{\alpha}{\mu}\bigg)k_{2}},
 \end{equation}
 in the limit of large $L$, which is.
 %We find $\rho_\text{LD}$ to be 
 independent of $\beta$. Considering Eq.~(\ref{ld density strong}) we find the following:
  \begin{eqnarray}
  &&\rho_\text{LD} \xrightarrow{\mu \rightarrow 0} 0, \label{ld-st-mu0}\\
  &&\rho_\text{LD} \xrightarrow{\mu \rightarrow \infty} \frac{\alpha k_{2}}{k_{1}+k_{2}}. \label{ld-st-mu-infty}
 \end{eqnarray}
  Thus, for $\mu \rightarrow \infty$, we have $\rho_\text{LD}\rightarrow\alpha/2$ when $k_{1}=k_{2}$ and $\rho_\text{LD}\rightarrow\alpha$ when $k_{1}=0$.

  Following the logic outlined in Sec.~\ref{ssd and pb}, we %calculate the reservoir populations $N_{1}$ and $N_{2}$. We 
  find
\begin{eqnarray}
  &&N_{1}=\frac{Lk_{2} \mu}{k_{1}+(1+\frac{\alpha}{\mu})k_{2}}, \label{N1 strong}\\
  &&N_{2}=\frac{Lk_{1} \mu}{k_{1}+(1+\frac{\alpha}{\mu})k_{2}}, \label{N2 strong}
 \end{eqnarray}
%As previously stated in Eq.~\ref{n1byn2-st}, the relative population of the reservoirs is determined solely by the ratio of particle exchange rates in the strong coupling limit: 
in agreement with $N_{1}/N_{2}=k_{2}/k_{1}$. This effectively eliminates one of the reservoirs in the strong coupling limit.

\begin{figure*}[htb]
 \includegraphics[width=\linewidth]{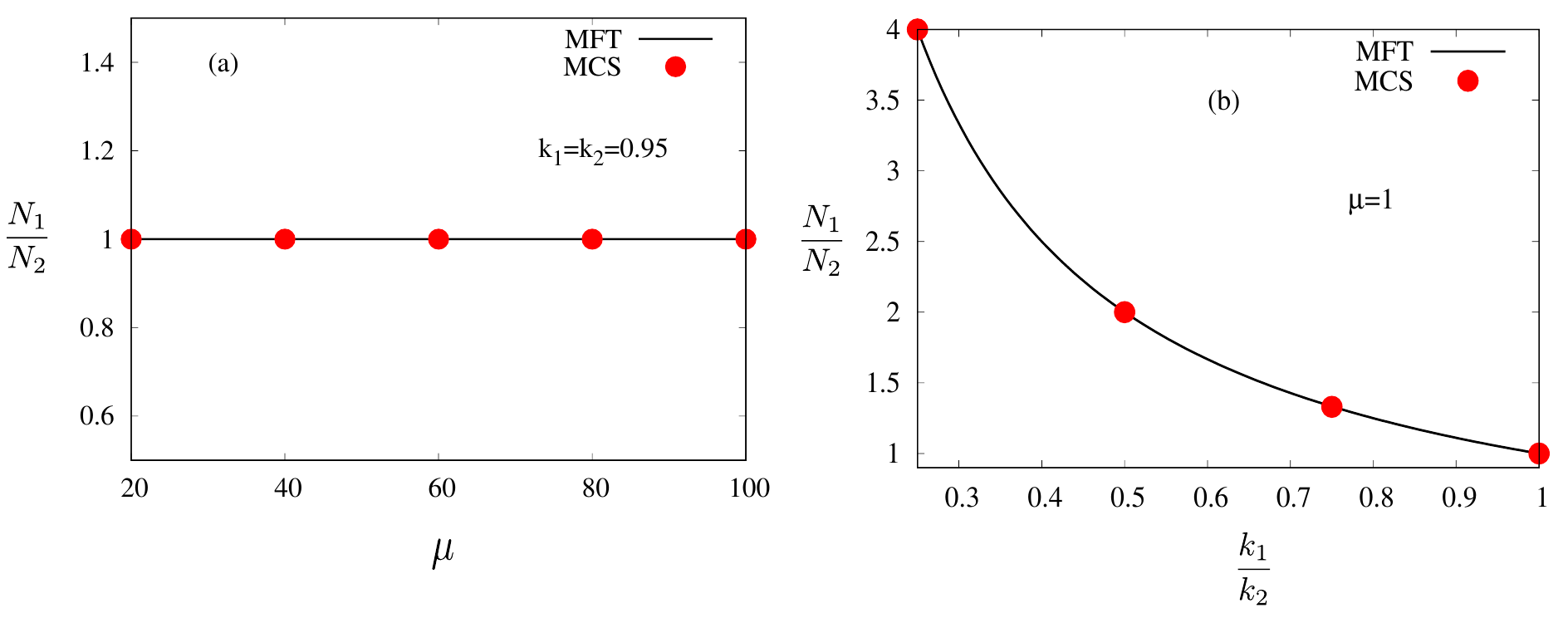}
 \\
\caption{(a) Plot showing dependence of the reservoir population ratio $N_{1}/N_{2}$ on $\mu$ for fixed exchange rate $k_1=k_2=0.95$ in the strong coupling limit of the model. (b) Plot presenting the variation of the reservoir population ratio $N_{1}/N_{2}$ with  $k_{1}/k_{2}$ for a fixed filling factor $\mu=1$ in the strong coupling limit. In these plots, the entry and exit rates are carefully selected to ensure that the TASEP lane remains in the LD phase.  The values of $\alpha=0.1$ and $\beta=1$ are used. The results obtained by MFT (see Eqs.~(\ref{N1 strong}) and (\ref{N2 strong})) and MCS are in good agreement. }
\label{n1/n2-vs-mu-and-k1/k2}
\end{figure*}

Similar to the weak coupling limit, LD phase can be present for any (positive) value of $\mu$ in the strong coupling limit. This is demonstrated in Fig.~\ref{pd strong}, where the LD phase can be found for small $\mu$ values as well as large $\mu$ values. Since $\rho_\text{LD}<1/2$, one gets the following as a condition for LD phase existence according to MFT:
\begin{equation}
 \label{alpha-less-than-ld-st}
 \alpha < \frac{k_{1}+k_{2}}{\bigg(2-\frac{1}{\mu}\bigg)k_{2}}.
\end{equation}
This is illustrated in Fig.~\ref{pd strong}. For instance, when $k_{1}=0$, $k_{2}=0.95$, and $\mu=1$ (Fig.~\ref{pd strong}(b)), the condition (\ref{alpha-less-than-ld-st}) infers an LD phase for $\alpha<1$. Similarly when $\mu$ is very large, say $\mu=1000$ (Fig.~\ref{pd strong}(d)), the condition (\ref{alpha-less-than-ld-st}), LD phase is limited up to $\alpha=0.5$ along the $\alpha$-axis.

%In the phase diagrams of Fig.~\ref{pd strong}, the occurrence of LD phase is shown for different $\mu$. Clearly, the LD phase occurs for all $\mu$ values, ranging from a low value (such as $\mu=0.3$) to a high value (such as $\mu=1000$). %{\cred As $\mu$ increases significantly when $k_{1}=0$, the LD-MC boundary gradually converges to the line $\alpha=1/2$ from above, consistent with the condition (\ref{ld-mu-up-pos-st})}.

 \subsubsection{{ High-density phase}}
 \label{hd phase strong}
 We again proceed with the reasoning presented in the weak coupling limit; see Sec.~\ref{ssd and pb}. Neglecting contributions that are $\mathcal{O}(1/L)$ in Eq.~(\ref{hd-quad}) in the thermodynamic limit, the HD phase density $\rho_\text{HD}$ in the strong coupling limit is
\begin{equation}
 \label{hd density strong}
  \rho_\text{HD} = \frac{k_{1}+(1-\beta)k_{2}}{\bigg(1+\frac{\beta}{\mu}\bigg)k_{1}+k_{2}},
 \end{equation}
 in the limit of large $L$. As expected, $\rho_\text{HD}$ is independent of $\alpha$. We find from Eq.~(\ref{hd density strong}) that
 \begin{equation}
  \rho_\text{HD} \xrightarrow{\mu \rightarrow \infty} \bigg(1-\frac{\beta k_{2}}{k_{1}+k_{2}}\bigg), \label{hd-st-mu-infty}
 \end{equation}
 which for the two cases $k_{1}=k_{2}$ and $k_{1}=0$, yields $\rho_\text{HD}\rightarrow (1-\beta/2)$ and $\rho_\text{HD}\rightarrow (1-\beta)$, respectively.
The reservoir populations in the HD phase are
 \begin{eqnarray}
  &&N_{1}=\frac{Lk_{2}(\mu-1+\beta)}{\bigg(1+\frac{\beta}{\mu}\bigg)k_{1}+k_{2}}, \label{N1-strong-hd}\\
  &&N_{2}=\frac{Lk_{1}(\mu-1+\beta)}{\bigg(1+\frac{\beta}{\mu}\bigg)k_{1}+k_{2}}, \label{N2-strong-hd}
 \end{eqnarray}
%Again, the ratio of reservoir populations depends only on the exchange rates: 
in agreement with $N_{1}/N_{2}=k_{2}/k_{1}$.

%To determine the lower threshold of $\mu$ below which the HD phase cannot be sustained due to particle shortage in the TASEP lane, we consider $\rho_\text{HD}>1/2$. However, no finite upper threshold of $\mu$ can be found for HD phase existence. Following is the range of $\mu$ over which HD phase appears:

%\begin{equation}
%\label{hd-mu-range}
%\bigg(\frac{k_{1}}{\frac{k_{1}+k_{2}}{\beta}-2 k_{2}}\bigg) < \mu < \infty.
%\end{equation}
 Similar to the weak coupling limit case, the HD phase exists for arbitrarily large $\mu$; see the phase diagrams in Fig.~\ref{pd strong}. When $\mu$ is sufficiently small, such as $\mu=0.3$, HD phase is absent due to an inadequate number of particles being available. {Since $\rho_\text{HD}>1/2$, we get the following condition for HD phase by using Eq.~(\ref{hd density strong}):} %{\bf why do you say above that is in the MC phase section below?}
\begin{equation}
 \label{beta-less-than-hd-st}
 \beta < \frac{k_{1}+k_{2}}{2k_{2}+\frac{k_{1}}{\mu}},
\end{equation}
which implies that, for a fixed value of $k_{1}$, $k_{2}$, and $\mu$ HD phase exists if $\beta$ satisfies the above condition. This can be readily seen in the phase diagrams of Fig.~\ref{pd strong} where we have taken $k_{1}=0,k_{2}=0.95$. In this case, HD phase lies under the line $\beta=1/2$ for any value of $\mu$.

%In the phase diagrams of Fig.~\ref{pd strong}, the occurrence of HD phase is displayed. When the value of $\mu$ is sufficiently small, such as $\mu=0.3$, HD phase is absent due to an inadequate number of particles being available. As $\mu$ is raised significantly to a larger value while keeping $k_{1}=0$, the HD-MC boundary gradually converges to $\beta=1/2$ line from below, consistent with the condition (\ref{hd-mu-low-pos-st}).

 \subsubsection{ Maximal current phase}
 \label{mc phase strong}

\begin{figure*}[htb]
 \includegraphics[width=\linewidth]{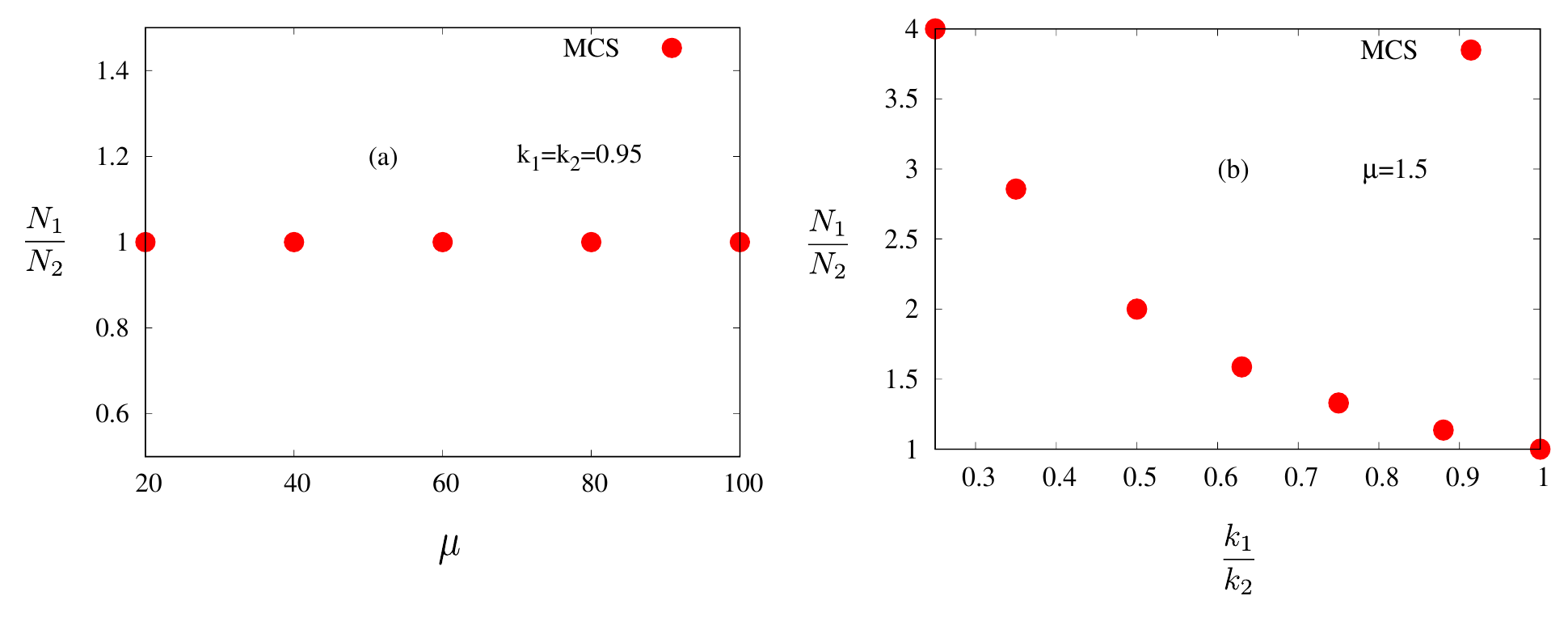}
 \\
\caption{{(a) Plot showing dependence of the reservoir population ratio $N_{1}/N_{2}$ on $\mu$ for fixed exchange rate $k_1=k_2=0.95$ in the strong coupling limit of the model. Entry and exit rate parameters $\alpha=\beta=3$ are chosen such that the system is in MC phase. (b) Plot of $N_{1}/N_{2}$ vs $k_{1}/k_{2}$ for a fixed filling factor $\mu=1.5$ in the strong coupling limit. $\alpha=1.2$ and $\beta=0.8$ are chosen so that the system is in the MC phase.}}
\label{pop_ratio_mc_st}
\end{figure*}

 In the MC phase, $\rho_\text{MC}=1/2$ in the bulk of the TASEP lane. Particle number conservation gives $N_{0}=N_{1}+N_{2}+L/2$ in this phase. The maximum current $J_{\text{max}}$ associated with the MC phase is given by $J_{\text{max}} =\rho(1-\rho)= 1/4$. Analogous to an open TASEP, we must have %following are the conditions for the existence of the  MC phase in the TASEP lane:
 \begin{eqnarray}
  &&\alpha_\text{eff}=\alpha\frac{N_{1}}{N_{0}}>\frac{1}{2},\\
  &&\beta_\text{eff}=\beta\bigg(1-\frac{N_{2}}{N_{0}}\bigg)>\frac{1}{2}
 \end{eqnarray}
 for the MC phase.
 The boundaries between the LD and MC phases, and between the HD and MC phases can be determined by %substituting the given values of 
 setting $\rho_\text{LD}$ and $\rho_\text{HD}=1/2$ into Eqs.~(\ref{ld density strong}) and (\ref{hd density strong}), respectively, giving %. We thus obtain following equations as the LD-MC and HD-MC boundaries respectly:
 \begin{eqnarray}
  &&\alpha=\frac{\mu(k_{1}+k_{2})}{(2\mu-1)k_{2}}, \label{ldmc strong}\\
  &&\beta=\frac{\mu(k_{1}+k_{2})}{k_{1}+2\mu k_{2}}, \label{hdmc strong}
 \end{eqnarray}
 as the conditions for the the LD-MC and HD-MC boundaries respectively.
  %Considering the non-negativity of 
  Since $\alpha>0$ from Eq.~(\ref{ldmc strong}) the MC phase should exist for any $\mu>1/2$ in the strong coupling limit. Furthermore, $\beta$ in Eq.~(\ref{hdmc strong}) is always positive. %This infers no upper threshold of $\mu$ for MC phase to exist. Following is the range:
  Thus
 \begin{equation}
  \label{mc-range-st}
  \frac{1}{2}<\mu<\infty
 \end{equation}
 for the existence of the MC phase, which is
%Following the same line of argument as given in the weak coupling limit, (\ref{mc-range-st}) is 
also the range of $\mu$ within which HD phase appears in the phase diagram, see Fig.~\ref{pd strong}.
See the phase diagrams in Fig.~\ref{pd strong}, where the MC phase does not appear for $\mu<1/2$, such as $\mu=0.3$, in agreement with the discussions above. Also the variation of $N_{1}/N_{2}$ either with $\mu$ keeping exchange rates fixed or with $k_{1}/k_{2}$ keeping $\mu$ fixed are depicted in Figs.~\ref{pop_ratio_mc_st}(a) and \ref{pop_ratio_mc_st}(b) respectively.

 \subsubsection{ Domain wall phase}
 \label{sp phase strong}

 %In addition to the above phases with uniform density, we further find a domain wall (DW) phase with a non-uniform density,  where the 
 In the domain wall (DW) phase, LD and HD phases are connected by a domain wall in the bulk of the TASEP lane. %The precise location of this DW can be obtained by particle number conservation, inferring it as an LDW. At this point, 
 We determine the position $x_{w}$ and height $\Delta$ of the LDW in a similar manner as done for the weak coupling limit case. By neglecting terms of order ${\cal O}(1/L)$ in the thermodynamic limit, we can solve the coupled Eqs.~(\ref{First coupled equation model 2}) and (\ref{Second coupled equation model 2}) to obtain expressions for $N_{1}/N_{0}$ and $x_{w}$. The solutions in the strong coupling limit are as follows:
 \begin{eqnarray}
  &&\frac{N_{1}}{N_{0}}=\frac{k_{2}}{k_{1}+\frac{\alpha}{\beta}k_{2}}, \label{n1/n0 strong}\\
  &&x_{w}=\frac{1-\bigg(\frac{k_{2}}{k_{1}+\frac{\alpha}{\beta}k_{2}}\bigg)\bigg[\mu\bigg(\frac{\alpha}{\beta}-1\bigg)+\alpha\bigg]}{1-2\alpha \frac{k_{2}}{k_{1}+\frac{\alpha}{\beta}k_{2}}}. \label{xw strong}
 \end{eqnarray}

 \begin{figure}[!h]
 \centering
 \includegraphics[width=\columnwidth]{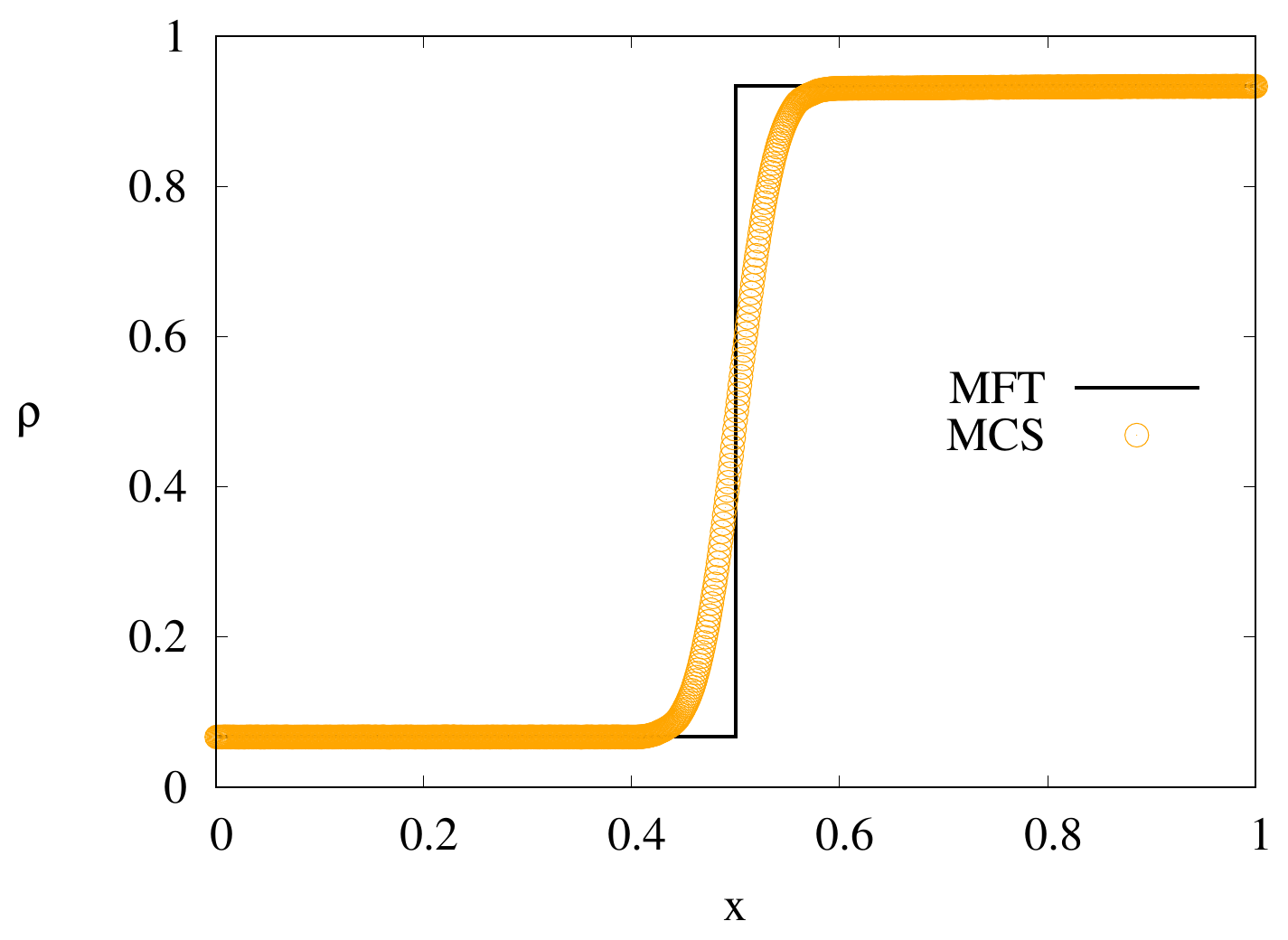}
 \caption{Steady-state density profile in the DW phase in the strong coupling limit of the model. Model parameters are: $\mu=3/2$, $k_{1}=k_{2}=0.7$, $\alpha=0.2$ and $\beta=0.1$. MCS results agree well with the MFT predictions given by Eqs.~(\ref{xw strong}), (\ref{rhold-dw-st}), and (\ref{rhohd-dw-st}).}
 \label{st_ldw}
 \end{figure}

 In the DW phase, the reservoir populations $N_{1}$ and $N_{2}$ are related by the condition for a DW, $\alpha_\text{eff}=\beta_\text{eff}$ or $\alpha N_{1}/N_{0}=\beta(1-N_{2}/N_{0})$, giving
 \begin{equation}
  \label{n2-st}
  \frac{N_{2}}{N_{0}}=1-\frac{\alpha}{\beta}\frac{N_{1}}{N_{0}}=1-\frac{k_{2}}{\frac{\beta}{\alpha}k_{1}+k_{2}}.
 \end{equation}
 The steady-state low and high densities in the DW phase are
\begin{align}
  &\rho_\text{LD}=\alpha \frac{N_{1}}{N_{0}}=\frac{\alpha k_{2}}{k_{1}+\frac{\alpha}{\beta}k_{2}}, \label{rhold-dw-st} \\
  & \rho_\text{HD}=1-\rho_\text{LD}=1-\frac{\alpha k_{2}}{k_{1}+\frac{\alpha}{\beta}k_{2}}. \label{rhohd-dw-st}
\end{align}
Then the domain wall height $\Delta$ is given by
 % Following is the expression for domain wall height ($\Delta$):
 \begin{equation}
 \label{delta strong}
 \Delta=\rho_\text{HD}-\rho_\text{LD}=1-\frac{2\alpha \beta k_{2}}{\beta k_{1}+\alpha k_{2}}.
 \end{equation}
The expression of $\Delta$ in the strong coupling limit, as given in Eq.~(\ref{delta strong}), depends only on parameters $\alpha$, $\beta$, $k_{1}$, and $k_{2}$. Notably, it is independent of the filling factor $\mu$.

\begin{figure*}[htb]
 \includegraphics[width=\columnwidth]{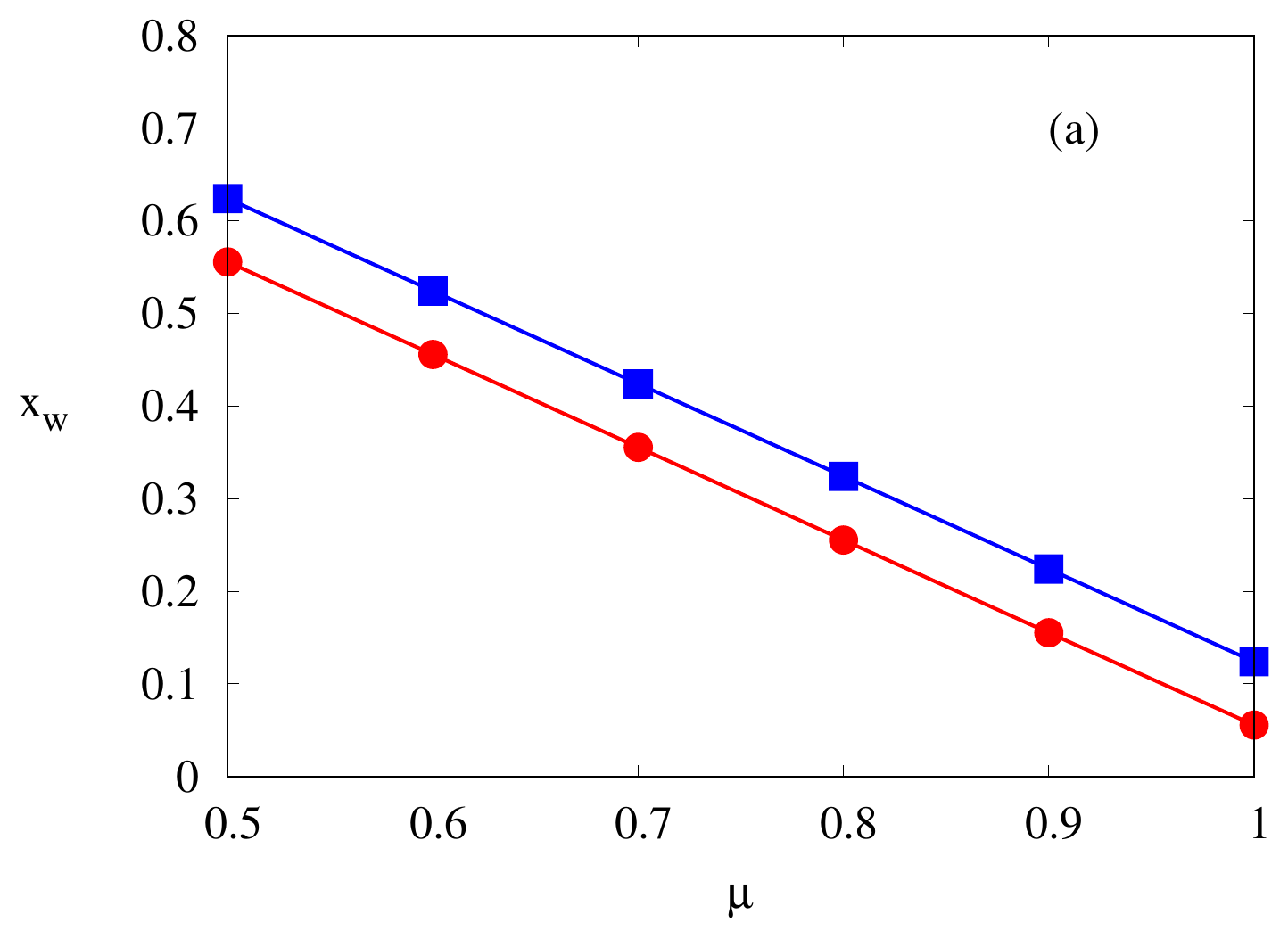}
 \hfill
 \includegraphics[width=\columnwidth]{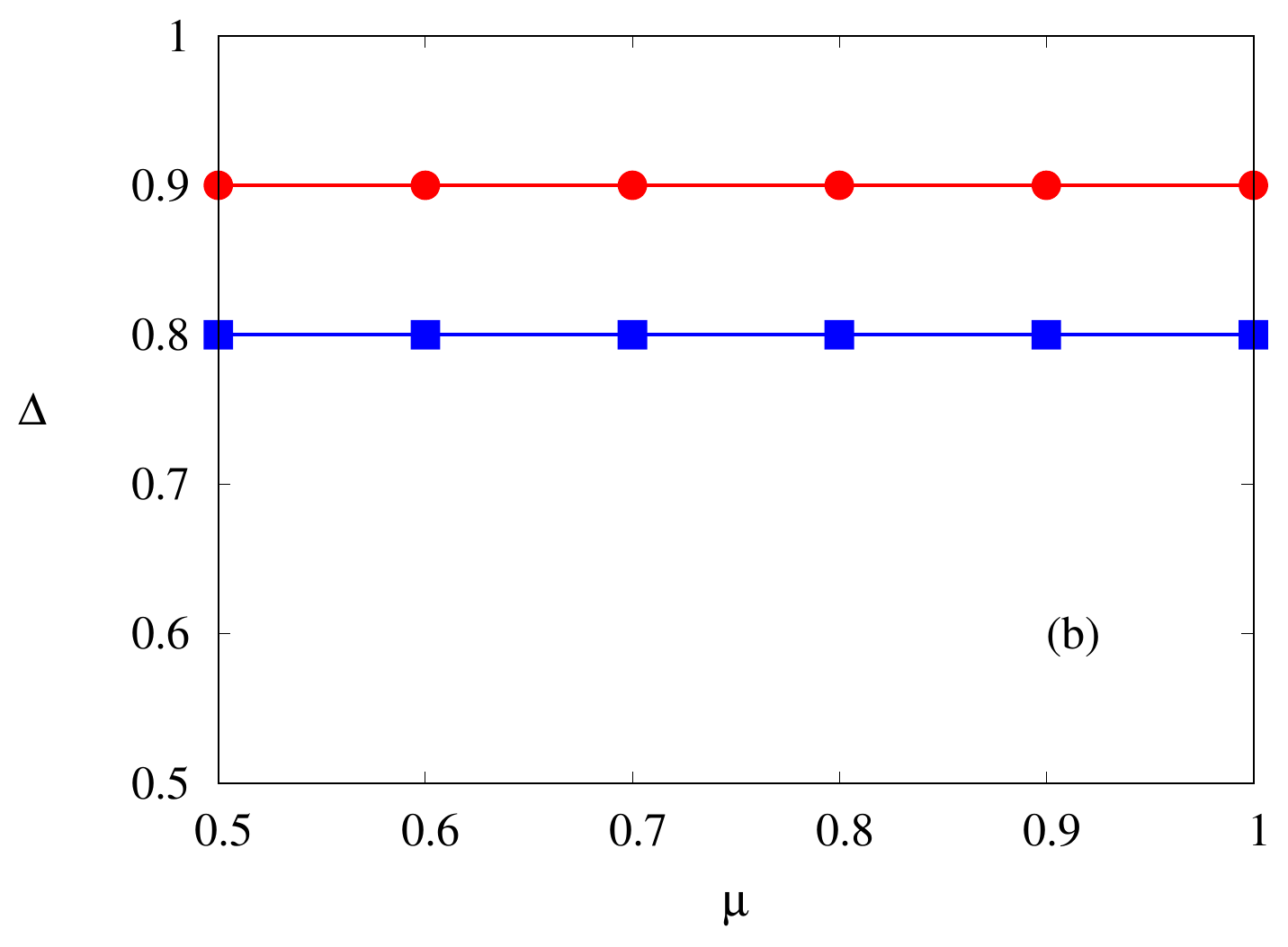}
 \\
\caption{(a) Plots of DW position ($x_{w}$) vs the filling factor ($\mu$) in the strong coupling limit of the model. The exchange rates are set to $k_{1}=0$ and $k_{2}=0.95$, while two different sets of entry and exit rate parameters are considered: $\alpha=0.5$ with $\beta=0.05$ (red circles) and $\alpha=0.5$ with $\beta=0.1$ (blue squares). Solid lines represent the predictions from MFT [see Eq.~(\ref{xw strong})], while the discrete points correspond to the results obtained through MCS. Within the given range of $\mu$, the $x_{w}$ linearly decreases with increasing $\mu$.
(b) Plots of DW height ($\Delta$) as a function of the filling factor ($\mu$) in the strong coupling limit of the model. The parameter values for this plot are the same as those used in the plot Fig.~\ref{xw-del-vs-mu-strong}(a), with matching colors and symbols. Notably, $\Delta$ remains unchanged as $\mu$ varies within the examined range, see Eq.~(\ref{delta strong}). Both MFT and MCS results are in good agreement.
}
\label{xw-del-vs-mu-strong}
\end{figure*}

\begin{figure*}[htb]
 \includegraphics[width=\columnwidth]{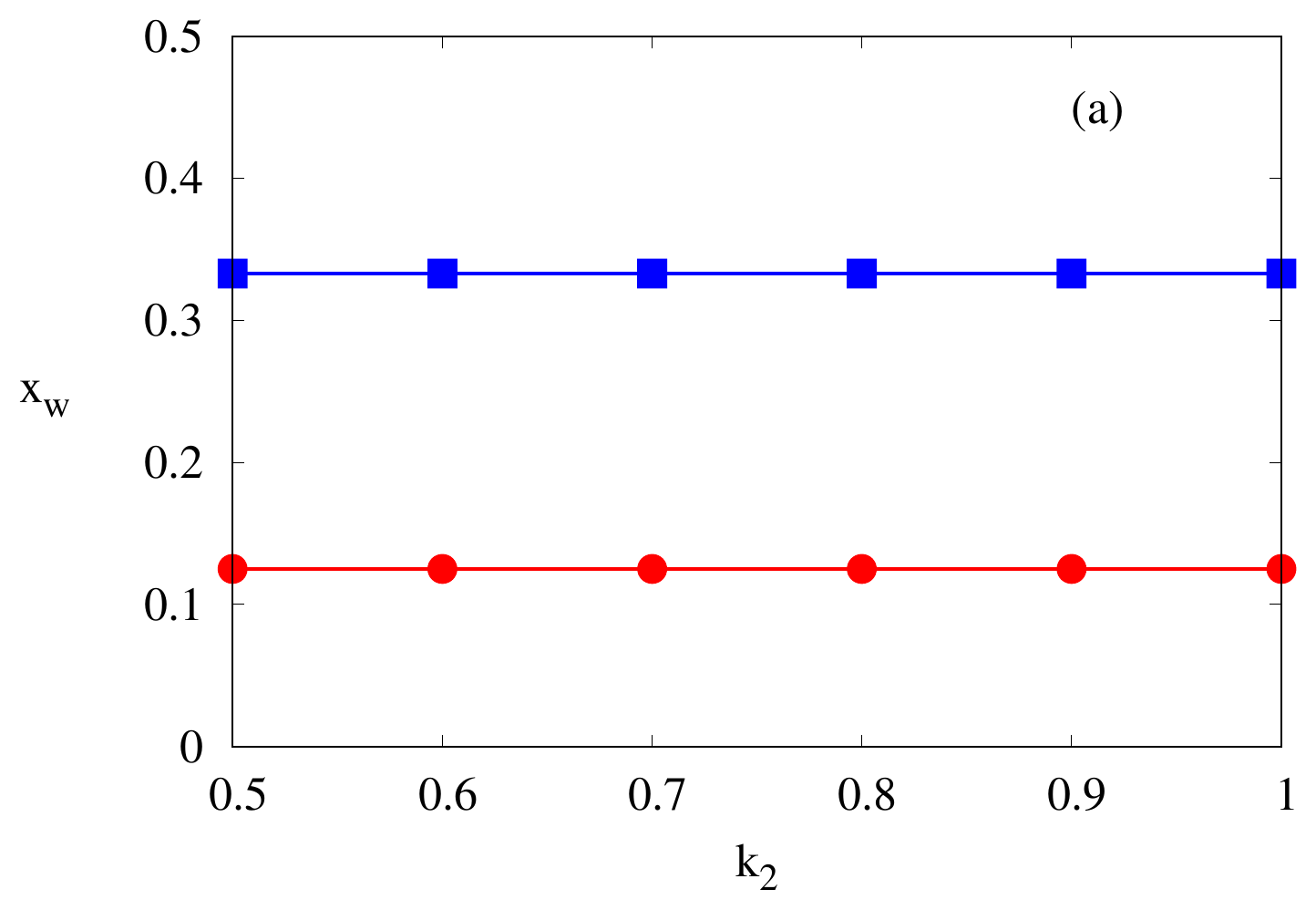}
 \hfill
 \includegraphics[width=\columnwidth]{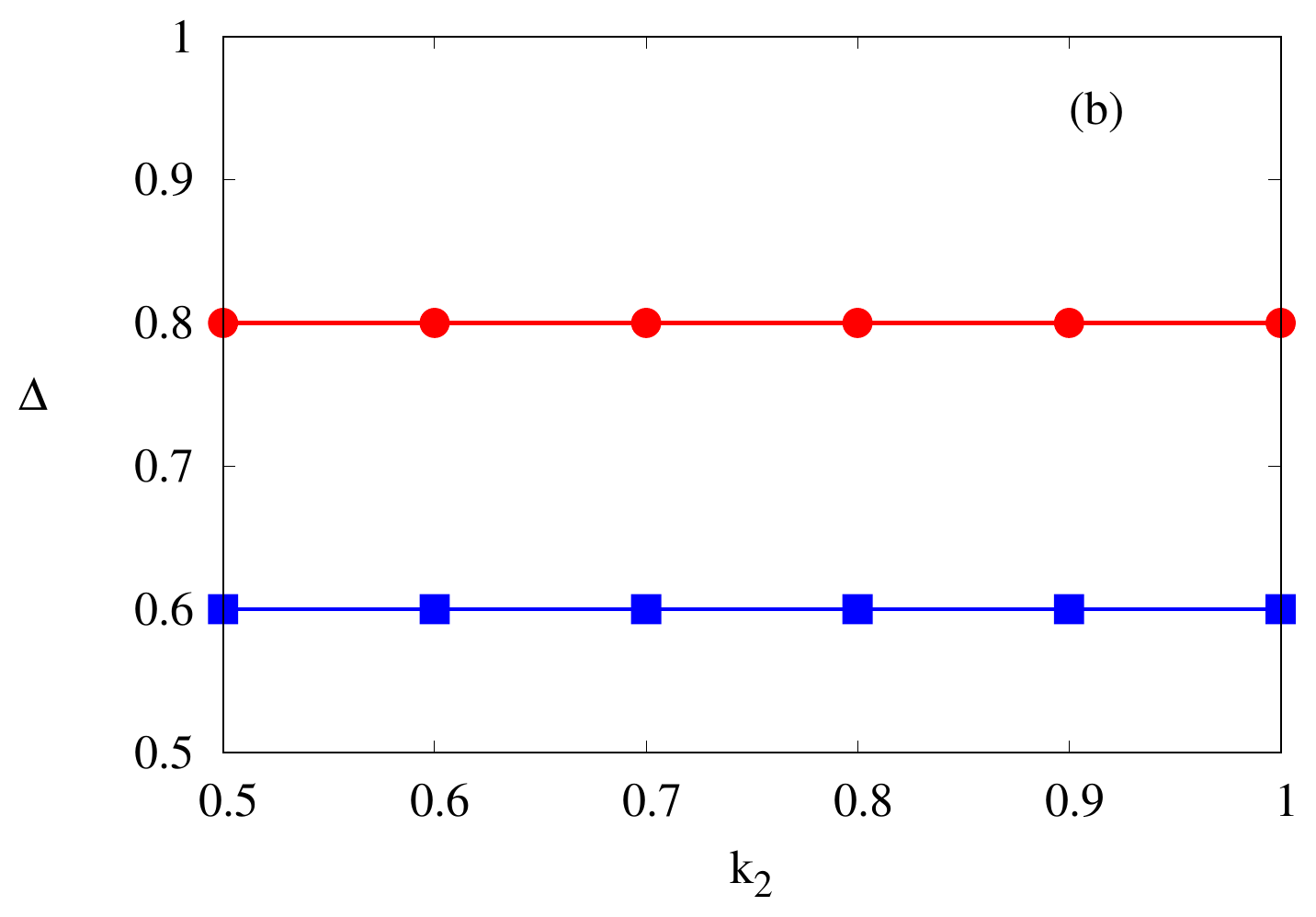}
 \\
\caption{(a) Plots of DW position ($x_w$) vs $k_{2}$ in the strong coupling limit of the model when $k_{1}=0$. The plot with red circles corresponds to $\alpha=0.5$ and $\beta=0.1$, while the plot with blue squares represents $\alpha=0.5$ and $\beta=0.2$. The position of DW does not vary with $k_{2}$, see Eq.~(\ref{xw strong}). (b) Plots of DW height ($\Delta$) vs $k_{2}$ in the strong coupling limit with $k_{1}=0$. The red circular and blue squared plots correspond to the same parameter values as in the Fig.~\ref{xw-del-vs-k20-strong}(a).  MCS outcomes satisfy MFT predictions, see Eq.~(\ref{delta strong}).}
\label{xw-del-vs-k20-strong}
\end{figure*}

In Fig.~\ref{xw-del-vs-mu-strong}, the plots illustrates a linear decrease in the position $x_{w}$ of the DW as $\mu$ increases, while the height $\Delta$ of the DW remains constant throughout the range of $\mu$ values. On the other hand, in Fig.~\ref{xw-del-vs-k20-strong}, both the position and height of the DW exhibit no significant variation with $k_{2}$.

Following the same arguments given in weak coupling limit of the model, the range of $\mu$ over which DW phase exists can be identified in strong coupling limit. Using the fact that DW position, $x_{w}$, should be between 0 and 1, we find this range as below:
\begin{equation}
 \label{range-of-mu-dw-st}
 \bigg(\frac{\alpha}{\frac{\alpha}{\beta}-1}\bigg)<\mu<\bigg(\frac{\frac{k_{1}}{k_{2}}+\frac{\alpha}{\beta}-\alpha}{\frac{\alpha}{\beta}-1}\bigg).
\end{equation}

Furthermore, the boundaries between the LD and DW, and between the HD and DW phases can be determined by substituting $x_{w}=1$ and $x_{w}=0$ respectively into Eq.~(\ref{xw strong}). The resulting equations for LD-DW and HD-DW phase boundaries are respectively:

 \begin{eqnarray}
  &&\mu\bigg(1-\frac{\alpha}{\beta}\bigg)+\alpha=0, \label{ldsp boundary strong}\\
  &&\bigg(\frac{k_{2}}{k_{1}+\frac{\alpha}{\beta}k_{2}}\bigg)\bigg[\mu\bigg(1-\frac{\alpha}{\beta}\bigg)-\alpha\bigg]+1=0. \label{hdsp boundary strong}
 \end{eqnarray}

\subsection{Delocalization of LDW}
\label{DDW formation}

\begin{figure*}[htb]
 \includegraphics[width=0.5\textwidth]{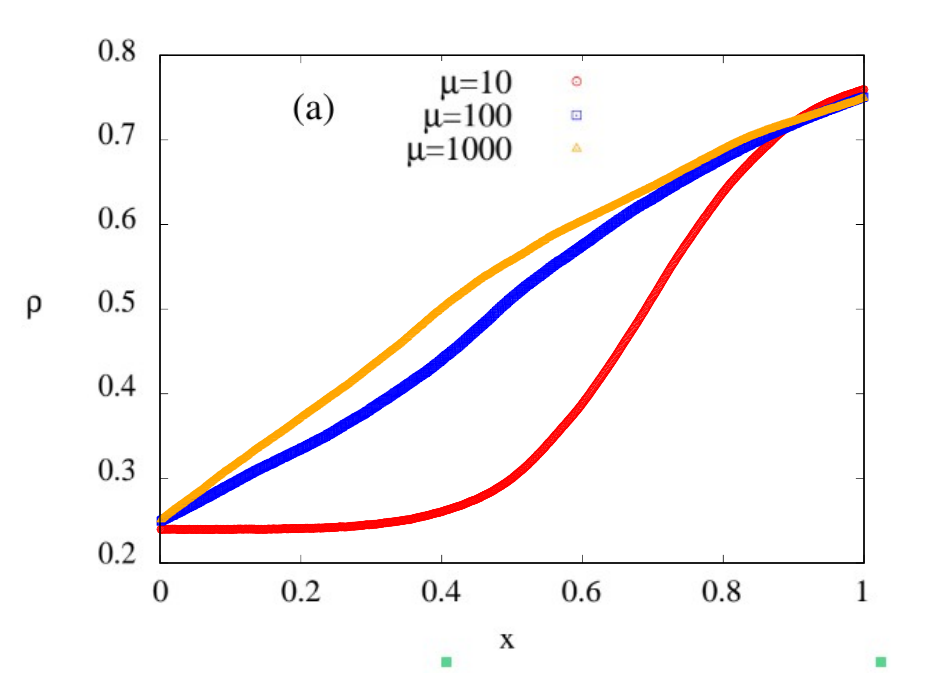}
 \hfill
 \includegraphics[width=0.48\textwidth]{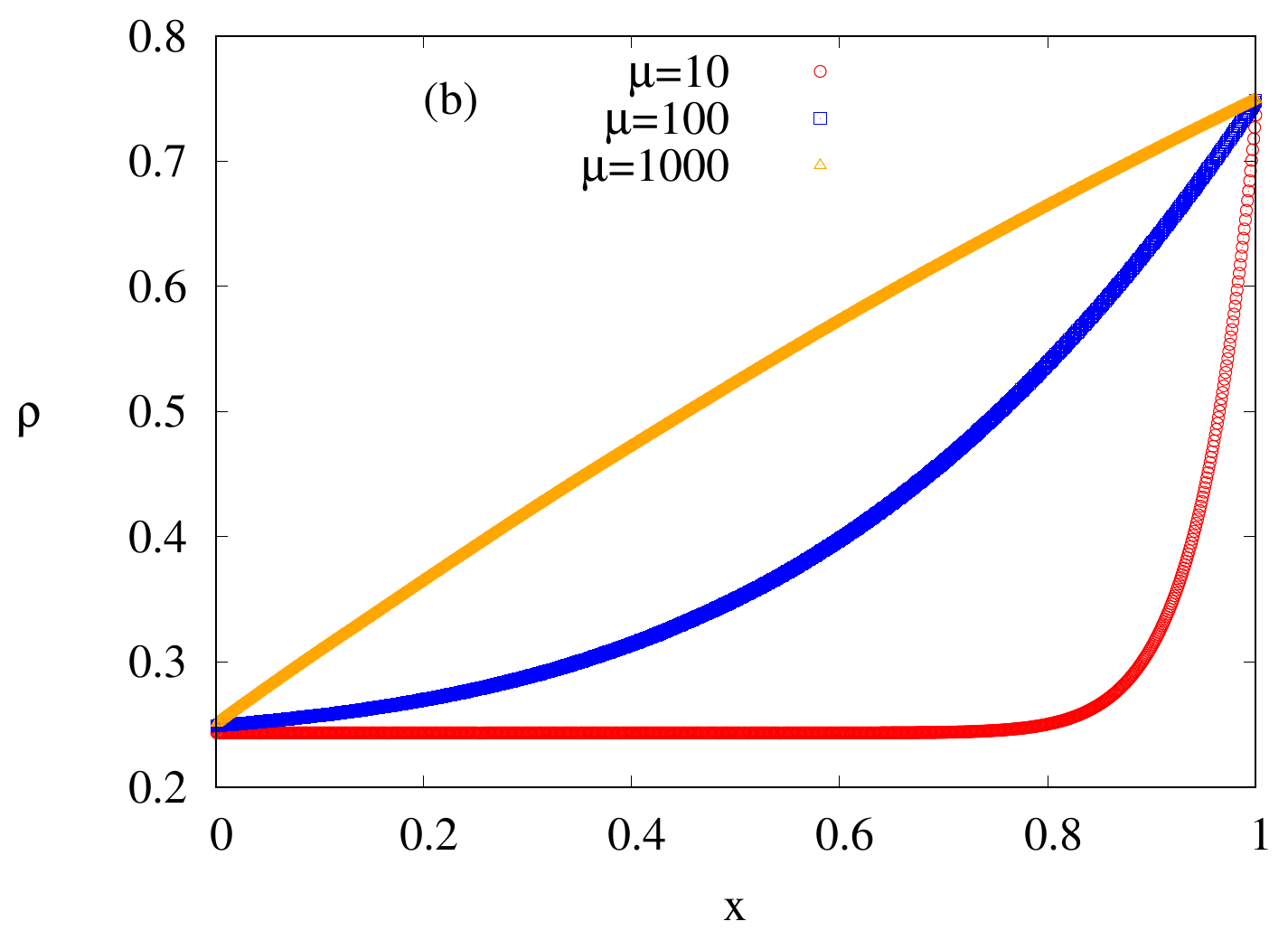}
 \\
\caption{Gradual delocalization of a domain wall with increase in $\mu$. (a) This corresponds to the weak coupling limit. Parameter values are: $\alpha=0.25$, $\beta=0.24$, $\mu=10$, $k_{10}=0$, and $k_{20}=0.95$ (red circles); $\alpha=0.25$, $\beta=0.2488$, $\mu=100$, $k_{10}=0$, and $k_{20}=0.95$ (blue squares); and $\alpha=0.25$, $\beta=0.24985$, $\mu=1000$, $k_{10}=0$, and $k_{20}=0.95$ (orange triangles). (b) This corresponds to the strong coupling limit. Parameter values are: $\alpha=0.25$, $\beta=0.24985$, $\mu=10$, $k_{10}=0$, and $k_{20}=0.95$ (red circles); $\alpha=0.25$, $\beta=0.24985$, $\mu=100$, $k_{10}=0$, and $k_{20}=0.95$ (blue squares); and $\alpha=0.25$, $\beta=0.24985$, $\mu=1000$, $k_{10}=0$, and $k_{20}=0.95$ (orange triangles). For both weak and strong coupling limits, system size $L=1000$ and averages over $2 \times 10^{9}$ times steps are done to get the MCS density profiles. Clearly, the LDW for smaller values of $\mu$ gets delocalized as $\mu \sim L$.}
\label{ddw-scaling}
\end{figure*}

{ Similar to the weak coupling limit, DWs in the strong coupling limit also gradually get delocalized with increasing $\mu$. This is clearly shown in Fig.~\ref{ddw-scaling} where one can see that the domain walls get completely delocalized with $\mu \sim L$, generally similar to what we have found in the weak coupling limit; see Sec.~\ref{deloc} of the main text for related heuristic arguments behind it.}

\subsection{Phase boundaries meet at a common point}
 \label{pb-strong-mcp}

 Similar to the weak coupling limit, for specific values of $\mu$ when all four phases (LD, HD, MC, and DW) exist, they intersect in the $\alpha-\beta$ plane at a single point. This point, which is named (four-phase) multicritical point, can be obtaind as the meeting point of all four phase boundaries in Eqs.~(\ref{ldmc strong}), (\ref{hdmc strong}), (\ref{ldsp boundary strong}), and (\ref{hdsp boundary strong}):
 \begin{equation}
  \label{mp strong}
  (\alpha_{c},\beta_{c})=\bigg(\frac{\mu(k_{1}+k_{2})}{(2\mu-1)k_{2}},\frac{\mu(k_{1}+k_{2})}{k_{1}+2\mu k_{2}}\bigg).
 \end{equation}
 Depending explicitly on $k_{1}$, $k_{2}$, and $\mu$ this unique point exists for any value of $\mu$ greater than 1/2, above which all four phases appear in the phase diagram, see Fig.~\ref{pd strong}. The distance between the origin (0,0) and the multicritical point ($\alpha_{c},\beta_{c}$) can be calculated using the formula:
 \begin{equation}
  \label{d strong}
  d=\sqrt{\bigg(\frac{\mu(k_{1}+k_{2})}{(2\mu-1)k_{2}}\bigg)^{2}+\bigg(\frac{\mu(k_{1}+k_{2})}{k_{1}+2\mu k_{2}}\bigg)^{2}}.
 \end{equation}
 { See section~\ref{phase-meet-weak} for similar results in the weak coupling case.}
 %As $\mu$ approaches 1/2, the distance $d$ tends to diverge. As $\mu \rightarrow \infty$, the multicritical point approaches ($1/2,1/2$) and its distance from the origin becomes $d=1/\sqrt{2}$, similar to an open TASEP.

 %\section{VARIATION OF THE RESERVOIR POPULATIONS}
 %\label{reser-popu-st}

%Fig.~\ref{n1/n2-vs-mu-and-k1/k2} illustrates the relationship between the reservoir population ratio ($N_{1}/N_{2}$) and either the filling factor ($\mu$) with constant exchange rates or the ratio of exchange rates ($k_{1}/k_{2}$) with a fixed $\mu$ in the strong coupling limit.

\section{Simulation algorithm}
\label{sim-algo}

In this study, the   results on the stationary densities and phase diagrams of the model were validated through Monte Carlo simulations. The simulation rules are as follows: (i) If the first site ($i=1$) of the TASEP lane ($T$) is empty, a particle from reservoir $R_{1}$ enters $T$ with a rate $\alpha_\text{eff}$; (ii) Particles in $T$ can hop with rate 1 to the next site in the bulk of $T$ provided that site is empty; (iii) Upon reaching the last site ($i=L$) of $T$, a particle exits with a rate $\beta_\text{eff}$ into reservoir $R_{2}$; (iv) Further, particles in the reservoirs ($R_{1}$ and $R_{2}$) can jump directly to the other reservoir with rates $k_{1}$ from $R_{1}$ to $R_{2}$ and $k_{2}$ from $R_{2}$ to $R_{1}$,%; and (v) If, during any iteration, $k_{1}N_{1}$ and/or $k_{2}N_{2}$ are greater than 1, the rates are normalised by dividing each rate by the maximum value between $k_{1}N_{1}$ and $k_{2}N_{2}$.
The above dynamical update rules were implemented with random sequential updates.  After sufficient number of iterations (typically $10 \%$ to $30 \%$ of the total number of time steps) were performed to reach a steady-state, time-averaged density profiles were obtained.

%\pagebreak

\end{document}